\documentclass[12pt]{article}
\pdfoutput=1
\usepackage{jheppub}
\usepackage{ytableau}
\usepackage{comment}
\usepackage[vcentermath]{youngtab}

\newcommand\be{\begin{equation}}
\newcommand\ee{\end{equation}}

\newcommand\Tr{\mathrm{Tr}}

\preprint{RUP-22-17, KIAS-P22059}

\title{$\mathcal{N}=2^{*}$ Schur indices}
\abstract{
We find closed-form expressions for the Schur indices of 4d $\mathcal{N}=2^{*}$ super Yang-Mills theory with unitary gauge groups for arbitrary ranks via the Fermi-gas formulation. 
They can be written as a sum over the Young diagrams associated with spectral zeta functions of an ideal Fermi-gas system. These functions are expressed in terms of the twisted Weierstrass functions, generating functions for quasi-Jacobi forms. 
The indices lie in the polynomial ring generated by the Kronecker theta function 
and the Weierstrass functions which contains the polynomial ring of the quasi-Jacobi forms. 
The grand canonical ensemble allows for another simple exact form of the indices as infinite series. 
In addition, we find that the unflavored Schur indices and their limits can be expressed in terms of several generating functions for combinatorial objects, 
including sum of triangular numbers, generalized sums of divisors and overpartitions. 
}
\author[a]{Yasuyuki Hatsuda}
\author[b]{and Tadashi Okazaki}

\emailAdd{yhatsuda@rikkyo.ac.jp, tokazaki@kias.re.kr}
\affiliation[a]{Department of Physics, Rikkyo University, Toshima, Tokyo 171-8501, Japan}
\affiliation[b]{
School of Physics, Korea Institute for Advanced Study,\\
85 Hoegi-ro, Cheongnyangri-dong, Dongdaemun-gu, Seoul 02455, Republic of Korea}
\begin{document}
\maketitle

\section{Introduction}
\label{sec_intro}
The superconformal index \cite{Romelsberger:2005eg,Kinney:2005ej} of four-dimensional field theories is a powerful tool to 
study the spectrum of superconformal field theories (SCFTs). 
It can be defined as a trace over the Hilbert space on $S^3$ in the radial quantization. 
Alternatively, it can be defined from the UV description as a partition function on $S^1\times S^3$ up to the Casimir energy factor \cite{Assel:2015nca}. 
It is a meaningful quantity even for $\mathcal{N}=2$ non-conformal field theories. 
While the generic superconformal index of $\mathcal{N}=2$ supersymmetric field theories involving three conformal fugacities enumerates the $1/8$ BPS local operators, 
it admits a limit of the fugacities which can count the $1/4$ BPS local operators, which is referred to as the Schur index \cite{Gadde:2011ik,Gadde:2011uv}. 
For a class $\mathcal{S}$ theory it can be interpreted as a correlation function of a 2d TQFT on a Riemann surface \cite{Gadde:2009kb,Gadde:2011ik}. 
According to the correspondence \cite{Beem:2013sza} between $\mathcal{N}=2$ SCFTs and vertex operator algebras (VOAs), 
the unflavored Schur index is equal to the vacuum character of the associated VOA. 
Also it has a nice modular property \cite{Razamat:2012uv} as a component of a vector valued modular function 
as it can solve a modular linear differential equation (MLDE) \cite{Beem:2017ooy,MR3890204}. 

In this paper we study the flavored Schur indices of 4d $\mathcal{N}=4$ super Yang-Mills (SYM) theories with unitary gauge groups. It can be also viewed as the Schur indices of $\mathcal{N}=2^{*}$ SYM theories as the flavor fugacity corresponds to the mass parameter for the adjoint hypermultiplet in the $\mathcal{N}=4$ vector multiplet. 
While an elementary evaluation of the indices for several low ranks is presented by performing the contour integrals in \cite{Pan:2021mrw}, we derive a complete closed-form expression of the indices for arbitrary ranks by applying the Fermi-gas method in \cite{Gaiotto:2020vqj} which generalizes the result \cite{Bourdier:2015wda} for the unflavored Schur index of $\mathcal{N}=4$ SYM theory.%
\footnote{Closed-form expressions for the unflavored Schur indices of $\mathcal{N}=4$ SYM theories are also found in \cite{Beem:2021zvt}.} 

The flavored Schur indices are given as a sum over the Young diagrams in such a way that the density matrix and its spectral zeta functions are given by the Kronecker theta function \cite{zbMATH02706826,MR1723749,MR1106744} and the twisted Weierstrass functions \cite{Dong:1997ea,Mason:2008zzb}. The twisted Weierstrass function plays a role of a generating function for quasi-Jacobi forms \cite{MR2796409}.  
The normalized $\mathcal{N}=2^{*}$ Schur indices are shown to lie in the polynomial ring 
in the Kronecker theta function and the Weierstrass functions which contains the polynomial ring of quasi-Jacobi forms.%

We also use the grand canonical ensemble to provide alternative closed-form expressions of the $\mathcal{N}=2^{*}$ Schur indices as infinite series. According to the observation in \cite{Buican:2020moo,Kang:2021lic}, we also obtain exact closed-form expressions for the unflavored indices of the $\mathcal{N}=2$ 
$\widehat{\Gamma}(SU(N))$ SCFTs \cite{DelZotto:2015rca,Xie:2016evu,Buican:2016arp,Closset:2020scj,Closset:2020afy} where $\Gamma$ is the simply laced Lie group of ADE type. 
While it is shown in \cite{Buican:2016arp} that the unflavored Schur index of the $\widehat{E}_6(SU(2))$ SCFT is given by the vacuum character of the $\mathcal{A}(6)$ algebra \cite{MR2766985,Feigin:2007sp}, we find alternative expressions from our closed-form formulae. 

In addition, we find that the unflavored Schur indices of $\mathcal{N}=4$ SYM theory with (special) unitary gauge groups are identified with MacMahon's generating functions $A_k(q)$ and $C_k(q)$ \cite{MR1576612} for the generalized sums of divisors up to overall powers of $q$ and the Ramanujan theta function. This generalizes the result in \cite{Kang:2021lic} for the case with odd rank gauge groups. This also shows that the normalized unflavored Schur indices of $\mathcal{N}=4$ SYM theory with (special) unitary gauge groups lie in the ring of quasi-modular forms \cite{MR1363056}. The unflavored $U(\infty)$ and $SU(\infty)$ Schur indices which are equivalent to the generating functions for overpartitions \cite{MR2034322} and $3$-colored partitions respectively.\footnote{The $SU(\infty)$ index is also discussed in \cite{Honda:2022hvy}. }
We investigate the asymptotic growth of the numbers of the BPS local operators by applying the Meinardus theorem \cite{MR62781}. 

\subsection{Structure}
The organization of the paper is as follows. 
In section \ref{sec_canonical} we introduce the flavored Schur indices of $\mathcal{N}=4$ SYM theories with unitary gauge group. 
In section \ref{sec_Fermigas} we formalize the Fermi-gas method which allows for a closed-form expression of the $\mathcal{N}=2^{*}$ Schur indices 
in terms of the twisted Weierstrass functions, the twisted Eisenstein series as well as the Jacobi theta functions. 
In section \ref{sec_gcanonical} we take the grand canonical ensemble to get another type of closed-form expressions as certain infinite series 
for the Schur indices of $\mathcal{N}=2^{*}$ SYM theory. We also present a closed-form expression of the unflavored Schur indices of $\widehat{\Gamma}(SU(N))$  SCFTs. 
In section \ref{sec_combi} we show that the unflavored Schur indices of $\mathcal{N}=4$ SYM theory with unitary gauge groups 
can be written in terms of MacMahon's generating functions $A_k$ and $C_k$ for the generalized sums of divisors. 
The large $N$ unflavored Schur indices of $\mathcal{N}=4$ $U(N)$ and $SU(N)$ SYM theories are identified with generating functions 
for overpartitions and $3$-colored partitions. The asymptotic growth of the numbers of the BPS local operators in the large $N$ limit are analyzed. 
In appendix \ref{app_notations} we summarize definitions and notations 
of the $q$-factorial, the Jacobi theta functions, the (twisted) Eisenstein series, and the (twisted) Weierstrass functions.  
In appendix \ref{app_other} we present further examples of the closed-from expressions of the Schur indices. 

\subsection{Future works}

\begin{itemize}

\item The Schur indices can be decorated by the line operators \cite{Gang:2012yr,Cordova:2016uwk,Neitzke:2017cxz,Gaiotto:2020vqj}. 
It is expected that the Schur correlation functions of the line operators in $\mathcal{N}=2^{*}$ $U(N)$ SYM theory enjoy a hidden triality symmetry as explicitly checked for some examples in \cite{Gaiotto:2020vqj}, which upon a twisted circle compactification can reduce to the triality symmetry \cite{Gaiotto:2020vqj,Hatsuda:2021oxa} of the sphere correlators of the 3d $\mathcal{N}=4$ $U(N)$ ADHM theory. We hope to report more details of the Schur correlators in the upcoming work \cite{HOline:2022}. 

\item The modular covariant combinations of the twisted Eisenstein series and derivatives 
acting on the Jacobi forms are given in \cite{Gaberdiel:2009vs} and those acting on the quasi-Jacobi forms are studies in \cite{MR4281261}. 
It would be intriguing to study differential equations satisfied by the $\mathcal{N}=2^{*}$ Schur indices 
and to examine their transformation laws by using our closed formulae. 

\item The coefficients of the supersymmetric indices can encode interesting combinatorial statistics.%
\footnote{See \cite{Okazaki:2022sxo,Hayashi:2022ldo} for the combinatorial interpretations for supersymmetric indices of M2-brane SCFTs. } It would be interesting to extend discussions for the unflavored Schur indices in section \ref{sec_combi} on combinatorial interpretations to the flavored case. 

\item It would be nice to investigate closed-form expressions of the Schur indices of $\mathcal{N}=2^{*}$ SYM theory as well as $\mathcal{N}=4$ SYM theory with other gauge groups. The Fermi-gas method may be useful by generalizing the Frobenius determinant identity \cite{Frobenius:1882uber} or Fay's trisecant identity \cite{MR0335789}. 

\item The high-temperature limit of the (grand) canonical flavored Schur indices may be investigated 
by applying the generalization of the Euler-Maclaurin summation formula \cite{https://doi.org/10.48550/arxiv.2109.10394}.%
\footnote{See \cite{Eleftheriou:2022kkv} for the analysis of the unflavored Schur indices. }

\item In the presence of boundaries or/and corners the indices encode the BPS boundary conditions with additional degrees of freedom as the half- and the quarter-indices \cite{Gaiotto:2019jvo}. It would be nice to find their closed-form expressions. 

\end{itemize}

\section{Canonical indices}
\label{sec_canonical}
In this section, we collect some basics on the flavored Schur index of $\mathcal{N}=4$ $U(N)$ SYM theory.
We begin with a matrix integral that exactly computes the flavored Schur index,%
\footnote{See \cite{Gaiotto:2019jvo} for the notation and definition of the flavored Schur index of $\mathcal{N}=4$ SYM theory.}
\begin{align}
\label{uN_GOindex}
\mathcal{I}^{U(N)}(t;q)
&=
\frac{1}{N!}\frac{(q)_{\infty}^{2N}}{(q^{\frac12} t^{\pm 2};q)_{\infty}^N}
\oint_{|\sigma_i|=1} \prod_{i=1}^N \frac{d\sigma_i}{2\pi i\sigma_i} 
\frac{
\prod_{i\neq j} \left( \frac{\sigma_i}{\sigma_j};q \right)_{\infty} \left( q\frac{\sigma_i}{\sigma_j};q \right)_{\infty}
}
{
\prod_{i\neq j} \left(q^{\frac12}t^{-2} \frac{\sigma_i}{\sigma_j};q \right)_{\infty} \left( q^{\frac12}t^{2} \frac{\sigma_i}{\sigma_j};q \right)_{\infty}
},
\end{align}
where the integration contour for gauge fugacities $\sigma_i=e^{2\pi i\alpha_i}$, $i=1,\cdots, N$ is taken as a unit torus $\mathbb{T}^N$. 
It is a formal Taylor series in $q^{1/2}=e^{\pi i\tau}$ and its coefficients are Laurent polynomials in a fugacity $t=e^{2\pi i\rho}$ with integer coefficients. 
$\tau$ is a complex structure of the torus on which the underlying VOA is supported \cite{Dedushenko:2019yiw,Pan:2019bor,Jeong:2019pzg} 
or the radius of $S^1$ relative to the radius of $S^3$ 
and the fugacity $t=e^{2\pi i \rho}$ couples to the difference of the Cartan generators of $SU(2)_C$ and $SU(2)_H$ subgroups of the R-symmetry group $SU(4)_{R}$. 

The index (\ref{uN_GOindex}) starts from $1+\cdots$ so that one can count the number of the quarter-BPS local operators as a protected quantity by reading off coefficients in its expansion. 
The index (\ref{uN_GOindex}) is manifestly invariant under the transformation 
\begin{align}
\label{Sym_t}
t\rightarrow t^{-1},
\end{align}
as there is a symmetry which exchanges the $SU(2)_H$ with the $SU(2)_C$. 
When $t=1$, the index (\ref{uN_GOindex}) reduces to the unflavored Schur index of $\mathcal{N}=4$ $U(N)$ SYM theory. 

\subsection{Examples}
For example, the flavored Schur index of $\mathcal{N}=4$ $U(1)$ SYM theory is
\begin{align}
\label{u1_GOindex1}
\mathcal{I}^{U(1)}(t;q)
&=\frac{(q)_{\infty}^2}{(q^{\frac12}t^2;q)_{\infty}(q^{\frac12}t^{-2};q)_{\infty}}
\nonumber\\
&=1+(\underbrace{t^{2}}_{X}+\underbrace{t^{-2}}_{Y})q^{1/2}
+(-1+\underbrace{t^4}_{X^2}+\underbrace{t^{-4}}_{Y^2})q
+(\underbrace{t^6}_{X^3}+\underbrace{t^{-6}}_{Y^3})q^{3/2}
\nonumber\\
&+(\underbrace{t^8}_{X^4}+\underbrace{t^{-8}}_{Y^4})q^{2}
+(\underbrace{t^{10}}_{X^5}-t^{2}-t^{-2}+\underbrace{t^{-10}}_{Y^5})q^{5/2}+\cdots
\end{align}
The terms of the form $q^{k/2}t^{2k}$ and $q^{k/2}t^{-2k}$ with $k=1,2,\cdots$ enumerate the half-BPS local operators $X^k$ and $Y^k$  
where $X$ and $Y$ are the adjoint scalar fields $X$ and $Y$ transforming as $({\bf 1},{\bf 3})$ and $({\bf 3}, {\bf 1})$ under the $SU(2)_C\times SU(2)_H$ respectively. 
For other terms there are implicit cancellations between bosonic and fermionic contributions. 
For example, we have
\begin{align}
\begin{array}{c|c|c|c}
\textrm{terms}&\textrm{coefficients}&\textrm{bosonic operators}&\textrm{fermionic operators} \\ \hline 
q&-1&XY&\lambda, \overline{\lambda} \\
q^{3/2}t^2&0&X^2 Y, \partial X&\lambda X, \overline{\lambda}X \\
q^2&0&\partial X Y, X\partial Y, X^2 Y^2, \lambda\overline{\lambda}&\partial\lambda,\partial\overline{\lambda},\lambda XY,\overline{\lambda}XY \\
q^2 t^4&0&X^3Y, X\partial X&\lambda X^2, \overline{\lambda}X^2 \\
q^{5/2}&0&X^4Y, \partial X^3&\lambda X^3, \overline{\lambda} X^3 \\
q^{5/2}t^2&-1&X^3Y^2, X^2\partial Y, X\partial X Y, &
\lambda X^2Y, \overline{\lambda}X^2Y, \\
&&\partial^2X, \lambda\overline{\lambda}X
&\lambda \partial X, \partial \lambda X,  \overline{\lambda} \partial X, \overline{\partial} \lambda X \\
\end{array}
\end{align}
where $\lambda$ denotes the 4d gauginos transforming as $({\bf 2}, {\bf 2})$ under the $SU(2)_C\times SU(2)_H$. 

The flavored Schur index of $\mathcal{N}=4$ $U(2)$ SYM theory has an expansion 
\begin{align}
\label{u2_GOindex1}
&\mathcal{I}^{U(2)}(t;q)
\nonumber\\
&=1+\Bigl(\underbrace{t^2}_{\Tr X}+\underbrace{t^{-2}}_{\Tr Y}\Bigr)q^{1/2}
+\Bigl(
\underbrace{2t^4}_{
\begin{smallmatrix}
\Tr (X^2),\\
(\Tr X)^2 \\ 
\end{smallmatrix}}
+\underbrace{2t^{-4}}_{
\begin{smallmatrix}
\Tr(Y^2),\\
(\Tr Y)^2\\
\end{smallmatrix}
}
\Bigr)q
+\Bigl(\underbrace{2t^6}_{
\begin{smallmatrix}\Tr(X^3)\\ \Tr(X^2)\Tr X\end{smallmatrix}
}
+\underbrace{2t^{-6}}_{
\begin{smallmatrix}\Tr(Y^3)\\ \Tr(X^2)\Tr Y\end{smallmatrix}
}
\Bigr)q^{3/2}
\nonumber\\
&+\Bigl(
\underbrace{3t^8}_{
\begin{smallmatrix}\Tr(X^4)\\ \Tr(X)^3\Tr X\\ \Tr(X^2)^2\end{smallmatrix}
}
+\underbrace{3t^{-8}}_{
\begin{smallmatrix}\Tr(Y^4)\\ \Tr(Y)^3\Tr Y\\ \Tr(Y^2)^2\end{smallmatrix}
}
\Bigr)q^2
+(\underbrace{3t^{10}}_{
\begin{smallmatrix}\Tr(X^5)\\ \Tr(X^4)\Tr X\\ \Tr(X^3)\Tr(X^2)\end{smallmatrix}
}+t^2+t^{-2}
+\underbrace{3t^{-10}}_{
\begin{smallmatrix}\Tr(Y^5)\\ \Tr(Y^4)\Tr Y\\ \Tr(Y^3)\Tr(Y^2)\end{smallmatrix}
}
)q^{5/2}+\cdots
\end{align}
The terms with $q^{k/2}t^{2k}$ (resp. $q^{k/2}t^{-2k}$), $k=1,2,\cdots$ count the number of the half-BPS local operators. 
It includes single and double trace operators consisting of the adjoint scalar field $X$ (resp. $Y$). 
Again cancellations of several bosonic and fermionic operators occur in the other terms. 

\subsection{Half-BPS indices}
$\mathcal{N}=4$ $U(N)$ SYM theory has the half-BPS local operators of a fixed scaling dimension $\Delta$ consisting of the adjoint scalar fields. 
For example, those operators consisting of $X$ take the form \cite{Corley:2001zk} 
\begin{align}
\label{1/2BPS_op}
\Tr (X^{l_1})^{k_1} 
\Tr (X^{l_2})^{k_2}
\cdots 
\Tr (X^{l_m})^{k_m}
\end{align}
obeying the condition 
\begin{align}
\Delta&=\sum_{i=1}^m l_i k_i. 
\end{align}
They can be labeled by the Young diagram of $\Delta$ boxes whose length is no greater than $N$. 
We can demonstrate this by deriving from the flavored index (\ref{uN_GOindex}) the half-BPS index that enumerates the half-BPS local operators consisting of $X$ by taking the following scaling limit: 
\begin{align}
\label{halfBPS_index}
\mathcal{I}^{U(N)}_{\frac12\textrm{BPS}}(\mathfrak{q})
&:=\lim_{\begin{smallmatrix}\mathfrak{q}=q^{1/4}t \textrm{: fixed}\\
q\rightarrow0 \end{smallmatrix}}
\mathcal{I}^{U(N)}(t;q)
=\prod_{n=1}^N \frac{1}{(1-\mathfrak{q}^n)}
\end{align}
where we have set $q$ to zero while keeping $\mathfrak{q}:=q^{1/4}t$ being constant.  
In fact, the half-BPS index (\ref{halfBPS_index}) is identified with a generating function for partitions with rows being no greater than $N$. 

\subsection{Large $N$ limit}
The ratio of the index to the large $N$ index can admit a giant graviton expansion \cite{Gaiotto:2021xce}%
\footnote{See also \cite{Arai:2020qaj,Imamura:2021ytr,Murthy:2022ien,Lee:2022vig,Imamura:2022aua} for the study of giant graviton expansion of the Schur indices. }
where each term in the expansion is identified with the index of giant gravitons in the holographic dual AdS space \cite{McGreevy:2000cw,Grisaru:2000zn,Hashimoto:2000zp}. 

Let us consider the large $N$ limit of the half-BPS index (\ref{halfBPS_index}). 
It is identified with a generating function 
for partitions of $n$ \cite{MR2445243}
\begin{align}
\label{largeN_halfBPS}
\mathcal{I}^{U(\infty)}_{\frac12\textrm{BPS}}(\mathfrak{q})
&=\prod_{n=1}^{\infty}\frac{1}{(1-\mathfrak{q}^n)}
=\sum_{n=0}^{\infty}p(n)\mathfrak{q}^n
\nonumber\\
&=1+\mathfrak{q}+2\mathfrak{q}^2+3\mathfrak{q}^3+5\mathfrak{q}^4+7\mathfrak{q}^5+11\mathfrak{q}^6
+15\mathfrak{q}^7+22\mathfrak{q}^8+\cdots
\end{align}
where $p(n)$ is the number of partitions of $n$. 
The asymptotic growth of the number of the half-BPS local operators 
is equivalent to that of $p(n)$. 
As $n\rightarrow \infty$ we have \cite{MR2280879} 
\begin{align}
\label{asym_1/2BPS}
p(n)&\sim \frac{1}{4n\sqrt{3}}
\exp\left[
\frac{\pi 2^{1/2}}{3^{1/2}}n^{1/2}
\right]. 
\end{align}
The exact numbers $p(n)$ of the operators 
and the values $p_{\mathrm{asymp}}(n)$ evaluated from (\ref{asym_1/2BPS}) are given by
\begin{align}
\label{asy_pn_table}
\begin{array}{c|c|c} 
n&p(n)&p_{\textrm{asymp}}(n) \\ \hline 
10&30&48.1043 \\
100&1.69230\times 10^{8}&1.99281\times 10^{8} \\
1000&2.31278\times 10^{31}&2.44020\times 10^{31} \\
5000&1.66801\times 10^{74}&1.70889\times 10^{74} \\
10000&3.57099\times 10^{106}&3.63281\times 10^{106} \\
\end{array}
\end{align}

The flavored index (\ref{uN_GOindex}) in the large $N$ limit can be evaluated as
\begin{align}
\label{largeN_GOindex}
&\mathcal{I}^{U(\infty)}(t;q)
\nonumber\\
&=
1+(t^2+t^{-2})q^{1/2}
+(2t^4+2t^{-4})q
+(3t^6+t^2+t^{-2}+3t^{-6})q^{3/2}
\nonumber\\
&+(2+5t^8+t^4+t^{-4}+5t^{-8})q^2
+(5t^{10}+t^6+t^2+t^{-2}+t^{-6}+5t^{-10})q^{5/2}
\nonumber\\
&
+(2+7t^{12}+2t^4+2t^{-4}+7t^{-12})q^3+\cdots.
\end{align}
We find that the large $N$ flavored index (\ref{largeN_GOindex}) is given by 
\begin{align}
\label{largeN_GOindex2}
\mathcal{I}^{U(\infty)}(t;q)
&=(q)_{\infty} 
\prod_{k=0}^{\infty}
\frac{
(q^{1+\frac{k+1}{2}} t^{\pm 2(k+1)};q)_{\infty}
}
{
(q^{\frac{k+1}{2}} t^{\pm 2(k+1)};q)_{\infty}
},
\end{align}
and that it coincides with 
\begin{align}
\label{graviton_index}
\frac{\prod_{n=1}^{\infty} (1-q^{n})}{\prod_{n=1}^{\infty} (1-t^{2n} q^{\frac{n}{2}}) (1-t^{-2n} q^{\frac{n}{2}} )}, 
\end{align}
which is obtained from the gravitational index in \cite{Kinney:2005ej}. 

We will discuss a combinatorial interpretation of the large $N$ indices in section \ref{sec_combi}. 

\section{Fermi-gas formulation}
\label{sec_Fermigas}
In this section, we explore closed-form expressions for the flavored Schur index of $\mathcal{N}=4$ $U(N)$ SYM theory.
To do so, we rewrite the matrix model \eqref{uN_GOindex} as the canonical partition function of an ideal Fermi-gas system.
For this purpose, it is more convenient to use another fugacity $\xi=q^{-1/2}t^2$ $=$ $e^{2\pi i\zeta}$ rather than $t$ itself.
With this choice of the fugacity, the flavored index (\ref{uN_GOindex}) reads
\begin{align}
\label{uN_findex}
\mathcal{I}^{U(N)}(\xi;q)=
&\frac{1}{N!}\frac{(q)_{\infty}^{2N}}{(\xi^{-1};q)_{\infty}^N(q\xi;q)_{\infty}^N}
\oint_{|\sigma_i|=1} \prod_{i=1}^N \frac{d\sigma_i}{2\pi i\sigma_i} 
\frac{
\prod_{i\neq j} \left( \frac{\sigma_i}{\sigma_j};q \right)_{\infty} \left( q\frac{\sigma_i}{\sigma_j};q \right)_{\infty}
}
{
\prod_{i\neq j} \left(\xi^{-1} \frac{\sigma_i}{\sigma_j};q \right)_{\infty} \left( q\xi \frac{\sigma_i}{\sigma_j};q \right)_{\infty}
}. 
\end{align}
The invariance under $t \to t^{-1}$ is now translated into
\begin{align}
\label{Sym_xi}
\xi\rightarrow q^{-1}\xi^{-1}. 
\end{align}
We note that 
the $q$-expansion of the index (\ref{uN_findex}) does not start from $1+\cdots$. 
In fact, the index (\ref{uN_findex}) can be interpreted as the Schur index of 4d $\mathcal{N}=2^{*}$ $U(N)$ SYM theory which is a non-conformal supersymmetric theory 
due to the introduction of a mass parameter for the adjoint hypermultiplet in $\mathcal{N}=4$ SYM theory corresponding to the chemical potential $\zeta$. 

\subsection{Kronecker theta function}
We rewrite the index (\ref{uN_findex}) as 
\begin{align}
\label{uN_findex2}
\mathcal{I}^{U(N)}(\xi;q)
&=\frac{(-1)^N\xi^{N^2/2}}{N!}
\oint_{|\sigma_i|=1} \prod_{i=1}^N \frac{d\sigma_i}{2\pi i\sigma_i} 
\frac{\theta'(1;q)^{N} \prod_{i<j} \theta(\frac{\sigma_i}{\sigma_j};q) \theta(\frac{\sigma_j}{\sigma_i};q)}
{\prod_{i,j}\theta(\frac{\sigma_i}{\sigma_j}\xi^{-1};q)},
\end{align}
where we define a new function
\begin{align}
\label{theta_DEF}
\theta(x;q)&:=\sum_{n\in \mathbb{Z}}(-1)^n x^{n+\frac12}q^{\frac{n^2+n}{2}}
\nonumber\\
&=(x^{\frac12}-x^{-\frac12}) \prod_{n=1}^\infty(1-q^n)(1-x q^n)(1-x^{-1}q^n)
\nonumber\\
&=-x^{-\frac12}(q;q)_{\infty} (x;q)_{\infty} (qx^{-1};q)_{\infty}. 
\end{align}
It is equal to the first Jacobi theta function (\ref{theta1}) up to a pre-factor $iq^{-1/8}$ 
\begin{align}
\theta(e^{2\pi i z};e^{2\pi i\tau})&=iq^{-\frac18} \vartheta_1 (z;\tau).
\end{align}
It has properties 
\begin{align}
\theta(e^{2\pi i (n+m\tau+z)};q)&=(-1)^{n+m} e^{-\pi i m^2 \tau-2\pi i mz} \theta(e^{2\pi iz};q), \\
\theta(x^{-1};q)&=-\theta(x;q), \\
\theta(x,e^{2\pi i(\tau+1)})&=\theta(x,e^{2\pi i\tau})
\end{align}
and satisfies a differential equation
\begin{align}
q\frac{\partial}{\partial q}\theta(x;q)
-\frac12 \left( x \frac{\partial}{\partial x} \right)^2 \theta(x;q)
+\frac18 \theta(x;q)&=0. 
\end{align}
Also we have
\begin{align}
\theta'(1;q)&:=\frac{\partial}{\partial x} \theta (x;q)\Bigl|_{x=1}=(q)_{\infty}^3. 
\end{align}

To proceed, let us introduce the Kronecker theta function \cite{zbMATH02706826,MR1723749,MR1106744,MR2796409},%
\footnote{
The Kronecker theta function $F(e^u,e^v;e^{2\pi i\tau})$ defined here is the same as the $F_{\tau}(u,v)$ in \cite{MR1106744}. 
}
\begin{align}
\label{Kronecker_fcn}
F(x,y;q)&:=\frac{\theta(xy;q) (q)_{\infty}^3}{\theta(x;q) \theta(y;q)}
=
-i\frac{\vartheta_1(z+w;\tau) \eta^3(\tau)}
{\vartheta_1(z;\tau) \vartheta_1(w;\tau)}
\end{align}
where $x=e^{2\pi iz}$ and $y=e^{2\pi iw}$. 
It satisfies the following relations 
\begin{align}
\label{Kronecker_sym}
F(x,y;q)&=F(y,x;q),\\
\label{Kronecker_flip}
F(x^{-1},y^{-1};q)&=-F(x,y;q),\\
\label{Kronecker_x1/x0}
F(x,x^{-1})&=0,\\
\label{Kronecker_ell1}
F(e^{2\pi i(n+m\tau+z)},y;q)&=e^{-2\pi i mw}F(e^{2\pi iz},y;q),\\
\label{Kronecker_ell2}
F(x,e^{2\pi i(n+m\tau+w)};q)&=e^{-2\pi imz}F(x,e^{2\pi iw};q), \\
F(qx^{-1},y;q)&=-y^{-1}F(x,y^{-1};q). 
\end{align}
It is a Jacobi form of weight $1$ which has the transformation laws
\begin{align}
\label{Kronecker_modular}
F\left(
e^{2\pi i\frac{z}{c\tau+d}}, e^{2\pi i\frac{w}{c\tau+d}}; e^{2\pi i \frac{a\tau+b}{c\tau+d}}
\right)
&=(c\tau+d) e^{\frac{2\pi i c zw}{c\tau+d}}
F\left(
e^{2\pi iz}, e^{2\pi iw}; e^{2\pi i\tau}
\right), \\
\label{Kronecker_elliptic}
F\left(e^{2\pi i(z+n\tau+s)};e^{2\pi i (w+m\tau+r)};e^{2\pi i\tau} \right)
&=e^{-2\pi i (mn \tau+mz+nw)}F(e^{2\pi iz};e^{2\pi iw};e^{2\pi i\tau})
\end{align}
for $\left( \begin{smallmatrix}a&b\\c&d\\ \end{smallmatrix}\right)$ $\in \Gamma=SL(2,\mathbb{Z})$ and $(m,r), (n,s)\in \mathbb{Z}^2$. 
It obeys a differential equation
\begin{align}
\label{Kronecker_diffeq}
q\frac{\partial}{\partial q}F(x,y;q)&=
xy\frac{\partial^2}{\partial x\partial y}F(x,y;q). 
\end{align}
It can be expressed as a double series 
\begin{align}
\label{Kronecker_double}
F(x,y;q)&=\frac{xy-1}{(x-1)(y-1)} 
-\sum_{m,n=1}^{\infty} 
(x^m y^n- x^{-m}y^{-n})q^{mn}. 
\end{align}
For $|q|<|x|<1$ it has the Fourier expansion:
\begin{align}
\label{Kronecker_series}
F(x,y;q)&=-\sum_{n\in \mathbb{Z}} \frac{x^n}{1-yq^n}. 
\end{align}
Notice that 
\begin{align}
F(e^{2\pi iz},y;q=e^{2\pi i\tau})&=P_1
\left[
\begin{matrix}
y^{-1}\\
1\\
\end{matrix}
\right](z,\tau)
\end{align}
where $P_1\left[
\begin{smallmatrix}
\theta \\
\phi \\
\end{smallmatrix}
\right](z,\tau)$ is the twisted Weierstrass function defined by (\ref{tP1}). 

This results from Ramanujan's $_1 {\psi}_{1}$ summation formula \cite{MR0004860}
\begin{align}
\label{ramanujan_sum}
\sum_{n\in \mathbb{Z}}
\frac{(a;q)_{n}}{(b;q)_{n}}z^{n}
&=\frac{
(q)_{\infty}(ba^{-1};q)_{\infty}
(az;q)_{\infty}(q a^{-1}z^{-1};q)_{\infty}
}
{
(b;q)_{\infty}
(qa^{-1};q)_{\infty}
(z;q)_{\infty}
(ba^{-1}z^{-1};q)_{\infty}
}
\end{align}
by setting $b=aq$. 

\subsection{Fermi-gas canonical partition function}
We are ready to go to the Fermi-gas formulation. We make use of the Frobenius determinant formula \cite{Frobenius:1882uber}%
\footnote{This is also obtained from the determinant formula \cite{Mason:2008zzb} which generalizes Fay's trisecant identity \cite{MR0335789}. }
\begin{align}
\label{Frobdet}
\frac{
(q)_{\infty}^{3N}\prod_{i<j}\theta(v_i v_j^{-1};q) \theta(w_j w_i^{-1};q)
}
{
\prod_{i,j}\theta(v_i w_j^{-1};q)
}
&=\frac{\theta(u;q)}{\theta(u\prod_i v_i w_i^{-1};q)}
\det_{i,j} F(v_i w_j^{-1},u;q), 
\end{align}
where $u=e^{2\pi i\nu}$ is an auxiliary fugacity. 
Then we can finally express the $\mathcal{N}=2^{*}$ Schur index (\ref{uN_findex2}) as \cite{Gaiotto:2020vqj}
\begin{align}
\label{uN_findex3}
\mathcal{I}^{U(N)}(\xi;q)
&=\frac{(-1)^N \xi^{N^2/2}}{N!}\frac{\theta(u;q)}{\theta(u\xi^{-N};q)}
\oint_{|\sigma_i|=1}d^N \sigma \det_{i,j} F\left(\frac{\sigma_i}{\sigma_j}\xi^{-1},u;q\right)
\nonumber\\
&=\frac{(-1)^N \xi^{N^2/2} \theta(u;q)}{\theta(u\xi^{-N};q)} \mathcal{Z}(N;u;\xi;q),
\end{align}
where
\begin{align}
\label{pfn_Fermi}
\mathcal{Z}(N;u;\xi;q):=
\frac{1}{N!}\oint_{|\sigma_i|=1}d^N \sigma \det_{i,j} F\left(\frac{\sigma_i}{\sigma_j}\xi^{-1},u;q\right).
\end{align}
Note that the function $\mathcal{Z}(N;u;\xi;q)$ depends on $u$, but the Schur index $\mathcal{I}^{U(N)}(\xi;q)$ does not. This means that we can choose $u$ freely to compute $\mathcal{I}^{U(N)}(\xi;q)$.
In fact, we will see that some specific choices of $u$ simplify exact expressions of $\mathcal{I}^{U(N)}(\xi;q)$.

The function $\mathcal{Z}(N;u;\xi;q)$ can be identified with a canonical partition function of Fermi-gas with $N$ particles on a circle. 
The one-particle density matrix which characterizes the Fermi-gas is given by
\begin{align}
\label{densmtx}
\rho_{0}(\alpha,\alpha';u;\xi;q)&=F(e^{2\pi i(\alpha-\alpha')}\xi^{-1},u;q)
\nonumber\\
&=-\frac{e^{2\pi ip(\alpha-\alpha')}\xi^{-p}}{1-uq^p}. 
\end{align}
In the Fermi-gas system $\alpha$ and $p$ can be thought of as the position and momentum operators. 

It is interesting to note that the canonical partition function satisfies a differential equation 
\begin{align}
\label{canonical_diff}
u\frac{\partial}{\partial u}\mathcal{Z}(N;u;\xi;q)
+\left[
P_1(\nu,\tau)-
P_1(\nu-N\zeta,\tau)
\right]\mathcal{Z}(N;u;\xi;q)&=0
\end{align}
where $P_{1}(z,\tau)$ is the Weierstrass function defined by (\ref{P1}). 

\subsection{Spectral zeta functions}
Our next strategy is to evaluate the canonical partition function $\mathcal{Z}(N;u;\xi;q)$ exactly.
It is known that $\mathcal{Z}(N;u;\xi;q)$ is given by so-called spectral zeta functions. 
The spectral zeta functions for the inverse of the density matrix of the Fermi-gas\footnote{Let $\lambda_n$ be eigenvalues of the inverse of the density matrix of the Fermi-gas. The spectral zeta functions are given by $Z_s=\sum_{n} \lambda_n^{-s}$.} are defined by
\begin{align}
\label{Z_l}
Z_{l}(u;\xi;q)&:=
\Tr (\rho_0^l)
=\int_{0}^{1} \prod_{i=1}^l d \alpha_l\ 
\rho_0(\alpha_1, \alpha_2) \cdots \rho_0(\alpha_l, \alpha_1)\nonumber\\
&=\sum_{p\in \mathbb{Z}} 
\left(
\frac{-\xi^{-p}}{1-uq^p}
\right)^l
\nonumber\\
&=
\frac{(-1)^{l-1}}{(l-1)!} \frac{\partial^{l-1}F ( q^{-(l-1)}\xi^{-l},u;q ) }
{\partial u^{l-1}}. 
\end{align}
Plugging the double series (\ref{Kronecker_double}) into (\ref{Z_l}), we get an explicit formula
\begin{align}
\label{Z_l_series}
&Z_l(u;\xi;q)
=
\frac{\delta_{l,1}}{(1-\xi)}+
\frac{1}{(u-1)^l}
\nonumber\\
&+
\sum_{m,n=1}^{\infty}
\left[
\left(
\begin{matrix}
n+l-2\\
l-1
\end{matrix}
\right)
q^{m(n+l-1)}\xi^{ml}u^{-n-l+1}
+(-1)^{l}
\left(
\begin{matrix}
n\\
l-1
\end{matrix}
\right)
q^{m(n-l+1)}\xi^{-ml}u^{n-l+1}
\right]
\nonumber\\
&=
\frac{\delta_{l,1}}{(1-\xi)}+
\frac{1}{(u-1)^l}
\nonumber\\
&+
\sum_{n=1}^{\infty}
\left[
\left(
\begin{matrix}
n+l-2\\
l-1\\
\end{matrix}
\right)
\frac{q^{n+l-1} \xi^l u^{-n-l+1}}{1-q^{n+l-1}\xi^{l}}
+(-1)^l
\left(
\begin{matrix}
n\\
l-1\\
\end{matrix}
\right)
\frac{q^{n-l+1} \xi^{-l} u^{n-l+1}}{1-q^{n-l+1} \xi^{-l}}
\right]. 
\end{align}
We note that the spectral zeta function for $l=1$ is identified with the Kronecker theta function (\ref{Kronecker_fcn}) 
or equivalently with the twisted Weierstrass function 
\begin{align}
\label{Z_1}
Z_1(u;\xi;q)
=\frac{(q)_{\infty}^3 \theta(\xi^{-1}u;q)}{\theta(\xi^{-1};q)\theta(u;q)}
=F(\xi^{-1},u;q)
=P_1 \left[
\begin{matrix}
\xi \\
1 \\
\end{matrix}
\right](\nu,\tau). 
\end{align}
From (\ref{Z_l}) and (\ref{Z_1}) the spectral zeta functions for $l>1$ can be simply obtained by differentiating $Z_1(u;\xi;q)$ with respect to $u$
\begin{align}
Z_{l}(u;\xi;q)
&=
\frac{(-1)^{l-1}}{(l-1)!} \frac{\partial^{l-1} Z_1 (u;q^{l-1}\xi^l;q) }
{\partial u^{l-1}}. 
\end{align}

Using a relation
\begin{align}
\label{xdx}
\frac{\partial^n}{\partial u^n}
&=u^{-n}\sum_{k=1}^n (2\pi i)^{-k}s(n,k)\frac{\partial^k}{\partial \nu^k},
\end{align}
where $u=e^{2\pi i\nu}$ and $s(n,k)$ are the Stirling numbers of the first kind, 
we can also write the spectral zeta functions for $l\ge2$ in terms of the twisted Weierstrass function (\ref{tPk})
\begin{align}
\label{Z_l_formula}
Z_l(u;\xi;q)&=\frac{u^{-(l-1)}}{(l-1)!}
\sum_{k=1}^{l-1}
k! |s(l-1,k) |
P_{k+1}\left[\begin{matrix}q^{l-1}\xi^l\\1\end{matrix}\right](\nu,\tau). 
\end{align}
where we have used a relation
\begin{align}
s(n,k)&=(-1)^{n-k} |s(n,k)| . 
\end{align}
For our purpose, the following representation of the twisted Weierstrass function is useful:
\begin{align}
P_{k}\left[\begin{matrix}\theta \\ 1\end{matrix}\right](z,\tau)
&=\frac{\theta}{1-\theta}\delta_{k1}+\frac{(-1)^k}{(k-1)!} {\rm Li}_{1-k}(x)\nonumber \\
&\quad+\frac{1}{(k-1)!}\sum_{n=1}^\infty n^{k-1}
\biggl[ \frac{(-1)^{k}\theta^{-1}x^n q^n}{1-\theta^{-1}q^n}
+\frac{\theta x^{-n} q^n}{1-\theta q^n} \biggr]
\end{align}
where $\theta \ne 1$, $x=e^{2\pi i z}$ and $q=e^{2\pi i \tau}$.

For small $l$, we have
\begin{align}
\label{Z_2}
Z_2(u;\xi;q)&=
\frac{1}{u}P_{2}\left[\begin{matrix} q\xi^2\\1\\ \end{matrix}\right](\nu,\tau), \\
\label{Z_3}
Z_3(u;\xi;q)&=\frac{1}{2u^2} \left(
P_2\left[
\begin{matrix}
q^2\xi^3\\
1\\
\end{matrix}
\right](\nu,\tau)
+2P_3\left[
\begin{matrix}
q^2\xi^3\\
1\\
\end{matrix}
\right](\nu,\tau)
\right), \\
\label{Z_4}
Z_4(u;\xi;q)&=\frac{1}{3u^3}
\left(
P_2\left[
\begin{matrix}
q^3\xi^4\\
1\\
\end{matrix}
\right](\nu,\tau)
+3
P_3\left[
\begin{matrix}
q^3\xi^4\\
1\\
\end{matrix}
\right](\nu,\tau)
+3
P_4\left[
\begin{matrix}
q^3\xi^4\\
1\\
\end{matrix}
\right](\nu,\tau)
\right),\\
\label{Z_5}
Z_5(u;\xi;q)&=\frac{1}{12u^4}
\Biggl(
3P_2\left[
\begin{matrix}
q^4\xi^5\\
1\\
\end{matrix}
\right](\nu,\tau)
+
11P_3\left[
\begin{matrix}
q^4\xi^5\\
1\\
\end{matrix}
\right](\nu,\tau)
\nonumber\\
&+
18P_4\left[
\begin{matrix}
q^4\xi^5\\
1\\
\end{matrix}
\right](\nu,\tau)
+
12P_5\left[
\begin{matrix}
q^4\xi^5\\
1\\
\end{matrix}
\right](\nu,\tau)
\Biggr). 
\end{align}

\subsection{Closed-form formula}
Now we are ready to give a first closed-form expression of the canonical Schur index 
by using the spectral zeta functions as building blocks. 
Let 
\begin{align}
\lambda=(\lambda_1^{m_1} \lambda_2^{m_2} \cdots \lambda_r^{m_r}),
\end{align}
with 
\begin{align}
&\sum_{i=1}^{r} m_{i} \lambda_{i}=N, \\
&\lambda_1>\lambda_2>\cdots >\lambda_r>\lambda_{r+1}=0
\end{align}
be a partition of integer $N$, i.e. the Young diagram of $N$ boxes. 
In terms of the spectral zeta functions $Z_{l}(u;\xi;q)$, we can express the canonical partition function as a sum over the Young diagrams
\begin{align}
\label{pfn_Fermi_closed}
\mathcal{Z}(N,u;\xi;q)
&=\sum_{\lambda}(-1)^{N-r} \prod_{i=1}^{r}
\frac{1}{\lambda_i^{m_i} (m_i !)}
Z_{\lambda_i}(u;\xi;q)^{m_i},
\end{align}
and the $\mathcal{N}=2^{*}$ $U(N)$ Schur index 
\begin{align}
\label{uN_findex_closed}
\mathcal{I}^{U(N)}(\xi;q)&=
\frac{(-1)^N \xi^{N^2/2} \theta(u;q)}{\theta(u\xi^{-N};q)}
\mathcal{Z}(N;u;\xi;q).
\end{align}
As we have seen in (\ref{Z_l_formula}), the spectral zeta functions \eqref{Z_l_formula} are given by polynomials in the twisted Weierstrass functions.
Thus the formula \eqref{uN_findex_closed} presents a systematic evaluation of the exact index.

The Schur indices of $\mathcal{N}=4$ $SU(N)$ SYM theory can be similarly written 
as a closed-fom according to a relation
\begin{align}
\label{u_su}
\mathcal{I}^{SU(N)}(\xi;q)&=
\frac{
\mathcal{I}^{U(N)}(\xi;q)
}
{
\mathcal{I}^{U(1)}(\xi;q)
}. 
\end{align}

Note that the R.H.S of (\ref{uN_findex_closed}) involves the auxiliary fugacity $u$.
This $u$ disappears in the final result of the index.
The cancellation of the $u$-dependence happens quite non-trivially.  
Instead, in this paper, we use the $u$-independence of the index, and fix $u$ to simplify the results.
There are several possibilities to do so. These possibilities lead to apparently different closed-form expressions.

\subsection{Quasi-Jacobi forms}
While the formulae (\ref{pfn_Fermi_closed}) and (\ref{uN_findex_closed}) state that the spectral zeta functions can be viewed as building blocks of the $\mathcal{N}=2^{*}$ $U(N)$ Schur index, they can also play a role of a generating function for quasi-Jacobi forms \cite{MR2796409}. 

A meromorphic quasi-Jacobi form of weight $k\in \mathbb{Z}$, index $m\in \mathbb{Z}$ and depth $(s,t)$ 
is a meromorphic function $\varphi(z,\tau):$ $\mathbb{C}\times \mathbb{H}\rightarrow \mathbb{C}$ 
obeying the following transformation laws \cite{MR2796409}: 
\footnote{Another definition of a quasi-Jacobi form is found in \cite{Kawai:2000px} 
and a definition of quasi-Jacobi form for higher rank is found in \cite{Krauel:2013lra,MR3956895}. }
\begin{align}
\label{QJacobi1}
\varphi\left(
\frac{z}{c\tau+d},\frac{a\tau+b}{c\tau+d}
\right)
&=\sum_{i\le s, j\le t}
(c\tau+d)^{k-i-j} \varphi_{i,j}(z,\tau)c^{i+j}z^{i}, \\
\label{QJacobi2}
\varphi(z+\lambda\tau+\mu,\tau)&=\sum_{i\le s}\varphi_{i,0}(z,\tau)\lambda^i
\end{align}
for $\left( \begin{smallmatrix}a&b\\c&d\\ \end{smallmatrix} \right)$ $\in$ $SL(2,\mathbb{Z})$ 
and $(\lambda,\mu)$ $\in \mathbb{Z}^2$ 
where $\varphi_{i,j}(z,\tau)$ and $\varphi_{i,0}(z,\tau)$ are meromorphic functions. 

Alternatively, the meromorphic quasi-Jacobi form is defined by introducing an almost meromorphic Jacobi form \cite{MR2796409}. 
An almost meromorphic Jacobi form of weight $k\in \mathbb{Z}$, index $m\in \mathbb{Z}$ and depth $(s,t)$ 
is a (real) meromorphic function with the form
\begin{align}
\label{AJacobi}
\Psi(z,\tau)&=\sum_{i,j\ge 0} 
\psi_{i,j}(z,\tau)\left(\frac{1}{\tau_2}\right)^i \left(
\frac{z_2}{\tau_2}
\right)^j
\end{align}
where $\tau=\tau_1+i\tau_2$, $z=z_1+iz_2$ and $\psi_{i,j}(z,\tau)$ are complex meromorphic functions, 
satisfying the transformation laws
\begin{align}
\Psi\left(
\frac{z}{c\tau+d},\frac{a\tau+b}{c\tau+d}
\right)&=(c\tau+d)^k e^{2\pi i \frac{cmz^2}{c\tau+d}}\Psi(z,\tau), \\
\Psi(z+\lambda\tau+\mu,\tau)&=e^{-2\pi im(\lambda^2+2\lambda z)}\Psi(z,\tau). 
\end{align}
Then a quasi-Jacobi form arises as a constant term $\psi_{0,0}(z,\tau)$ $=$ $\varphi(z,\tau)$ 
with respect to $1/\tau_2$ and $z_2/\tau_2$ of the almost meromorphic form. 

The algebra of quasi-Jacobi forms which is bigger than the algebra of Jacobi forms is the algebra of functions on $\mathbb{C}\times \mathbb{H}$ 
generated by the Eisenstein series $G_2(\tau)$ and the Weierstrass functions $P_k(z,\tau)$ \cite{MR2796409}. 
Alternatively, it is generated by the coefficients of the Kronecker theta function \cite{MR2796409}. 

We note that the basic spectral zeta function $Z_1(u;\xi;q)$ is equivalent to the Kronecker theta function 
so that it can be expanded as
\begin{align}
\label{Z_1_Gk}
Z_1(u;\xi;q)
&=\frac{1}{2\pi i\nu}-\sum_{k=1}^{\infty}G_k(\zeta,\tau)(2\pi i\nu)^{k-1}
\end{align}
where $G_k(z,\tau)$ is the twisted Eisenstein series (\ref{tGk_ztau}). 
Also the logarithm of the spectral zeta function has an expansion \cite{Bringmann:2020ndq}
\begin{align}
\label{logG_Fk}
\log \left(
2\pi i\nu Z_1(u;\xi;q)
\right)
&=\log(2\pi i\nu)+P_0(\nu,\tau)+P_0(\zeta,\tau)-P_0(\zeta+\nu,\tau)
\nonumber\\
&=\sum_{k=1}^{\infty} F_k(\zeta,\tau)(2\pi i\nu)^k, 
\end{align}
where $P_k(z,\tau)$ is the Weierstrass function (\ref{Pk}) and 
the function $F_k(z,\tau)$ is given by 
\begin{align}
\label{Fk}
F_k(z,\tau)&=\frac{(-1)^{k+1}}{k}\left( P_k(z,\tau)-G_k(\tau) \right)
\end{align}
where $G_k(\tau)$ is the Eisenstein series (\ref{Gk}). 
As $F_k(z;\tau)$ and $G_2(\tau)$ form the ring of quasi-Jacobi forms \cite{MR2796409}, 
the spectral zeta function $Z_1(u;\xi;q)$ plays a role of a generating function for quasi-Jacobi forms. 

Similarly, when one expands the other spectral zeta functions $Z_{l}(u;\xi;q)$ with $l>1$ with respect to $\nu$, 
each coefficient appears as a quasi-Jacobi form since the $Z_{l}(u;\xi;q)$ can be obtained from (\ref{Z_l}) by differentiating $Z_1(u;\xi;q)$ with respect to $u$ 
so that it can be expressed 
as a polynomial in the $Z_1(u;\xi;q)$ and the Weierstrass functions $P_{k}(\nu,\tau)$ which belong to the algebra of the quasi-Jacobi forms \cite{MR2796409}. 

Since the canonical partition functions $\mathcal{Z}(N;u;\xi;q)$ are given by polynomials in the spectral zeta functions $Z_l(u;\xi;q)$, 
each coefficient in the Taylor expansion of the canonical partition function in $\nu$ 
appears as a quasi-Jacobi form. 

Furthermore, it can be shown from (\ref{Z_l_formula}), (\ref{pfn_Fermi_closed}) and (\ref{uN_findex_closed}) 
that the normalized $\mathcal{N}=2^{*}$ $U(N)$ Schur index
\begin{align}
\widetilde{\mathcal{I}}^{U(N)}(\xi;q) := \xi^{-N^2/2}\mathcal{I}^{U(N)}(\xi;q)
\end{align}
is the meromorphic function on $\mathbb{C}\times \mathbb{H}$ which lies in the polynomial ring generated by 
the Kronecker theta function and the Weierstrass functions. 
It contains the ring of quasi-Jacobi forms of weight $k< N$ generated by the Weierstrass functions as well as the Eisenstein series $G_2(\tau)$. 

\subsection{Examples}

\subsubsection{$N=1$}
For $N=1$ the canonical partition function is simply 
\begin{align}
\label{fermi_pfn1}
\mathcal{Z}(1;u;\xi;q)&=Z_{1}(u;\xi;q)=P_1\left[
\begin{matrix}
\xi\\
1\\
\end{matrix}
\right]
(\nu,\tau).
\end{align}
The $\mathcal{N}=2^{*}$ $U(1)$ Schur index then reads
\begin{align}
\label{u1_findex2}
\mathcal{I}^{U(1)}(\xi;q)
=-\xi^{1/2}\frac{\theta(u;q)}{\theta(u\xi^{-1};q)} \mathcal{Z}(1;u;\xi;q)
=\xi^{1/2}\frac{(q)_{\infty}^3}{\theta(\xi;q)}.
\end{align}
In this case, one can easily check the cancellation of the $u$-dependence.
We can also express it in terms of the Dedekind $\eta$-function,
\begin{align}
\mathcal{I}^{U(1)}(\xi;q)=-i\xi^{1/2}\frac{\eta^3(\tau)}{\vartheta_1(\zeta;\tau)},
\end{align}
where  
\begin{align}
\eta(\tau)&:=q^{\frac{1}{24}}\prod_{n=1}^{\infty}(1-q^n).
\end{align}

We observe that the normalized index
\begin{align}
\label{u1_findex3}
\widetilde{\mathcal{I}}^{U(1)}(\xi;q)
:=\xi^{-1/2}\mathcal{I}^{U(1)}(\xi;q)=\frac{(q)_{\infty}^3}{\theta(\xi;q)}
=
-i\frac{\eta^3(\tau)}{\vartheta_1(\zeta;\tau)},
\end{align}
can be written as 
\begin{align}
\label{u1_findex3a}
\widetilde{\mathcal{I}}^{U(1)}(\xi;q)
&=\exp\left[
\frac{G_2(\tau)}{2} (2\pi i\zeta)^2
\right]\sigma^{-1}(\zeta,\tau)
\nonumber\\
&=
\frac{1}{2\pi i\zeta}\exp
\left[
\sum_{k=1}^{\infty}\frac{G_{2k}(\tau)}{2k} (2\pi i\zeta)^{2k}
\right],
\end{align}
where $G_k(\tau)$ is the Eisenstein series (\ref{Gk}) and $\sigma(x,\tau)$ is the Weierstrass $\sigma$-function (\ref{Weier_s}). 
This is the inverse of the elliptic prime form (\ref{P0_1}) on the elliptic curve with modulus $\tau$ \cite{MR2352717}. 

The normalized index (\ref{u1_findex3}) is a Jacobi form of weight $1$ and index $-\frac12$ obeying the transformation laws
\begin{align}
\label{u1mod1}
\widetilde{\mathcal{I}}^{U(1)}(e^{2\pi i \frac{\zeta}{c\tau+d}};e^{2\pi i \frac{a\tau+b}{c\tau+d}})
&=(c\tau+d) e^{-\frac{\pi i c \zeta^2}{c\tau+d}} \widetilde{\mathcal{I}}^{U(1)}(e^{2\pi i\zeta},e^{2\pi i\tau}),\\
\label{u1mod2}
\widetilde{\mathcal{I}}^{U(1)}(e^{2\pi i (\zeta+\lambda\tau+\mu)},e^{2\pi i\tau})
&=(-1)^{\lambda+\mu}\xi^{\lambda}q^{\frac{\lambda^2}{2}}\widetilde{\mathcal{I}}^{U(1)}(e^{2\pi i\zeta},e^{2\pi i\tau}). 
\end{align}

One can also write it as
\begin{align}
\widetilde{\mathcal{I}}^{U(1)}(\xi;q)&=F(\xi^{1/2};\xi^{-1};q)
\end{align}
in terms of the Kronecker theta function. 

\subsubsection{$N=2$}
For $N \geq 2$, it is particularly convenient to set $u=\xi$ at the beginning because $Z_1(\xi; \xi; q)$ vanishes in this choice.
The relation between the canonical partition function and the spectral zeta functions is drastically simplified.

We first see the formula for generic values of $u$.
For $N=2$ the canonical partition function has two contributions 
from two partitions $\tiny{\yng(1,1)}$ and $\tiny{\yng(2)}$, which correspond to the partitions $\lambda=(1^2), (2^1)$, respectively. Therefore we have
\begin{align}
\label{fermi_pfn2}
\mathcal{Z}(2;u;\xi;q)&=\frac12 Z_1(u;\xi;q)^2-\frac12 Z_2(u;\xi;q)
\nonumber\\
&= \frac12 \left(
P_1\left[
\begin{matrix}
\xi\\
1\\
\end{matrix}
\right]^2 (\nu,\tau)
-\frac{1}{u}
P_2 \left[
\begin{matrix}
q\xi^2\\
1\\
\end{matrix}
\right](\nu,\tau)
\right)
\nonumber\\
&=\frac12 
\left(
P_1
\left[
\begin{matrix}
\xi\\
1\\
\end{matrix}
\right]^2(\nu,\tau)
-P_2
\left[
\begin{matrix}
\xi^2\\
1\\
\end{matrix}
\right](\nu,\tau)
+
P_1
\left[
\begin{matrix}
\xi^2\\
1\\
\end{matrix}
\right](\nu,\tau)
\right)
\nonumber\\
&=\frac12 
\left(
P_1\left[
\begin{matrix}
\xi\\
1\\
\end{matrix}
\right]^2 (\nu,\tau)
-
P_1
\left[
\begin{matrix}
\xi^2\\
1\\
\end{matrix}
\right](\nu,\tau)
\left(
P_1(\nu,\tau)
-P_1(\nu-2\zeta,\tau)
-1
\right)
\right). 
\end{align}

The Schur index can be written in terms of the twisted Weierstrass function (\ref{tPk}). 
While the canonical partition function in (\ref{fermi_pfn2}) has two contributions, 
it is simplified by choosing $u=\xi$,
\begin{align}
\mathcal{Z}(2;\xi;\xi;q)&=-\frac12 Z_2(\xi;\xi;q).
\end{align}
Then the Schur index is simply given by 
\begin{align}
\label{I_2}
\mathcal{I}^{U(2)}(\xi;q)
&=\frac{\xi}{2}P_2\left[
\begin{matrix}
\xi^2 q\\
1\\
\end{matrix}
\right](\zeta,\tau)
\nonumber\\
&=
\xi^2 
P_1\left[
\begin{matrix}
\xi^2\\
1\\
\end{matrix}
\right](\zeta,\tau)
\left(
P_1(\zeta,\tau)-\frac12
\right). 
\end{align}
The invariance under the transformation (\ref{Sym_xi}) follows from 
the transformations 
\begin{align}
P_2\left[
\begin{matrix}
\theta^{-1}\\
1\\
\end{matrix}
\right](-z,\tau)
&=
P_2\left[
\begin{matrix}
\theta\\
1\\
\end{matrix}
\right](z,\tau), \\
P_2\left[
\begin{matrix}
\theta\\
1\\
\end{matrix}
\right](z+\tau,\tau)
&=\theta
P_2\left[
\begin{matrix}
\theta\\
1\\
\end{matrix}
\right](z,\tau),
\end{align}
of the twisted Weierstrass function. 

The normalized $U(2)$ index can be rewritten as
\begin{align}
\widetilde{\mathcal{I}}^{U(2)}(\xi;q)&=
\xi^{-2}\mathcal{I}^{U(2)}(\xi;q)
\nonumber\\
&=F(\xi;\xi^{-2};q) 
\left(P_1(\zeta,\tau)-\frac12 \right)
\end{align}
in terms of the Kronecker theta function and the Weierstrass function. 

\subsubsection{$N=3$}
When $N=3$, the canonical partition function has three contributions 
corresponding to three Young diagrams $\tiny{\yng(1,1,1)}$, $\tiny{\yng(2,1)}$ and $\tiny{\yng(3)}$
\begin{align}
\label{fermi_pfn3}
\mathcal{Z}(3;u;\xi;q)
&=
\frac16 \left(
Z_1(u;\xi;q)^3-3Z_1(u;\xi;q)Z_2(u;\xi;q)+2Z_3(u;\xi;q)
\right). 
\end{align}
In terms of the twisted Weierstrass functions, it is given by
\begin{align}
\label{fermi_pfn3_2}
&
\mathcal{Z}(3;u;\xi;q)
\nonumber\\
&=\frac{1}{6}
\Biggl(
P_1\left[
\begin{matrix}
\xi\\
1\\
\end{matrix}
\right](\nu,\tau)
-\frac{3}{u}P_1\left[
\begin{matrix}
\xi\\
1\\
\end{matrix}
\right](\nu,\tau)
P_2\left[
\begin{matrix}
q\xi^2\\
1\\
\end{matrix}
\right](\nu,\tau)
+\frac{1}{u^2}P_2\left[
\begin{matrix}
q^2\xi^2\\
1\\
\end{matrix}
\right](\nu,\tau)
+\frac{2}{u^2}
P_3\left[
\begin{matrix}
q^2\xi^3\\
1\\
\end{matrix}
\right](\nu,\tau)
\Biggr)
\nonumber\\
&=\frac{1}{6}
\Biggl[
P_1\left[
\begin{matrix}
\xi\\
1\\
\end{matrix}
\right]^3(\nu,\tau)
-P_1\left[
\begin{matrix}
\xi\\
1\\
\end{matrix}
\right](\nu,\tau)
P_1\left[
\begin{matrix}
\xi^2\\
1\\
\end{matrix}
\right](\nu,\tau)
\left(
P_1(\nu,\tau)-P_1(\nu-2\zeta,\tau)-1
\right)
\nonumber\\
&+
P_1\left[
\begin{matrix}
\xi^3\\
1\\
\end{matrix}
\right](\nu,\tau)
\Bigl(
\left(P_1(\nu,\tau)-P_1(\nu-3\zeta,\tau)\right)^2
-3\left(P_1(\nu,\tau)-P_1(\nu-3\zeta,\tau)\right)
\nonumber\\
&+\left(P_2(\nu,\tau)-P_2(\nu-3\zeta,\tau)\right)
+2
\Bigr)
\Biggr]. 
\end{align}

As a specialized canonical partition function at $u=\xi$ is simply given by
\begin{align}
\mathcal{Z}(3;\xi;\xi;q)&=\frac13 Z_3(\xi;\xi;q),
\end{align}
we can express the $\mathcal{N}=2^{*}$ $U(3)$ Schur index as
\begin{align}
\label{I_3}
\mathcal{I}^{U(3)}(\xi;q)
&=\frac{\xi^{5/2}}{6}
\frac{\theta(\xi)}{\theta(\xi^2)}
\left(
P_2\left[
\begin{matrix}
q^2\xi^3\\
1\\
\end{matrix}
\right](\zeta,\tau)
+2P_3\left[
\begin{matrix}
q^2\xi^3\\
1\\
\end{matrix}
\right](\zeta,\tau)
\right)
\end{align}
in terms of the twisted Weierstrass functions. 

We can write the normalized $U(3)$ index as
\begin{align}
\widetilde{\mathcal{I}}^{U(3)}(\xi;q)
&=
\xi^{-9/2}\mathcal{I}^{U(3)}(\xi;q)
\nonumber\\
&=\frac{1}{6}
F(\xi^{2/3};\xi^{-3};q)
\Biggl(
(P_1(\zeta,\tau)+P_1(2\zeta,\tau)-2)^2
\nonumber\\
&+(P_1(\zeta,\tau)+P_1(2\zeta,\tau)-2)
+(P_2(\zeta,\tau)-P_2(2\zeta,\tau))
\Biggr)
\end{align}
in terms of the Kronecker theta function and the Weierstrass functions. 

\subsubsection{$N=4$}
For $N=4$, the canonical partition function has five contributions 
corresponding to five Young diagrams $\tiny{\yng(1,1,1,1)}$, $\tiny{\yng(2,1,1)}$, $\tiny{\yng(3,1)}$, $\tiny{\yng(2,2)}$ and $\tiny{\yng(4)}$. 
We get
\begin{align}
\label{fermi_pfn4}
\mathcal{Z}(4;u;\xi;q)
&=
\frac{1}{24}
\Biggl(
Z_1(u;\xi;q)^4-6Z_1(u;\xi;q)^2Z_2(u;\xi;q)
\nonumber\\
&+8Z_1(u;\xi;q)Z_3(u;\xi;q)+3Z_2(u;\xi;q)^2-6Z_4(u;\xi;q)
\Biggr). 
\end{align}
It is given by
\begin{align}
\mathcal{Z}(4;u;\xi;q)
&=
\frac{1}{24}
\Biggl[
P_1 \left[
\begin{matrix}
\xi\\
1\\
\end{matrix}
\right]^4 (\nu,\tau)
-\frac{6}{u}
P_1\left[
\begin{matrix}
\xi\\
1\\
\end{matrix}
\right]^2 (\nu,\tau)
P_2\left[
\begin{matrix}
q \xi^2\\
1\\
\end{matrix}
\right](\nu,\tau)
\nonumber\\
&+
\frac{4}{u^2}
P_1\left[
\begin{matrix}
\xi\\
1\\
\end{matrix}
\right](\nu,\tau)
\left(
P_2\left[
\begin{matrix}
q^2\xi^3\\
1\\
\end{matrix}
\right](\nu,\tau)
+
2P_3\left[
\begin{matrix}
q^2\xi^3\\
1\\
\end{matrix}
\right](\nu,\tau)
\right)
+\frac{1}{u^2}
P_2\left[
\begin{matrix}
q\xi^2\\
1\\
\end{matrix}
\right]^2 (\nu,\tau)
\nonumber\\
&-\frac{2}{u^3}
\left(
P_2\left[
\begin{matrix}
q^3\xi^4\\
1\\
\end{matrix}
\right](\nu,\tau)
+3P_3\left[
\begin{matrix}
q^3\xi^4\\
1\\
\end{matrix}
\right](\nu,\tau)
+3P_4\left[
\begin{matrix}
q^3\xi^4\\
1\\
\end{matrix}
\right](\nu,\tau)
\right)
\Biggr]. 
\end{align}
Now we have
\begin{equation}
\begin{aligned}
\mathcal{Z}(4;\xi;\xi;q)&=-\frac14 Z_4(\xi;\xi;q)+\frac{1}{8}Z_2(\xi;\xi;q)^2. 
\end{aligned}
\end{equation}
Making use of the twisted Weierstrass functions, we can express the Schur index in terms of the twisted Weierstrass function:
\begin{align}
\label{I_4}
\mathcal{I}^{U(4)}(\xi;q)
&=
-\frac{\xi^5}{24}
\frac{\theta(\xi)}{\theta(\xi^3)}
\Biggl(
3\xi P_2\left[
\begin{matrix}
q\xi^2\\
1\\
\end{matrix}
\right]^2(\zeta,\tau)
-2P_2\left[
\begin{matrix}
q^3\xi^4\\
1\\
\end{matrix}
\right](\zeta,\tau)
\nonumber\\
&
-6P_3\left[
\begin{matrix}
q^3\xi^4\\
1\\
\end{matrix}
\right](\zeta,\tau)
-6P_4\left[
\begin{matrix}
q^3\xi^4\\
1\\
\end{matrix}
\right](\zeta,\tau)
\Biggr). 
\end{align}

The normalized $U(4)$ Schur index is given by a polynomial in 
the Kronecker theta function and Weierstrass functions
\begin{align}
\widetilde{\mathcal{I}}^{U(4)}(\xi;q)
&=\xi^{-8}\mathcal{I}^{U(4)}(\xi;q)
\nonumber\\
&=\frac{1}{24}
\Biggl[
3F(\xi;\xi^{-2};q)F(\xi^3;\xi^{-2};q)
\Bigl(2P_1(\zeta,\tau)-1\Bigr)^2
\nonumber\\
&+F(\xi^2;\xi^{-4};q)
\Biggl\{
\Bigl(P_1(\zeta,\tau)+P_1(3\zeta,\tau)\Bigr)^3
-6\Bigl(P_1(\zeta,\tau)+P_1(3\zeta,\tau)\Bigr)^2
\nonumber\\
&+
3\Bigl(
P_1(\zeta,\tau)+P_1(3\zeta,\tau)
\Bigr)
\Bigl(
P_2(\zeta,\tau)-P_2(3\zeta,\tau)
\Bigr)
\nonumber\\
&+11\Bigl(
P_1(\zeta,\tau)+P_1(3\zeta,\tau)
\Bigr)
-6\Bigl(
P_2(\zeta,\tau)-P_2(3\zeta,\tau)
\Bigr)
\nonumber\\
&+2\Bigl(
P_3(\zeta,\tau)+P_3(3\zeta,\tau)
\Bigr)-6
\Biggr\}
\Biggr]. 
\end{align}

\subsubsection{$N=5$}
When $N=5$, there are seven Young diagrams $\tiny{\yng(1,1,1,1,1)}$, $\tiny{\yng(2,1,1,1)}$, $\tiny{\yng(3,1,1)}$, $\tiny{\yng(3,2)}$, $\tiny{\yng(2,2,1)}$, $\tiny{\yng(4,1)}$ and $\tiny{\yng(5)}$ which contribute to the canonical partition function. 
It is given by
\begin{align}
\label{fermi_pfn5}
&\mathcal{Z}(5;u;\xi;q)
=
\frac{1}{120}
\Biggl(
Z_1(u;\xi;q)^5
-10Z_1(u;\xi;q)^3Z_2(u;\xi;q)
\nonumber\\
&
+20Z_1(u;\xi;q)^2Z_3(u;\xi;q)
-20Z_2(u;\xi;q)Z_3(u;\xi;q)
\nonumber\\
&
+15Z_1(u;\xi;q)Z_2(u;\xi;q)^2
-30Z_1(u;\xi;q)Z_4(u;\xi;q)
+24Z_5(u;\xi;q)
\Biggr).
\end{align}
From (\ref{Z_1}) and (\ref{Z_2})-(\ref{Z_5}) we find
\begin{align}
\label{fermi_pfn5_P}
\mathcal{Z}(5;u;\xi;q)
&=
\frac{1}{120}
\Biggl[
P_1\left[
\begin{matrix}
\xi\\
1\\
\end{matrix}
\right]^5-\frac{10}{u}
P_1\left[
\begin{matrix}
\xi\\
1\\
\end{matrix}
\right]^3
P_2
\left[
\begin{matrix}
q\xi^2\\
1\\
\end{matrix}
\right]
\nonumber\\
&+\frac{1}{u^2}
\left(
15 P_1\left[
\begin{matrix}
\xi\\
1\\
\end{matrix}
\right]
P_2\left[
\begin{matrix}
q\xi^2\\
1\\
\end{matrix}
\right]^2
+10
P_1\left[
\begin{matrix}
\xi\\
1\\
\end{matrix}
\right]^2
\left(
P_2\left[
\begin{matrix}
q^2\xi^3\\
1\\
\end{matrix}
\right]
+2P_3\left[
\begin{matrix}
q^2\xi^3\\
1\\
\end{matrix}
\right]
\right)
\right)
\nonumber\\
&+
\frac{1}{u^3}
\Biggl(
-10
P_2\left[
\begin{matrix}
q\xi^2\\
1\\
\end{matrix}
\right]
\left(
P_2\left[
\begin{matrix}
q^2\xi^3\\
1\\
\end{matrix}
\right]
+2P_3\left[
\begin{matrix}
q^2\xi^3\\
1\\
\end{matrix}
\right]
\right)
\nonumber\\
&-10P_1\left[
\begin{matrix}
\xi\\
1\\
\end{matrix}
\right]
\left(
P_2\left[
\begin{matrix}
q^3\xi^4\\
1\\
\end{matrix}
\right]
+3P_3\left[
\begin{matrix}
q^3\xi^4\\
1\\
\end{matrix}
\right]
+P_4\left[
\begin{matrix}
q^3\xi^4\\
1\\
\end{matrix}
\right]
\right)
\Biggr)
\nonumber\\
&+\frac{1}{u^4}
\left(
6P_2\left[
\begin{matrix}
q^4\xi^5\\
1\\
\end{matrix}
\right]
+22P_3\left[
\begin{matrix}
q^4\xi^5\\
1\\
\end{matrix}
\right]
+36P_4\left[
\begin{matrix}
q^4\xi^5\\
1\\
\end{matrix}
\right]
+24P_5\left[
\begin{matrix}
q^4\xi^5\\
1\\
\end{matrix}
\right]
\right)
\Biggr],
\end{align}
where $P_k\left[
\begin{smallmatrix}
\theta \\
1\\
\end{smallmatrix}
\right]$ is an abbreviation for $P_k\left[
\begin{smallmatrix}
\theta \\
1\\
\end{smallmatrix}
\right](\nu,\tau)$. 

When we set $u$ to $\xi$, the canonical partition function is drastically simplified as
\begin{align}
\mathcal{Z}(5;\xi;\xi;q)&=
-\frac{1}{6}Z_2(\xi;\xi;q)Z_3(\xi;\xi;q)
+\frac{1}{5}Z_5(\xi;\xi;q). 
\end{align}
Thus we can write the $U(5)$ Schur index as
\begin{align}
\label{I_5}
&\mathcal{I}^{U(5)}(\xi;q)=
-\frac{\xi^{17/2}}{120}
\frac{\theta(\xi)}{\theta(\xi^4)}
\Biggl[
10\xi P_2\left[
\begin{matrix}
q\xi^2\\
1\\
\end{matrix}
\right](\zeta,\tau)
\left(
P_2\left[
\begin{matrix}
q^2\xi^3\\
1\\
\end{matrix}
\right](\zeta,\tau)
+
2P_3\left[
\begin{matrix}
q^2\xi^3\\
1\\
\end{matrix}
\right](\zeta,\tau)
\right)
\nonumber\\
&-6P_2\left[
\begin{matrix}
q^4\xi^5\\
1\\
\end{matrix}
\right](\zeta,\tau)
-22P_3\left[
\begin{matrix}
q^4\xi^5\\
1\\
\end{matrix}
\right](\zeta,\tau)
-36P_4\left[
\begin{matrix}
q^4\xi^5\\
1\\
\end{matrix}
\right](\zeta,\tau)
-24P_5\left[
\begin{matrix}
q^4\xi^5\\
1\\
\end{matrix}
\right](\zeta,\tau)
\Biggr]
\end{align}
in terms of the twisted Weierstrass functions. 

We can write the normalized $U(5)$ index as polynomial in 
the Kronecker theta function and the Weierstrass functions
\begin{align}
\widetilde{\mathcal{I}}^{U(5)}
&=\frac{1}{120}
\Biggl[
-10F(\xi^2;\xi^{-4};q)F(\xi^{3/2};\xi^{-3};q) 
\Bigl(2P_1(\zeta,\tau)-1\Bigr)
(p_{1,2}^2+p_{2,2}-3p_{1,2}+2)
\nonumber\\
&+F(\xi^{5/2};\xi^{-5};q)
(
p_{1,4}^4-10p_{1,4}^3+6p_{1,4}^2p_{2,4}+35p_{1,4}^2+3p_{2,4}^2-10p_{1,4}p_{2,4}+8p_{1,4}p_{3,4}
\nonumber\\
&-50p_{1,4}+35p_{2,4}-20p_{3,4}+6p_{4,4}+24
)
\Biggr]
\end{align}
where 
\begin{align}
p_{k,l}&=P_{k}(\zeta,\tau)+(-1)^{k+1}P_{k}(l \zeta,\tau). 
\end{align}

There are many other exact expressions.
We will present some of them in appendix~\ref{app_other}.

\section{Grand canonical indices}
\label{sec_gcanonical}
For the purpose of obtaining the closed-form expressions of the Schur indices, 
the Fermi-gas formulation naturally suggests us to consider the grand canonical ensemble.

\subsection{Fermi-gas grand partition function}
Let us define a grand canonical partition function by
\begin{align}
\label{Xi_DEF}
\Xi(\mu;u;\xi;q)&=1+\sum_{N=1}^{\infty} \mathcal{Z}(N;u;\xi;q) \mu^N,
\end{align}
where $\mu=e^{2\pi i\eta}$ is the fugacity and $\eta$ plays a role of the chemical potential. 
According to the Fredholm determinant, we can express the grand canonical partition function in terms of the spectral zeta functions \cite{Gaiotto:2020vqj}
\begin{align}
\label{Xi_1}
\Xi(\mu;u;\xi;q)
&=\exp\left[
-\sum_{l=1}^{\infty}\frac{(-\mu)^l}{l} Z_{l}(u;\xi;q)
\right]
\nonumber\\
&=
\prod_{p\in \mathbb{Z}}\left(
1-\frac{\mu\xi^{-p}}{1-uq^p}
\right)
=\prod_{p\in \mathbb{Z}}\frac{1-uq^{p}-\mu \xi^{-p}}{1-uq^{p}},
\end{align}
where we have plugged the infinite sum (\ref{Z_l}) into the spectral zeta functions. 

The Schur index is equal to the canonical partition function up to the quotient of the theta functions
\begin{align}
\label{eq:I_N}
\mathcal{I}^{U(N)}(\xi;q)=\Lambda(N;u;\xi;q)\mathcal{Z}(N;u;\xi;q).
\end{align}
where
\begin{align}
\label{TQ}
\Lambda(N;u;\xi;q)&:=(-1)^N\xi^{N^2/2}\frac{\theta(u;q)}{\theta(u\xi^{-N};q)}
=(-1)^N \xi^{N^2/2} \frac{\vartheta_1(\nu;\tau)}{\vartheta_1(\nu-N\zeta;\tau)}. 
\end{align}
We have 
\begin{align}
\label{TQ_prop}
\Lambda(N;\xi^{N/2};\xi;q)&=(-1)^{N+1}\xi^{N^2/2},\\
\label{TQ_sym}
\Lambda(N;u^{-1};q^{-1}\xi^{-1};q)&=(-u)^{-N}\Lambda(N;u;\xi;q). 
\end{align}
Since the $\mathcal{N}=2^{*}$ Schur index is invariant under the transformation (\ref{Sym_xi}) 
and the function (\ref{TQ}) obeys the equation (\ref{TQ_sym}),  
the canonical partition function satisfies  
\begin{align}
\label{canonical_sym}
\mathcal{Z}(N;u^{-1};q^{-1}\xi^{-1};q)&=(-u)^{N} \mathcal{Z}(N;u;\xi;q). 
\end{align}

The grand canonical partition function (\ref{Xi_1}) is not invariant under the transformation (\ref{Sym_xi}). 
Instead, it is shown to be invariant under the following extended transformations: 
\begin{align}
\label{Sym_xiumu}
\xi&\rightarrow q^{-1}\xi^{-1},\nonumber\\
u&\rightarrow u^{-1},\nonumber\\
\mu&\rightarrow -\mu u^{-1}.
\end{align}
Note that this is consistent with the transformation laws (\ref{canonical_sym}) of the canonical partition function. 

\subsection{Closed-form formula}
Making use of Cauchy's integral theorem, the canonical partition function is recovered from the grand canonical partition function as
\begin{align}
\label{pfn_Fermi_closed2}
\mathcal{Z}(N;u;\xi;q)&=
\oint \frac{d\mu}{2\pi i\mu^{N+1}}
\Xi(\mu;u;\xi;q)
\nonumber\\
&=
(-1)^N\sum_{\begin{smallmatrix}
p_1,\cdots,p_N\in \mathbb{Z}\\
p_1<\cdots<p_N\\ \end{smallmatrix}}
\prod_{i=1}^{N}
\frac{\xi^{-p_i}}{1-uq^{p_i}}. 
\end{align}
If we set $u=\xi^{N/2}$, then the prefactor $\Lambda(N; \xi^{N/2};\xi;q)$ in \eqref{eq:I_N} becomes $(-1)^{N+1}\xi^{N^2/2}$.
In this choice, the expression (\ref{pfn_Fermi_closed2}) leads to another closed-form expression of the canonical $\mathcal{N}=2^{*}$ Schur index
\begin{align}
\label{uN_findex_closed2}
\mathcal{I}^{U(N)}(\xi;q)
&=
-
\sum_{\begin{smallmatrix}
p_1,\cdots,p_N\in \mathbb{Z}\\
p_1<\cdots<p_N\\ \end{smallmatrix}}
\prod_{i=1}^{N}
\frac{\xi^{-p_i+\frac{N^2}{2}}}{1-\xi^{\frac{N}{2}} q^{p_i}}. 
\end{align}
In other words, the matrix integral (\ref{uN_findex}) is equal to the infinite series (\ref{uN_findex_closed2}). 

For example, we have 
\begin{align}
\label{u1_findex_closed2}
\mathcal{I}^{U(1)}(\xi;q)&=
-\sum_{p\in \mathbb{Z}}
\frac{\xi^{-p+\frac12}}{1-\xi^{\frac12} q^p}
,\\
\label{u2_findex_closed2}
\mathcal{I}^{U(2)}(\xi;q)&=
-\sum_{\begin{smallmatrix}
p_1, p_2\in \mathbb{Z}\\
p_1<p_2\\
\end{smallmatrix}
}
\frac{\xi^{-p_1 -p_2+2}}{(1-\xi q^{p_1}) (1-\xi q^{p_2})}, 
\\
\label{u3_findex_closed2}
\mathcal{I}^{U(3)}(\xi;q)&=
-\sum_{\begin{smallmatrix}
p_1, p_2, p_3 \in \mathbb{Z}\\
p_1<p_2<p_3\\
\end{smallmatrix}
}
\frac{\xi^{-p_1 -p_2-p_3+\frac{9}{2}}}{(1-\xi^{\frac32} q^{p_1}) (1-\xi^{\frac32} q^{p_2}) (1-\xi^{\frac32} q^{p_3})},
\\
\label{u4_findex_closed2}
\mathcal{I}^{U(4)}(\xi;q)&=
-\sum_{\begin{smallmatrix}
p_1, p_2, p_3, p_4 \in \mathbb{Z}\\
p_1<p_2<p_3<p_4\\
\end{smallmatrix}
}
\frac{\xi^{-p_1 -p_2-p_3-p_4+8}}{(1-\xi^{2} q^{p_1}) (1-\xi^{2} q^{p_2}) (1-\xi^{2} q^{p_3})(1-\xi^{2} q^{p_4})},\\
\label{u5_findex_closed2}
\mathcal{I}^{U(5)}(\xi;q)&=
-\sum_{\begin{smallmatrix}
p_1,p_2,p_3,p_4,p_5\in \mathbb{Z}\\
p_1<p_2<p_3<p_4<p_5\\
\end{smallmatrix}}
\frac{\xi^{-p_1 -p_2-p_3-p_4-p_5+\frac{25}{2}}}{(1-\xi^{\frac{5}{2}} q^{p_1}) (1-\xi^{\frac{5}{2}} q^{p_2}) (1-\xi^{\frac{5}{2}} q^{p_3})(1-\xi^{\frac{5}{2}} q^{p_4})(1-\xi^{\frac{5}{2}} q^{p_5})}. 
\end{align}
We have confirmed that the $q$-series of these results agree with 
the expansions of the matrix integral (\ref{uN_findex}) as well as those of our previous results in \eqref{I_2}-\eqref{I_5}, as expected.

We can also obtain several closed formulae for the unflavored indices. 
For $\mathcal{N}=4$ $U(2k+\delta)$ SYM theory with $\delta=0,1$,
 the unflavored Schur index is given by
\begin{align}
\label{uN_ufindex_new}
\mathcal{I}^{U(2k+\delta)}(q)&=
(-1)^k \sum_{
\begin{smallmatrix}
p_1,\cdots, p_{2k+\delta}\in \mathbb{Z} \\
p_1<\cdots<p_{2k+\delta}\\
\end{smallmatrix}
}
\prod_{i=1}^{2k+\delta}
\frac{q^{\frac{p_i}{2} -\frac{(2k+\delta) (2k+\delta-1)}{8}}}{1-q^{p_i+\frac14}}. 
\end{align}
Note that from (\ref{u_su}) the unflavored Schur indices of $\mathcal{N}=4$ $SU(2k+\delta)$ SYM theory are simply given by 
$\mathcal{I}^{SU(N)}(q)$ $=$ $\mathcal{I}^{U(N)}(q)\frac{(q^{1/2};q)_{\infty}^2}{(q;q)_{\infty}^2}$. 

Other formulae for the unflavored Schur indices can be derived by observing the fact \cite{Buican:2020moo,Kang:2021lic} that 
the unflavored Schur indices of 4d $\mathcal{N}=2$ $\widehat{\Gamma}(SU(N))$ SCFTs \cite{DelZotto:2015rca,Xie:2016evu,Buican:2016arp,Closset:2020scj,Closset:2020afy} can be realized by specializing the fugacities of the flavored Schur indices of $\mathcal{N}=4$ $SU(N)$ SYM theories. 

The 4d $\mathcal{N}=2$ SCFTs $\widehat{\Gamma}(G)$ are labeled by two simply laced Lie groups $\Gamma$ and $G$ of ADE type. 
For $\Gamma$ $=$ $D_4$, $E_6$, $E_7$ or $E_8$ 
and $\mathrm{gcd}(h_G^{\vee}, m^{\vee}_{\Gamma,\mathrm{max}})$ $=$ $1$ 
where $h_G^{\vee}$ is the dual Coxeter number of $G$ 
and $m^{\vee}_{\Gamma,\mathrm{max}}$ is the largest comark for the associated affine Dynkin diagram $\widehat{\Gamma}$, 
the SCFTs have equal central charges $a=c$ and their unflavored Schur indices can be obtained from 
those of $\mathcal{N}=4$ SYM theory as \cite{Buican:2020moo,Kang:2021lic} 
\footnote{
The fugacity $t^2$ here is the fugacity $x$ in \cite{Buican:2020moo,Kang:2021lic}. 
}
\begin{align}
\mathcal{I}^{\widehat{\Gamma}(G)}(q)&=\mathcal{I}^{G}
\left(\xi=q^{\frac{1-m^{\vee}_{\Gamma,\mathrm{max}}}{2}}; q^{m^{\vee}_{\Gamma,\mathrm{max}}}
\right). 
\end{align}
The largest comarks for $\Gamma$ $=$ $D_4$, $E_6$, $E_7$ and $E_8$ are 
\begin{align}
m^{\vee}_{D_4,\mathrm{max}}&=2,& 
m^{\vee}_{E_6,\mathrm{max}}&=3,& 
m^{\vee}_{E_7,\mathrm{max}}&=4,& 
m^{\vee}_{E_8,\mathrm{max}}&=6. 
\end{align}
The unflavored Schur indices of $\widehat{D}_4(SU(N))$ SCFTs are 
$\mathcal{I}^{\widehat{D}_4(SU(N))}(q)$ $=$ $\mathcal{I}^{U(N)}(q^2)\frac{(q;q^2)_{\infty}^2}{(q^2;q^2)_{\infty}^2}$ as argued in \cite{Kang:2021lic}. 
In addition, we can get from (\ref{pfn_Fermi_closed2}) the closed-form expressions for the unflavored Schur indices of the $\widehat{E}_k(SU(N))$ SCFTs
\begin{align}
\label{E6}
\mathcal{I}^{\widehat{E}_6(SU(N))}(q)&=
-\frac{\theta(q;q^3) \theta(q^{\frac32};q^3)}
{(q^3;q^3)_{\infty}^3 \theta(q^{\frac32+N};q^3)}
\sum_{
\begin{smallmatrix}
p_1,\cdots, p_{N}\in \mathbb{Z} \\
p_1<\cdots<p_{N}\\
\end{smallmatrix}
}
\prod_{i=1}^{N}
\frac{q^{p_i-\frac{N^2-1}{2}}}
{1-q^{3p_i+\frac32}},\\
\label{E7}
\mathcal{I}^{\widehat{E}_7(SU(N))}(q)&=
-\frac{\theta(q;q^4) \theta(q^{\frac43};q^4)}
{(q^4;q^4)_{\infty}^3 \theta(q^{\frac43+N};q^4)}
\sum_{
\begin{smallmatrix}
p_1,\cdots, p_{N}\in \mathbb{Z} \\
p_1<\cdots<p_{N}\\
\end{smallmatrix}
}
\prod_{i=1}^{N}
\frac{q^{p_i-\frac{N^2-1}{2}}}
{1-q^{4p_i+\frac43}},\\
\label{E8}
\mathcal{I}^{\widehat{E}_8(SU(N))}(q)&=
-\frac{\theta(q;q^6) \theta(q^{\frac65};q^6)}
{(q^6;q^6)_{\infty}^3 \theta(q^{\frac65+N};q^6)}
\sum_{
\begin{smallmatrix}
p_1,\cdots, p_{N}\in \mathbb{Z} \\
p_1<\cdots<p_{N}\\
\end{smallmatrix}
}
\prod_{i=1}^{N}
\frac{q^{p_i-\frac{N^2-1}{2}}}
{1-q^{6p_i+\frac65}}. 
\end{align}
We have checked that the equations (\ref{E6}), (\ref{E7}) and (\ref{E8}) precisely agree with the $q$-expansion 
evaluated in \cite{Buican:2016arp,Kang:2021lic}. 

We note that one can find identities of $q$-series by extracting other closed-form expressions from our formulae. 
For the $\widehat{E}_6(SU(2))$ SCFT it is argued in \cite{Buican:2016arp} that 
the unflavored Schur index is given by 
\begin{align}
\label{E6_su2}
\mathcal{I}^{\widehat{E}_6(SU(2))}(q)&=
\mathrm{ch}^{\mathcal{A}(6)}(-1;q)
\nonumber\\
&=\frac{1}{(q;q)_{\infty}}\sum_{n=1}^{\infty} (-1)^{n-1}n 
\left(
q^{\frac{3n^2-n-2}{2}}
-q^{\frac{3n^2+n-2}{2}}
\right). 
\end{align}
where 
\begin{align}
\mathrm{ch}^{\mathcal{A}(6)}(z;q)
&:=\Tr_{\mathcal{A}(6)}q^{L_0}z^{2A}
\nonumber\\
&=\frac{1}{(q;q)_{\infty}}
\sum_{n=1}^{\infty} \sum_{j=-\frac{n-1}{2}}^{\frac{n-1}{2}}
\left(
q^{\frac{3n^2-n-2}{2}}
-q^{\frac{3n^2+n-2}{2}}
\right)z^{2j}
\end{align}
is the vacuum character of the $\mathcal{A}(6)$ algebra \cite{MR2766985,Feigin:2007sp} 
and $A$ is an $\mathfrak{sl}(2)$ charge. 
From (\ref{E6}) we can write it as
\begin{align}
\label{E6_su2_1}
\mathcal{I}^{\widehat{E}_6(SU(2))}(q)&=
-\frac{\theta(q;q^3)\theta(q^{\frac32};q^3)}
{(q^3;q^3)_{\infty}^3\theta(q^{\frac72};q^3)}
\sum_{\begin{smallmatrix}p_1,p_2\in \mathbb{Z} \\
p_1<p_2 \end{smallmatrix}}
\frac{q^{p_1-p_2-\frac32}}{(1-q^{3p_1+\frac32}) (1-q^{3p_1+\frac32})}. 
\end{align}
On the other hand, from the canonical partition function we also find
\begin{align}
\label{E6_su2_2}
\mathcal{I}^{\widehat{E}_6(SU(2))}(q)&=
-\frac{(q^2;q^2)_{\infty} (q;q^3)_{\infty}}{(q^3;q^3)_{\infty}^2}
\sum_{\begin{smallmatrix}p_1,p_2\in \mathbb{Z} \\
p_1<p_2 \end{smallmatrix}}
\frac{q^{2p_1+2p_2-4}}{(1-q^{3p_1-2}) (1-q^{3p_2-2})}. 
\end{align}
These expressions are equal to 
\begin{align}
\label{E6_su2_3}
\mathcal{I}^{\widehat{E}_6(SU(2))}(q)&=
q^{-1}
\Bigl(
G_1(\tau,3\tau)-\frac12
\Bigr), 
\end{align}
which results from the formulae (\ref{u1_findex2}) and (\ref{I_2}). 
This reduces to a rather simple expression
\begin{align}
\mathcal{I}^{\widehat{E}_6(SU(2))}(q)&=
\sum_{n=1}^{\infty}
\frac{q^{n-1}}{1+q^n+q^{2n}}. 
\end{align}
It would be nice to rederive our formulae (\ref{E6})-(\ref{E8}) from the vacuum characters of the assoicated VOAs 
and to address them from their holographic duals.

\subsection{Unflavored Schur index}
Let us define the grand canonical unflavored Schur index by
\begin{align}
\label{Omega_DEF}
\Omega(\mu;q)=\sum_{N=0}^{\infty} \frac{\mathcal{I}^{U(N)} (q)}{\mathcal{I}^{U(\infty)} (q)}
q^{\frac{N^2}{8}}
\mu^N. 
\end{align}
It is proposed in \cite{Arai:2020qaj} that 
the factor $\mathcal{I}^{U(N)} (q)/\mathcal{I}^{U(\infty)}(q)$ can be interpreted as 
the contributions from the D3-branes wrapping supersymmetric cycles. 
We find that 
\begin{align}
\label{Omega_1}
\Omega(\mu;q)=
1+2\sum_{N=1}^{\infty}T_{N}\left(\frac{\mu}{2}\right)
q^{\frac{N^2}{8}},
\end{align}
where $T_n(x)$ are the $n$-th Chebyshev polynomials of the first kind which are defined by the relation
\begin{align}
\label{Chebyshev}
T_{n}(\cos x)&=\cos(n x). 
\end{align}
The expression turns out to be consistent with the result in \cite{MR3028756} and 
the expressions of the canonical unflavored Schur indices in terms of MacMahon's generating functions from the generalized sums of divisors discussed in section \ref{sec_combi}. 

Note that the grand canonical unflavored Schur index (\ref{Omega_1}) is simply the Jacobi theta function
\begin{align}
\label{Omega_2}
\Omega(\mu;q)&=
\vartheta_3\left(
\frac{ \cos^{-1}(\mu/2) }{2\pi}; \frac{\tau}{4}
\right),
\end{align}
by definition and that the large $N$ index is given by (\ref{large_uN_ufindex2}). 
Thus the full grand canonical unflavored Schur index is given by
\begin{align}
\label{full_Omega}
\widetilde{\Omega}(\mu;q)&:=\sum_{N=0}^{\infty}\mathcal{I}^{U(N)}(q)q^{\frac{N^2}{8}}\mu^N
\nonumber\\
&=\frac{
\vartheta_3\left(
\frac{ \cos^{-1}(\mu/2) }{2\pi}; \frac{\tau}{4}
\right)
}{\vartheta_4}. 
\end{align}
Making use of the identity \cite{MR4286926}
\begin{align}
\vartheta_3(z;\tau)&=\vartheta_3(2z,4\tau)+\vartheta_2(2z,4\tau), 
\end{align}
the expression (\ref{full_Omega}) is shown to agree with the result in \cite{Bourdier:2015wda}. 

\subsection{Half-BPS index}
Also we can consider the grand canonical half-BPS index. 
It is given by
\begin{align}
\label{gHS}
\Xi_{\frac12\textrm{BPS}}(\mu;\mathfrak{q})
&=1+\sum_{N=1}^{\infty} \mathcal{I}^{U(N)}_{\frac12\textrm{BPS}}(\mathfrak{q})\mu^N
\nonumber\\
&=\sum_{N=0}^{\infty} \frac{\mu^N}{(\mathfrak{q};\mathfrak{q})_{N}}
=\frac{1}{(\mu;\mathfrak{q})_{\infty}},
\end{align}
where we have used the identity
\begin{align}
\frac{1}{(x;q)_{\infty}}&=\sum_{n=0}^{\infty}\frac{x^n}{(q;q)_{n}}. 
\end{align}
When the resulting function (\ref{gHS}) is expanded as
\begin{align}
\frac{1}{(\mu;\mathfrak{q})_{\infty}}&=
\sum_{n=0}^{\infty}\sum_{N=0}^{\infty}p(n,N)\mathfrak{q}^{n}\mu^N, 
\end{align}
then the coefficient $p(n,N)$ is the number of partitions of $n$ with $N$ parts \cite{MR1634067}. 
It implies the correspondence between 
the $1/2$ BPS local operators of dimension $n$ in $\mathcal{N}=4$ $U(N)$ SYM theory 
and partitions of $n$ with $N$ parts. 

\section{Combinatorics}
\label{sec_combi}

\subsection{Unflavored Schur indices}
When $t\rightarrow 1$, the flavored Schur index (\ref{uN_GOindex}) reduces to the unflavored Schur index of $\mathcal{N}=4$ $U(N)$ SYM theory
\begin{align}
\label{uN_ufindex}
\mathcal{I}^{U(N)}(q)&=
\frac{q^{-\frac{N^2}{8}} \eta^{3N}(\tau)}{N!}\int_{0}^{1}\prod_{i=1}^{N} d\alpha_i 
\frac{\prod_{i<j} \vartheta_1^2 (\alpha_i-\alpha_j)}{\prod_{i,j} \vartheta_4^2(\alpha_i-\alpha_j)}. 
\end{align}
It can be expressed as a closed-form \cite{Bourdier:2015wda}
\begin{align}
\label{uN_ufindex2}
\mathcal{I}^{U(N)}(q)&=
\frac{1}{\vartheta_4}\sum_{n=0}^{\infty} 
(-1)^n \left[
\left(\begin{matrix}
N+n\\
N\\
\end{matrix}
\right)
+
\left(\begin{matrix}
N+n-1\\
N\\
\end{matrix}
\right)
\right]q^{\frac{n^2+Nn}{2}}
\nonumber\\
&=\frac{1}{N!}\frac{1}{\vartheta_4}
\sum_{n=0}^{\infty} 
(-1)^n
\frac{(N+2n) (N+n-1)!}{n!}q^{\frac{n^2+Nn}{2}}. 
\end{align}
In the following we discuss that there is another expression of the unflavored Schur indices of 4d $\mathcal{N}=4$ SYM theory with unitary gauge groups in terms of several generating functions for partitions. 

\subsection{Sum of triangular numbers}
The unflavored Schur index of 4d $\mathcal{N}=4$ $U(1)$ SYM theory is 
\begin{align}
\label{u1_ufindex}
\mathcal{I}^{U(1)}(q)&=
\frac{(q)_{\infty}^2}{(q^{\frac12};q)_{\infty}^2}
=\frac{(q)_{\infty}^4}{(q^{\frac12};q^{\frac12})_{\infty}^2}
=e^{-\frac{\pi i\tau}{4}} 
\frac{\eta(\tau)^4}{\eta(\tau/2)^2}
\nonumber\\
&=e^{-\frac{\pi i\tau}{4}} \frac{\eta(\tau)^3}{\vartheta_4}
=\frac12 e^{-\frac{\pi i\tau}{4}} \vartheta_2 \vartheta_3
\nonumber\\
&=1+2q^{1/2}+q+2q^{3/2}+2q^2+3q^3+2q^{7/2}+2q^{9/2}+\cdots.
\end{align}
The equality in the second line follows from the identity (\ref{theta234}). 

We can also write the $U(1)$ index (\ref{u1_ufindex}) as
\begin{align}
\label{u1_ufindex2}
\mathcal{I}^{U(1)}(q)&=
\psi(q^{1/2})^2,
\end{align}
where
\begin{align}
\label{ramanujan_psi}
\psi(q)&
=f(q,q^3)
=\sum_{n=0}^{\infty} q^{\frac{n(n+1)}{2}}
=\frac12 \sum_{n\in \mathbb{Z}} q^{\frac{n(n+1)}{2}}
\nonumber\\
&=(-q;q)_{\infty}(q^2;q^2)_{\infty}
=\frac{(q^2;q^2)_{\infty}}{(q;q^2)_{\infty}}
=\frac12 q^{-1/8}\vartheta_2(\tau)
\nonumber\\
&=1+q+q^3+q^6+q^{10}+q^{15}+q^{21}+q^{28}+q^{36}+\cdots ,
\end{align}
and 
\begin{align}
f(a,b):=\sum_{n\in \mathbb{Z}}a^{\frac{n(n+1)}{2}}b^{\frac{n(n-1)}{2}},
\end{align}
is the Ramanujan theta function \cite{MR1117903}. 
The function (\ref{ramanujan_psi}) is a generating function for the triangular numbers, 
the number of solution of 
\begin{align}
\frac{x(x+1)}{2}&=n,
\end{align}
where $x$ is a non-negative integer. 
More generally, a function 
\begin{align}
\psi(q^{1/2})^k&=
\sum_{n=0}t_k(n)q^\frac{n}{2},
\end{align}
is known as a generating function for the representations $t_k(n)$ of $n$ as a sum of the $k$ triangular numbers, 
the number of solutions of 
\begin{align}
\frac{x_1(x_1+1)}{2}+\frac{x_2(x_2+1)}{2}+\cdots +\frac{x_k(x_k+1)}{2},
\end{align}
where $x_i$ are non-negative integers. 
For $k=2$ we have 
\begin{align}
\begin{array}{c|c|c}
n&\frac{x_1(x_1+1)}{2}+\frac{x_2(x_2+1)}{2}&t_k(n) \\ \hline 
0&0+0&1 \\
1&0+1, 1+0&2\\
2&1+1&1 \\
3&0+3, 3+0&2\\
4&1+3. 3+1&2\\
5&-&0\\
6&0+6, 6+0, 3+3&3\\
7&1+6, 6+1&2 \\
8&-&0 \\
9&3+6, 6+3&2 \\
10&0+10,10+0 &2\\
\end{array}
\end{align}
Thus we have the following correspondence 
between the BPS local operators of dimension $n/2$ in $\mathcal{N}=4$ $U(1)$ SYM theory counted by the unflavored Schur index 
and the representations of integer $n$ as a sum of two triangular numbers. 

We also note that the $U(1)$ index can be expanded as
\begin{align}
\label{u1_ufindex3}
\mathcal{I}^{U(1)}(q)
&=\sum_{n\in \mathbb{Z}}
\frac{(-1)^n q^{\frac12 n(n+1)}}{1-q^{n+\frac12}}
\nonumber\\
&=
\sum_{n\in \mathbb{Z}}\frac{q^{\frac{n}{2}}}{1-q^{n+\frac14}},
\end{align}
where the first expansion is obtained from the identity \cite{MR2180457} involving the generating function for 
the sums of triangular numbers. 
The second expression follows from the closed-form expression (\ref{uN_ufindex_new}). 

\subsection{Generalized sum of divisors}
Next consider the unflavored Schur index of $U(2)$ SYM theory. 
From the formula (\ref{I_2}) it reads
\begin{align}
\label{u2_ufindex}
\mathcal{I}^{U(2)}(q)&=
\frac12 \sum_{n\in \mathbb{Z}}\frac{nq^{\frac{n-1}{2}}}{1-q^n}
=\sum_{n>0}\frac{nq^{\frac{n-1}{2}}}{1-q^n}. 
\end{align}
Noticing that 
\begin{align}
\sum_{n=0}^{\infty} \frac{zq^n}{(1-zq^n)^2}
=\sum_{n=0}^{\infty} \sum_{m=1}^{\infty} m(zq^n)^m
=\sum_{m=1}^{\infty} \frac{mz^m}{1-q^m}, 
\end{align}
the index (\ref{u2_ufindex}) can be rewritten as
\begin{align}
\mathcal{I}^{U(2)}(q)&=
\sum_{m>0}\frac{q^{m-1}}{(1-q^{m-\frac12})^2}=C_1(q^{\frac12})q^{-\frac12},
\end{align}
where
\begin{align}
C_1(q)&=\sum_{m>0} \frac{q^{2m-1}}{(1-q^{2m-1})^2}=\sum_{n}c_{1,n}q^n,
\end{align}
is MacMahon's generating function \cite{MR1576612} for the sum
\begin{align}
c_{1,n}&=\sum_{\lambda} s,
\end{align}
over all partitions $\lambda$ of $n$ into 
\begin{align}
n=s (2m-1),
\end{align}
where $s$ and $m$ are positive integers. 

For $U(2k)$ with $k>1$ we find that 
\begin{align}
\mathcal{I}^{U(2k)}(q)&=
C_k(q^{\frac12})q^{-\frac{k^2}{2}},
\end{align}
where 
\begin{align}
\label{MacMahon_Ck}
C_k(q)&=\sum_{0<m_1<\cdots<m_k}
\frac{q^{2m_1+\cdots +2m_k -k}}
{(1-q^{2m_1 -1})^2 \cdots (1-q^{2m_k -1})^2}. 
\end{align}
is known as a generating function for the generalized sum of divisors functions or partitions \cite{MR1576612}. 
If one expands the function (\ref{MacMahon_Ck}) as
\begin{align}
\label{MacMahon_Ck_exp}
C_k(q)&=\sum_{n=1}^{\infty} c_{k,n} q^n, 
\end{align}
then the coefficients $c_{n,k}$ are given by
\begin{align}
c_{k,n}&=\sum_{\lambda} s_1 \cdots s_k,
\end{align}
where the sum is taken over partitions $\lambda$ of $n$ into
\begin{align}
n=s_1 (2m_1-1)+s_2 (2m_2-1)+\cdots +s_k (2m_k-1),
\end{align}
with the restriction $0<m_1<m_2<\cdots <m_k$ and $s_i>0$. 
For example, for $\mathcal{N}=4$ $U(2)$ SYM theory 
each coefficient in the expansion of Schur index is the sum over all divisors whose conjugate is an odd number. 

For $N=2k+1$, $k\ge 1$ we find that 
\begin{align}
\mathcal{I}^{U(2k+1)}(q)&=
\frac{(q;q)_{\infty}^3 (-q^{\frac12};q^{\frac12})_{\infty}}
{(q^{\frac12};q^{\frac12})_{\infty}}A_{k}(q) q^{-\frac{k(k+1)}{2}}
\nonumber\\
&=
\frac{(q)_{\infty}^2}{(q^{\frac12};q)_{\infty}^2}A_{k}(q) q^{-\frac{k(k+1)}{2}},
\end{align}
where 
\begin{align}
\label{MacMahon_Ak}
A_k(q)&=\sum_{0<m_1<\cdots<m_k}
\frac{q^{m_1+\cdots +m_k}}{(1-q^{m_1})^2\cdots (1-q^{m_k})^2},
\end{align}
is also a generating function for the generalized sum of divisors functions or partitions \cite{MR1576612}. 
When we write 
\begin{align}
A_k(q)&=\sum_{n=1}^{\infty} a_{k,n}q^n,
\end{align}
then the coefficients $a_{n,k}$ are
\begin{align}
a_{k,n}&=\sum s_1 \cdots s_k,
\end{align}
where the sum is taken over partitions of $n$ into
\begin{align}
n=s_1 m_1+s_2m_2+\cdots s_k m_k,
\end{align}
with the condition $0<m_1<\cdots <m_k$ and $s_i>0$. 
For example, when $k=1$ it is a generating function for the sum of divisors function
\begin{align}
\label{divisor_s}
\sigma_k(n)&=\sum_{d|n}d^k,
\end{align}
in that
\begin{align}
A_1(q)
&=\sum_{m=1}^{\infty} \frac{q^m}{(1-q^m)^2}
=\sum_{n=1}^{\infty}\sigma_1(n)q^n. 
\end{align}

Making use of the relation (\ref{u_su}) we have%
\footnote{
Our result for the unflavored $\mathcal{N}=4$ $SU(2k)$ Schur indices differs from the result in \cite{Huang:2022bry} 
by the extra powers of $q$ and the Ramanujan theta function. 
}
\begin{align}
\label{su2k_ufindex}
\mathcal{I}^{SU(2k)}(q)&=\frac{(q^{\frac12};q)_{\infty}^2}{(q)_{\infty}^2}
C_k(q^{\frac12}) q^{-\frac{k^2}{2}},\\
\label{su2k+1_ufindex}
\mathcal{I}^{SU(2k+1)}(q)
&=A_k(q)q^{-\frac{k(k+1)}{2}}. 
\end{align}
We see that 
the relation (\ref{su2k+1_ufindex}) for odd ranks of gauge groups which was found in \cite{Kang:2021lic} can be generalized to arbitrary ranks of unitary gauge groups. 

It follows from (\ref{u1_ufindex3}) that the unflavored $U(1)$ index satisfies a differential equation
\begin{align}
\label{u1_diff}
q\frac{d}{dq}\log \mathcal{I}^{U(1)}(q)
&=
-3A_1(q)+C_1(q^{\frac12}). 
\end{align}
Also the functions $A_k(q)$ and $C_k(q)$ obey \cite{MR3028756}
\begin{align}
\label{Ak_diff}
A_k(q)&=\frac{1}{2k (2k+1)} 
\left[
(6A_1(q)+k(k-1))A_{k-1}(q)
-2q\frac{d}{dq}A_{k-1}(q)
\right],\\
\label{Ck_diff}
C_k(q)&=\frac{1}{2k(2k-1)}
\left[
(2C_1(q)+(k-1)^2)C_{k-1}(q)
-q\frac{d}{dq}C_{k-1}(q)
\right]. 
\end{align}
From relations
\begin{align}
A_1(q)&=\frac{1-E_2(\tau)}{24},\\
C_1(q)&=\frac{E_2(2\tau)-E_2(\tau)}{24},
\end{align}
where $E_k(\tau)$ is the Eisenstein series (\ref{Ek}), 
together with the recurrence relations (\ref{Ak_diff}), (\ref{Ck_diff}), 
it is shown \cite{MR3028756} that the functions $A_k(q)$ and $C_k(q)$ are the quasi-modular forms 
since the ring of the quasi-modular forms is generated by the Eisenstein series $E_2(\tau)$, $E_4(\tau)$ and $E_6(\tau)$ of 
weights $2$, $4$ and $6$ or by $q\frac{d}{dq}$ and $E_2(\tau)$ \cite{MR1363056}. 
This shows that the normalized unflavored Schur indices of $\mathcal{N}=4$ SYM theory of (special) unitary gauge groups belong to the algebra of quasi-modular forms. 

For example, we have
\begin{align}
\mathcal{I}^{SU(5)}(q)&=\frac{q^{-3}}{5760}\left[ 27+5(E_2(\tau)-6)E_2(\tau)-2E_4(\tau) \right], \\
\mathcal{I}^{U(4)}(q)&=\frac{q^{-2}}{3456}\Bigl[
-E_2(\tau)^2+2E_4(\tau)-2E_2(\tau)(E_2(\tau/2)-6)
\nonumber\\
&+2E_2(\tau/2)(E_2(\tau/2)-6)-E_4(\tau/2)
\Bigr],\\
\mathcal{I}^{SU(7)}(q)&=\frac{q^{-6}}{2903040}
\Bigl[
2025-210E_4(\tau)
\nonumber\\
&
+7E_2(\tau)
\Bigl(
-333-5(-15+E_2(\tau))E_2(\tau)
+6E_4(\tau)
\Bigr)-16E_6(\tau)
\Bigr], 
\end{align}
\begin{align}
\mathcal{I}^{U(6)}(q)&=
\frac{q^{-9}}{1244160}
\Bigl[
-6E_2^3(\tau/2)-4E_6(\tau/2)+6E_2(\tau/2)^2\left( 20+E_2(\tau) \right)
\nonumber\\
&-3E_4(\tau/2)(20+E_2(\tau))+3E_2(\tau/2)
\Bigl(
192-20E_2(\tau)+E_2(\tau)^2-6E_4(\tau)
\Bigr)
\nonumber\\
&
+120E_4(\tau)+3E_2(\tau/2)
\Bigl(
3E_4(\tau/2)+
(-40+E_2(\tau))E_2(\tau)
\nonumber\\
&-2(96+E_4(\tau))
\Bigr)
+16E_6(\tau)
\Bigr],
\end{align}
\begin{align}
\mathcal{I}^{SU(9)}(q)&=\frac{q^{-{10}}}{278691840}
\Bigl[
33075-980E_2(\tau)^3+35E_2(\tau)^4
\nonumber\\
&
+E_2(\tau)^2(9870-84E_4(\tau))
-12E_4(\tau)(329+E_4(\tau))-448E_6(\tau)
\nonumber\\
&
+4E_2(\tau)(-9687+294E_4(\tau)+16E_6(\tau))
\Bigr], \\
\mathcal{I}^{U(8)}(q)&=\frac{q^{-{16}}}{836075520}
\Bigl[
24E_2(\tau/2)^4-15E_4(\tau/2)^4-672E_6(\tau/2)
\nonumber\\
&+62208E_2(\tau)-16E_6(\tau/2)E_2(\tau)-7056E_2(\tau)^2+504E_2(\tau)^3
\nonumber\\
&-15E_2(\tau)^4-24E_2(\tau/2)^3(42+E_2(\tau))+14112E_4(\tau/2)
\nonumber\\
&
-3024E_2(\tau)E_4(\tau)+180E_2(\tau)^2E_4(\tau)+156E_4(\tau)^2
\nonumber\\
&+6E_4(\tau/2)
\Bigl(
(-84+E_2(\tau))E_2(\tau)-2(588+E_4(\tau))
\Bigr)
\nonumber\\
&+4E_2(\tau/2)
\Bigl\{
16E_6(\tau/2)+9E_4(\tau/2)(42+E_2(\tau))
\nonumber\\
&
-3\Bigl(
5184+E_2(\tau)(1176-42E_2(\tau)+E_2(\tau)^2-6E_4(\tau))
+84E_4(\tau)
\Bigr)
-16E_6(\tau)
\Bigr\}
\nonumber\\
&+64(42-5E_2(\tau))E_6(\tau)
\Bigr]. 
\end{align}

\subsection{Overpartitions}
In the large $N$ limit, the unflavored Schur index (\ref{uN_ufindex}) of $\mathcal{N}=4$ $U(N)$ SYM theory is given by
\begin{align}
\label{large_uN_ufindex}
\mathcal{I}^{U(\infty)}(q)
&=1+2q^{1/2}+4q+8q^{3/2}+14q^2+24q^{5/2}+40q^3+64q^{7/2}+\cdots.
\end{align}
We observe that the large $N$ index (\ref{large_uN_ufindex}) is identified with a generating function for the overpartition \cite{MR2034322}
\begin{align}
\label{large_uN_ufindex2}
\mathcal{I}^{U(\infty)}(q)
&=\frac{1}{\vartheta_4}
=\frac{1}{(q)_{\infty} (q^{\frac12};q)_{\infty}^2}
\nonumber\\
&=\prod_{n=1}^{\infty}\frac{1+q^{\frac{n}{2}}}{1-q^{\frac{n}{2}}}
=\prod_{n=1}^{\infty}\frac{1}{(1-q^{\frac{n}{2}}) (1-q^{n-\frac12})}
\nonumber\\
&
=\sum_{n=0}\overline{p}(n)q^{\frac{n}{2}}. 
\end{align}
Here $\overline{p}(n)$ is the number of overpartitions where an overpartition of $n$ is a partition of $n$ 
in which the first occurrence of a number may be overlined. 
By convention we define $\overline{p}(0)=1$. 
The overpartition of $n$ is equivalent to the $(2,1)$-colored partition of $n$ \cite{MR4254766}. 
It is also given by \cite{MR4364153}
\begin{align}
\overline{p}(n)&=
\sum (1+s_1) (1+s_3) \cdots (1+s_{2[n/2]-1}),
\end{align}
where the sum is taken over partitions of $n$ into 
\begin{align}
n&=s_1+s_2\cdot 2+\cdots s_n\cdot n. 
\end{align}
The asymptotic growth of the number of operators or 
equivalently that of overpartitions is evaluated from the Meinardus Theorem \cite{MR62781, MR1634067}. 
It is given by
\begin{align}
\label{asym_uN_ufindex}
\overline{p}(n)&\sim 
\frac{1}{8n} \exp\left[\pi n^{1/2} \right]. 
\end{align}
where the sum 
The exact numbers $\overline{p}(n)$ of the operators and the values $\overline{p}_{\textrm{asymp}}(n)$ evaluated from (\ref{asym_uN_ufindex}) are listed as follows: 
\begin{align}
\label{asym_uN_table}
\begin{array}{c|c|c}
n&\overline{p}(n)&\overline{p}_{\textrm{asymp}}(n) \\ \hline 
10&232&257.8973 \\
100&5.32874\times 10^{10}&5.50394\times 10^{10} \\
1000&1.72936\times 10^{39}&1.74694\times 10^{39} \\
5000&7.4466\times 10^{91}&7.480269\times 10^{91} \\
10000&3.41319\times 10^{131}&3.42409\times 10^{131} \\
\end{array}
\end{align}
Here we propose a refined formula of the asymptotic growth of overpartitions 
\begin{align}
\label{asym_uN_ufindex2}
\overline{p}(n)&\sim 
\frac{1}{8n}
\left( 1-\frac{1}{\pi n^{1/2}} \right)
\exp\left[
\pi n^{1/2}
\right],
\end{align}
which gives rise to a more accurate answer. 
For $n=10$ and $100$ numerical values of the formulae (\ref{asym_uN_ufindex}) are $231.9...$ and $5.328742437341...\times 10^{10}$, 
which are enough to get the exact numbers $232$ and $5.32874243734\times 10^{10}$. 
For $n=$ $1000$, $5000$ and $10000$ the asymptotic formula (\ref{asym_uN_ufindex}) is exact up to $34$, $63$ and $109$ digits!%
\footnote{
We note that there is another closed-form expression \cite{MR2195564} 
\begin{align}
\label{asym_pn3}
\overline{p}(n)&\sim \frac{1}{4n} \left(
\cosh \pi n^{1/2}- \frac{\sinh \pi n^{1/2}}{\pi n^{1/2}}
\right). 
\end{align}
}

Similarly, in the large $N$ limit the Schur index of $\mathcal{N}=4$ $SU(N)$ SYM theory is given by
\begin{align}
\label{large_suN_ufindex}
\mathcal{I}^{SU(\infty)}(q)&
=1+3q+9q^2+22q^3+51q^4+108q^5+\cdots.
\end{align}
This is simply given by (also see \cite{Honda:2022hvy})
\begin{align}
\mathcal{I}^{SU(\infty)}(q)&=\frac{1}{(q)_{\infty}^3}
=\sum_{n=0}p_{3}(n)q^{n},
\end{align}
where $p_3(n)$ is the number of the 3-colored partitions of $n$ \cite{MR1634067}, 
i.e. partitions of $n$ in which each part has one of $3$ colors. 

The asymptotic growth is evaluated from the Meinardus Theorem \cite{MR62781, MR1634067} as
\begin{align}
\label{asy_tn}
p_3(n)&\sim \frac{1}{2^{7/2}n^{3/2}}\exp\left[ 2^{1/2} \pi n^{1/2} \right]. 
\end{align}
We show the exact numbers $p_3(n)$ of the operators and the values $p_{\textrm{$3$ asymp}}(n)$ evaluated from (\ref{asy_tn}) 
\begin{align}
\label{asy_q3_table}
\begin{array}{c|c|c} 
n&p_3(n)&p_{\textrm{$3$ asymp}}(n) \\ \hline 
10&2640&3532.4 \\
100&1.58733\times 10^{15}&1.74418\times 10^{15} \\
1000&2.81917\times 10^{55}&2.90510\times 10^{55} \\
5000&6.75666\times 10^{129}&6.84818\times 10^{129} \\
10000&7.83819\times 10^{185}&7.91314\times 10^{185} \\
\end{array}
\end{align}
Unlike the $U(\infty)$ index, 
corrections to the asymptotic growth (\ref{asy_tn}) of the $SU(\infty)$ index does not seem to terminate by simply adding the next leading term. 

The leading exponential growth of the numbers of BPS local operators counted by the Schur indices 
in $\mathcal{N}=4$ SYM theory is the same as the growth (\ref{asym_1/2BPS}) for the partitions 
and that of operator degeneracy in two-dimensional CFTs \cite{Cardy:1991kr}. 

\subsection*{Acknowledgements}
The authors would like to thank Martin Halln\"{a}s, Kimyeong Lee, Sungjay Lee, Yu Nakayama, Hjalmar Rosengren for useful discussions and comments. 
The work of Y.H. is supported
in part by JSPS KAKENHI Grant No. 18K03657 and 22K03641.
The work of T.O. was supported by KIAS Individual Grants (PG084301) at Korea Institute for Advanced Study. 

\appendix

\section{Definitions and notations}
\label{app_notations}

\subsection{$q$-shifted factorial}
We have introduced the following notation by defining $q$-shifted factorial
\begin{align}
\label{qpoch_def}
(a;q)_{0}&:=1,\qquad
(a;q)_{n}:=\prod_{k=0}^{n-1}(1-aq^{k}),\qquad 
(q)_{n}:=\prod_{k=1}^{n}(1-q^{k}),\quad 
\quad  n\ge1,
\nonumber \\
(a;q)_{\infty}&:=\prod_{k=0}^{\infty}(1-aq^{k}),\qquad 
(q)_{\infty}:=\prod_{k=1}^{\infty} (1-q^k), 
\nonumber\\
(a^{\pm};q)_{\infty}&:=(a;q)_{\infty}(a^{-1};q)_{\infty},
\end{align}
where $a$ and $q$ are complex variables. 

\subsection{Jacobi theta functions}
The Jacobi theta functions are defined by
\begin{align}
\label{theta1}
\vartheta_1(z;\tau)
&=\sum_{n\in \mathbb{Z}}(-1)^{n-\frac12} q^{\frac12 (n+\frac12)^2}x^{n+\frac12}
=-\sum_{n\in \mathbb{Z}} e^{\pi i\tau (n+\frac12)^2+2\pi i(z+\frac12)(n+\frac12)}, \\
\label{theta2}
\vartheta_2 (z;\tau)
&=\sum_{n\in \mathbb{Z}} q^{\frac12 (n+\frac12)^2}x^{n+\frac12}
=\sum_{n\in \mathbb{Z}} e^{\pi i\tau (n+\frac12)^2+2\pi iz(n+\frac12)}, \\
\label{theta3}
\vartheta_3(z;\tau)
&=\sum_{n\in \mathbb{Z}} q^{\frac{n^2}{2}} x^n
=\sum_{n\in \mathbb{Z}} e^{\pi i\tau n^2+2\pi inz},\\
\label{theta4}
\vartheta_4(z;\tau)
&=\sum_{n\in \mathbb{Z}} (-1)^n q^{\frac{n^2}{2}} x^n
=\sum_{n\in \mathbb{Z}} e^{\pi in^2+2\pi i(z+\frac12) n}
\end{align}
with $q=e^{2\pi i\tau}$, $x=e^{2\pi iz}$, $\tau\in \mathbb{H}$ and $z\in \mathbb{C}$. 
The Jacobi theta function $\vartheta_1(z;\tau)$ is a holomorphic Jacobi form of weight $\frac12$ and index $\frac12$
\begin{align}
\label{theta1_tras1}
\vartheta_1\left(\frac{z}{\tau};-\frac{1}{\tau} \right)
&=-i\sqrt{\frac{\tau}{i}}e^{\frac{\pi iz^2}{\tau}}\vartheta_1(z;\tau),\\
\label{theta1_tras2}
\vartheta_1(z+1,\tau)&=-\vartheta_1(z,\tau),\\
\label{theta1_tras3}
\vartheta_1(z+\tau,\tau)&=-e^{-2\pi iz-\pi i\tau}\vartheta_1(z,\tau). 
\end{align}

We define  $\vartheta_i=\vartheta_i(0;\tau)$ to be the Jacobi theta functions with argument $z=0$. 
Then one can write 
\begin{align}
\vartheta_1&=0,\\
\vartheta_2&=2q^{1/8} (q)_{\infty} (-q;q)_{\infty}^2
=\frac{2\eta(2\tau)^2}{\eta(\tau)}, \\
\vartheta_3&=(q)_{\infty} (-q^{\frac12};q)_{\infty}^2
=\frac{\eta(\tau)^5}{\eta(\tau/2)^2\eta(2\tau)^2}, \\
\vartheta_4&=(q)_{\infty} (q^{\frac12};q)_{\infty}^2
=\frac{\eta(\tau/2)^2}{\eta(\tau)}, 
\end{align}
where $\vartheta_i=\vartheta_i(0;\tau)$. 
We have \cite{MR4286926}
\begin{align}
\label{theta234}
\vartheta_2 \vartheta_3 \vartheta_4&=2 \eta(\tau)^3, \\
\label{theta234_4}
\vartheta_4^4&=\vartheta_3^4-\vartheta_2^4, \\
\label{theta34_2}
\vartheta_2(\tau)\vartheta_3(\tau)&=\frac12 \vartheta_2(\tau/2)^2, \\
\vartheta_3(\tau)\vartheta_4(\tau)&=\vartheta_4^2(2\tau). 
\end{align}

\subsection{Eisenstein series}

\subsubsection{Classical Eisenstein series}
For $k\ge 2$ even the Eisenstein series is defined by 
\footnote{The $G_k(\tau)$ defined here are rescaled by $(2\pi i)^{-k}$. They the same as the $G_k(\tau)$ in \cite{Bringmann:2020ndq}, 
the $(2\pi i)^{-k}G_k(\tau)$ in \cite{MR1027834,Dong:1997ea}, the $E_k(\tau)$ in \cite{Mason:2008zzb}.}
\begin{align}
\label{Gk}
G_k(\tau)&=-\frac{B_k}{k!}+\frac{2}{(k-1)!}\sum_{n=1}^{\infty}\frac{n^{k-1}q^n}{1-q^n}
\nonumber\\
&=-\frac{B_k}{k!}+\frac{2}{(k-1)!}\sum_{n=1}^{\infty}\sigma_{k-1}(n)q^n,
\end{align}
where $B_k$ are the $k$-th Bernoulli numbers and $\sigma_k(n)$ is the divisor function (\ref{divisor_s}). 
We also define $G_0=-1$. 

We have
\begin{align}
G_{k}(\tau)&=
\frac{1}{(2\pi i)^k}
\sum_{(m,n)\in \mathbb{Z}^2\setminus\{(0,0)\}}
\frac{1}{(m+n\tau)^k}. 
\end{align}

For $k\ge 4$ even the Eisenstein series is a holomorphic modular form of weight $k$ on $SL(2,\mathbb{Z})$ 
whereas for $k=2$ it is a quasi-modular form \cite{MR1363056} of weight $2$
\begin{align}
\label{Gk_modular}
G_k\left( \frac{a\tau+b}{c\tau+d} \right)&=(c\tau+d)^k G_k(\tau)-\delta_{k,2}\frac{c(c\tau+d)}{2\pi i}. 
\end{align}

Alternatively we introduce 
\begin{align}
\label{Ek}
E_{k}(\tau)&=-\frac{k!}{B_k} G_k(\tau)
=1-\frac{2k}{B_{k}}\sum_{n=1}^{\infty}\sigma_{k-1}(n)q^n. 
\end{align}
It follows that
\begin{align}
\label{E2_exp}
E_2(\tau)&=P(q)
=24 q\frac{d}{dq}\log \eta(\tau),\\
\label{E22_exp}
2E_2(\tau)-E_2(\tau/2)
&=\frac{\eta(\tau)^{20}}{\eta(\tau/2)^8\eta(2\tau)^8}
+16\frac{\eta(2\tau)^8}{\eta(\tau)^4}
=\vartheta_2^4+\vartheta_3^4,\\
\label{E4_exp}
E_4(\tau)&=Q(q)
=\frac{\eta(\tau)^{16}}{\eta(2\tau)^8}
+2^8\frac{\eta(2\tau)^{16}}{\eta(\tau)^8}
=\frac12 (\vartheta_2^8+\vartheta_3^8+\vartheta_4^8),\\ 
\label{E6_exp}
E_6(\tau)&=R(q)=\frac{\eta(\tau)^{24}}{\eta(2\tau)^{12}}-2^5\cdot 3\cdot 5 \eta(2\tau)^{12}
\nonumber\\
&-2^9\cdot 3\cdot 11\frac{\eta(2\tau)^{12} \eta(4\tau)^8}{\eta(\tau)^8}
+2^{13}\frac{\eta(4\tau)^{24}}{\eta(2\tau)^{12}},\\
\label{E8_exp}
E_8(\tau)&=Q^2(q)=\frac{1}{2}(\vartheta_2^{16}+\vartheta_3^{16}+\vartheta_4^{16}), \\
\label{E10_exp}
E_{10}(\tau)&=Q(q)R(q),\\
\label{E12_exp}
E_{12}(\tau)&=\frac{1}{691}
\left(
Q(q)^3+250R(q)^2
\right), \\
\label{E14_exp}
E_{14}(\tau)&=Q(q)^2 R(q), 
\end{align}
where $P(q)$, $Q(q)$ and $R(q)$ are the functions introduced by Ramanujan \cite{MR2280861} (also see \cite{MR1837822}). 
We have differential equations 
\begin{align}
q\frac{dE_2(\tau)}{dq}&=\frac{1}{12}(E_2(\tau)^2-E_4(\tau)), \\
q\frac{dE_4(\tau)}{dq}&=\frac{1}{3}(E_2(\tau)E_4(\tau)-E_6(\tau)), \\
q\frac{dE_6}{d\tau}&=\frac12(E_2(\tau)E_6(\tau)-E_4(\tau)^2). 
\end{align}

The modular invariants of an elliptic curve are given the Eisenstein series \cite{MR1027834}
\begin{align}
\label{g2}
g_2(\tau)&=60\sum_{(m,n)\in \mathbb{Z}^2\setminus\{(0,0\}}
\frac{1}{(m+n\tau)^4}
=60(2\pi i)^4 G_4(\tau),\\
\label{g3}
g_3(\tau)&=140\sum_{(m,n)\in \mathbb{Z}^2\setminus\{(0,0\}}
\frac{1}{(m+n\tau)^6}
=140(2\pi i)^6 G_6(\tau).
\end{align}
\subsubsection{Twisted Eisenstein series}
The twisted Eisenstein series is defined by  
\footnote{
The $G_k\left[\begin{smallmatrix}\theta\\ \phi\\ \end{smallmatrix}\right](\tau)$ defined here are the same as 
the $E_k\left[\begin{smallmatrix}\theta\\ \phi\\ \end{smallmatrix}\right](\tau)$ in \cite{Mason:2008zzb}, $Q_k(\phi,\theta^{-1},\tau)$ in \cite{Dong:1997ea}. 
}
\begin{align}
\label{tGk}
&G_k\left[
\begin{matrix}
\theta\\
\phi\\
\end{matrix}
\right](\tau)=
-\frac{B_k(\lambda)}{k!}
\nonumber\\
&+\frac{1}{(k-1)!} {\sum_{n=0}^{\infty}}'
\frac{(n+\lambda)^{k-1} \theta^{-1} q^{n+\lambda}}{1-\theta^{-1}q^{n+\lambda}}
+\frac{(-1)^k}{(k-1)!}\sum_{n=1}^{\infty}\frac{(n-\lambda)^{k-1}\theta q^{n-\lambda}}{1-\theta q^{n-\lambda}},
\end{align}
where $\phi=e^{2\pi i\lambda}$ and the sum ${\sum}'$ implies that it omits $n=0$ if $(\theta,\phi)=(1,1)$. 

In particular, for $\phi=1$, we define 
\footnote{The $G_k(z,\tau)$ defined here are the same as 
the $(2\pi i)^{-k} \hat{G}_k(e^{2\pi i\tau}, e^{2\pi iz})$ $=\hat{\mathcal{G}}_k(e^{2\pi i\tau}, e^{2\pi iz})$ in \cite{Gaberdiel:2009vs}, 
the $(-1)^{k+1}J_k/k!$ in \cite{https://doi.org/10.48550/arxiv.1209.5628}, 
the $\widetilde{G}_k(z,\tau)$ in \cite{Bringmann:2020ndq}.}
\begin{align}
\label{tGk_ztau}
G_k(z,\tau)
&=G_{k}\left[
\begin{matrix}
x^{-1}\\
1\\
\end{matrix}
\right](\tau)
\nonumber\\
&=
-\frac{B_k}{k!}
+\frac{1}{(k-1)!}\left[
{\sum_{n=0}^{\infty}}'
\frac{n^{k-1}xq^n}{1-xq^n}
+(-1)^k 
\sum_{n=1}^{\infty}
\frac{n^{k-1}x^{-1}q^n}{1-x^{-1}q^n}
\right],
\end{align}
where $x=e^{2\pi iz}$ and ${\sum_{n=0}}'$ indicates that 
$n=0$ is excluded in the sum when $x=1$. 
We also define $G_0(z,\tau)=-1$. 
The twisted Eisenstein series is related to the ordinary Eisenstein series by 
\begin{align}
G_k\left[
\begin{matrix}
1\\
1\\
\end{matrix}
\right](\tau)
=G_k(0,\tau)
&=\begin{cases}
G_k(\tau)&\textrm{for $k$ even}\cr 
\frac12 \delta_{1,k}&\textrm{for $k$ odd}
\end{cases}. 
\end{align}

We have 
\begin{align}
G_k\left[
\begin{matrix}
\theta\\
\phi\\
\end{matrix}
\right](\tau)&=
\frac{1}{(2\pi i)^k}\sum_{m\in \mathbb{Z}}\theta^m 
\left[
\sum_{\begin{smallmatrix}
n\in \mathbb{Z}\\
(m,n)\neq (0,0)
\end{smallmatrix}
}
\frac{\phi^n}{(m\tau+n)^k}
\right],
\end{align}
for $\phi\neq 1$ and 
\begin{align}
G_k\left[
\begin{matrix}
\theta\\
\phi\\
\end{matrix}
\right](\tau)&=
\frac{1}{(2\pi i)^k}\sum_{n\in \mathbb{Z}}\phi^n 
\left[
\sum_{\begin{smallmatrix}
m\in \mathbb{Z}\\
(m,n)\neq (0,0)
\end{smallmatrix}
}
\frac{\theta^m}{(m\tau+n)^k}
\right],
\end{align}
for $\theta\neq 1$. 

The twisted Eisenstein has properties 
\begin{align}
\label{tGk_prop1}
G_k(-z,\tau)&=(-1)^k G_k(z,\tau),\\
\label{tGk_prop2}
G_k(z+\lambda\tau+\mu,\tau)&=\sum_{m=0}^{k}\frac{1}{m!}(-\lambda)^m G_{k-m}(z,\tau),\qquad 
\textrm{for $\lambda, \mu \in \mathbb{Z}$},\\
\label{tGk_prop3}
G_k\left(
\frac{z}{c\tau+d},\frac{a\tau+b}{c\tau+d}
\right)
&=(c\tau+d)^k \sum_{m=0}^{k} \frac{1}{m!} \left( \frac{cz}{c\tau+d} \right)^m 
G_{k-m}(z,\tau), \\
\label{tGk_prop4}
\frac{\partial}{\partial z}G_k(z,\tau)&=\frac{1}{k}\frac{\partial}{\partial \tau}G_k(z,\tau). 
\end{align}

The twisted Eisenstein series can be written in terms of the Weierstrass functions (\ref{Pk}) as well as the Eisenstein series (\ref{Gk}). 

For $k=1$ it is equivalent to the Weierstrass function (\ref{P1})
\begin{align}
G_1(z,\tau)&=-P_1(z,\tau). 
\end{align}
It is a quasi-Jacobi form of weight $1$, index $0$
\begin{align}
G_1\left(
\frac{1}{c\tau+d},\frac{a\tau+b}{c\tau+d}\right)
&=(c\tau+d)G_1(z,\tau)+2\pi icz. 
\end{align}

For $k=2$ we have 
\begin{align}
G_2(z,\tau)&=\frac12 \left(
P_2(z,\tau)-P_1(z,\tau)^2-G_2(\tau)
\right). 
\end{align}

Also the twisted Eisenstein series can be expressed in terms of the Eisenstein series and the theta function (\ref{theta_DEF})
\begin{align}
\label{tGk_ztau_theta}
&G_k(z,\tau)
\nonumber\\
&=
-\sum_{m=0}^{[\frac{k}{2}]}
\left[
\sum_{\begin{smallmatrix}
\{m_p\}\\
\sum_{p\ge 1}(2p)\cdot m_p=2m
\end{smallmatrix}
}
\prod_{p=1}^{\infty}\frac{G_{2p}^{m_p} (\tau)}{(2p)^{m_p}\cdot m_p!}
\right]
\frac{1}{(k-2m)!} 
\sum_{l=1}^{k-2m} S(k-2m,l)x^l \frac{\theta^{(l)}(x;q)}{\theta(x;q)},
\end{align}
where $S(k-2m,l)$ are the Stirling numbers of the second kind. 

For example, 
\begin{align}
\label{tG1_ztau}
G_1(z,\tau)
&=-x\frac{\theta'(x)}{\theta(x)},\\
\label{tG2_ztau}
G_2(z,\tau)&=-\frac12 \left(
x\frac{\theta'(x)}{\theta(x)}
+x^2\frac{\theta''(x)}{\theta(x)}
\right)-\frac12 G_2(\tau),\\
\label{tG3_ztau}
G_3(z,\tau)&=-\frac16 
\left( 
x\frac{\theta'(x)}{\theta(x)}
+3x^2 \frac{\theta''(x)}{\theta(x)}
+x^3\frac{\theta'''(x)}{\theta(x)}
\right)
-\frac12 G_{2}(\tau)x\frac{\theta'(x)}{\theta(x)},
\end{align}
\begin{align}
\label{tG4_ztau}
G_4(z,\tau)&=
-\frac{1}{24}\left(
x\frac{\theta'(x)}{\theta(x)}
+7x^2 \frac{\theta''(x)}{\theta(x)}
+6x^3 \frac{\theta'''(x)}{\theta(x)}
+x^4\frac{\theta''''(x)}{\theta(x)}
\right)
\nonumber\\
&
-\frac12 G_2(\tau)
\left(
\frac12 \left(
x\frac{\theta'(x)}{\theta(x)}
+x^2\frac{\theta''(x)}{\theta(x)}
\right)
\right)
-\frac14\left(G_4(\tau)+\frac12 G_2(\tau)^2 \right), \\
\label{tG5_ztau}
G_5(z,\tau)&=
-\frac{1}{120}
\Biggl(
x\frac{\theta'(x)}{\theta(x)}
+15x^2 \frac{\theta''(x)}{\theta(x)}
+25x^3 \frac{\theta'''(x)}{\theta(x)}
+10x^4 \frac{\theta''''(x)}{\theta(x)}
+x^5\frac{\theta'''''(x)}{\theta(x)}
\Biggr)
\nonumber\\
&
-\frac12 G_2(\tau)
\Biggl(
\frac16 
\left(x\frac{\theta'(x)}{\theta(x)}
+3x^2 \frac{\theta''(x)}{\theta(x)}
+x^3\frac{\theta'''(x)}{\theta(x)}
\right)
\Biggr)
\nonumber\\
&-\frac14 \left(
G_4(\tau)+\frac12 G_2^2(\tau)
\right)
x\frac{\theta'(x)}{\theta(x)},
\end{align}
\begin{align}
\label{tG6_ztau}
G_6(z,\tau)&=
-\frac{1}{720}
\Biggl(
x\frac{\theta'(x)}{\theta(x)}
+31x^2\frac{\theta''(x)}{\theta(x)}
+90x^3\frac{\theta'''(x)}{\theta(x)}
\nonumber\\
&+65x^4\frac{\theta^{(4)}(x)}{\theta(x)}
+15x^5\frac{\theta^{(5)}(x)}{\theta(x)}
+x^6\frac{\theta^{(6)}(x)}{\theta(x)}
\Biggr)
\nonumber\\
&-\frac12 G_2(\tau) \left(
\frac{1}{24}
\left(
x\frac{\theta'(x)}{\theta(x)}
+7x^2 \frac{\theta''(x)}{\theta(x)}
+6x^3 \frac{\theta'''(x)}{\theta(x)}
+x^4\frac{\theta''''(x)}{\theta(x)}
\right)
\right)
\nonumber\\
&-\frac14 \left(
G_4(\tau)+\frac12 G_2^2(\tau)
\right)
\left(\frac12
\left(
x\frac{\theta'(x)}{\theta(x)}
+x^2 \frac{\theta''(x)}{\theta(x)}
\right)
\right)
\nonumber\\
&-\frac16 \left(
G_6(\tau)+\frac34 G_4(\tau)G_2(\tau)+\frac18G_2^3(\tau)
\right). 
\end{align}
Conversely, ratios of the theta functions and their derivatives are given by the twisted Eisenstein series
\begin{align}
\label{xdx_G}
x\frac{\theta'(x)}{\theta(x)}&=-G_1(z,\tau), \\
\label{xdx2_G}
x^2\frac{\theta''(x)}{\theta(x)}&=-2G_2(z,\tau)-G_1(z,\tau)-G_2(\tau),\\
\label{xdx3_G}
x^3\frac{\theta'''(x)}{\theta(x)}&=-6G_3(z,\tau)+6G_2(z,\tau)+(2-3G_2(\tau))G_1(z,\tau)+3G_2(\tau), \\
\label{xdx4_G}
x^4\frac{\theta''''(x)}{\theta(x)}&=-24G_4(z,\tau)+36G_3(z,\tau)+2G_2(z,\tau)(-11+6G_2(\tau))
\nonumber\\
&+6G_1(z,\tau)(-1+3G_2(\tau))+G_2(\tau)^2-11G_2(\tau)-6G_4(\tau). 
\end{align}

\subsection{Weierstrass functions}

\subsubsection{Classical Weierstrass functions}
We define the Weierstrass functions 
\footnote{
The $P_k(z,\tau)$ is the same as the $P_k(z,\tau)$ in \cite{Bringmann:2020ndq}, 
the $(-2\pi i)^{-k}P_k(e^{2\pi iz},q)$ in \cite{MR1317233} and 
the $(2\pi i)^{-k} E_k(z,\tau)$ in \cite{MR1723749,MR2796409}. 
}
\begin{align}
\label{P1}
P_1(z,\tau)&:=-\sum_{n\in \mathbb{Z}\setminus \{0\}}\frac{x^n}{1-q^n}-\frac12,
\end{align}
and 
\begin{align}
\label{Pk}
P_k(z,\tau)&:=
\frac{(-1)^{k-1}}{(k-1)!} \frac{1}{(2\pi i)^{k-1}} \frac{\partial^{k-1}}{\partial z^{k-1}}
P_1(z,\tau)
\nonumber\\
&=\frac{(-1)^{k}}{(k-1)!} \sum_{n\in \mathbb{Z}\setminus \{0\}} 
\frac{n^{k-1}x^{n}}{1-q^{n}}. 
\end{align}

We have \cite{MR1723749}
\begin{align}
\label{P1_1}
P_1(z,\tau)&=\frac{1}{2\pi iz}-\sum_{k=1}^{\infty}G_k(\tau)(2\pi iz)^{k-1}, \\
\label{Pk_1}
P_k(z,\tau)&=\frac{1}{(2\pi iz)^k}+\sum_{l=k}^{\infty} 
\left(
\begin{matrix}
l-1\\
k-1\\
\end{matrix}
\right)
G_l(\tau) (2\pi iz)^{l-k}, \\
P_k(z,\tau)&=\frac{1}{(2\pi i)^k}\sum_{(m,n)\in \mathbb{Z}^2}\frac{1}{(z+m\tau+n)^k}. 
\end{align}

The $P_1(z,\tau)$ is a quasi-Jacobi form of weight $1$, index $0$ and depth $(1,0)$ \cite{MR2796409}
\begin{align}
P_1\left( \frac{z}{c\tau+d},\frac{a\tau+b}{c\tau+d} \right)
&=(c\tau+d)P_1(z,\tau)-2\pi i c z,\\
P_1(z+\lambda\tau+\mu,\tau)
&=P_1(z,\tau). 
\end{align}
It can be also expressed as
\begin{align}
\label{P1_2}
P_1(z,\tau)&=\frac{1}{2\pi i}\zeta(z,\tau)-2\pi i zG_2(\tau),
\end{align}
where 
\begin{align}
\zeta(z,\tau)&=\frac{1}{z}+\sum_{(m,n)\in \mathbb{Z}^2\setminus \{(0,0)\}}
\left(
\frac{1}{z-(m\tau+n)}
+\frac{1}{m\tau+n}
+\frac{z}{(m\tau+n)^2}
\right),
\end{align}
is the Weierstrass $\zeta$-function \cite{MR808396}. 

The $P_2(z,\tau)$ is expressed as
\begin{align}
P_2(z,\tau)&=\sum_{n\in \setminus \{0\}}
\frac{nx^n}{1-q^n}
\nonumber\\
&=-
\left(
x\frac{\theta'(x)}{\theta(x)}+x^2 \frac{\theta''(x)\theta(x)-{\theta'(x)}^2}{\theta^2(x)}
\right). 
\end{align}
We have
\begin{align}
\label{P2_1}
P_2(z,\tau)&=\frac{1}{(2\pi i z)^2}
+\sum_{k=2}^{\infty} (k-1) G_k(\tau)  (2\pi i z)^{k-2}. 
\end{align}
The $P_2(z,\tau)$ is a quasi-Jacobi form of weight $2$, index $0$ and depth $(0,1)$ \cite{MR2796409}
\begin{align}
P_2\left(
\frac{z}{c\tau+d},\frac{a\tau+b}{c\tau+d}
\right)&=(c\tau+d)^2 P_2(z,\tau)-\frac{c(c\tau+d)}{2\pi i},\\
P_2(z+\lambda\tau+\mu,\tau)&=P_2(z,\tau),
\end{align}
for $\lambda, \mu\in \mathbb{Z}$. 
It can be written as 
\begin{align}
\label{P2_2}
P_2(z,\tau)&=\frac{1}{(2\pi i)^2}\wp(z,\tau)+G_2(\tau),
\end{align}
where 
\begin{align}
\label{Weier_p}
\wp(z,\tau)&=\frac{1}{z^2}+\sum_{(m,n)\in \mathbb{Z}^2\setminus \{(0,0)\}}
\left(
\frac{1}{(z-(m\tau+n))^2}-\frac{1}{(m\tau+n)^2}
\right)
\nonumber\\
&=\left(
\pi \vartheta_2\vartheta_3 \frac{\vartheta_4(z;\tau)}{\vartheta_1(z;\tau)}
\right)^2
-\frac{\pi^2}{3}(\vartheta_2^4+\vartheta_3^4),
\end{align}
is the Weierstrass $\wp$-function with period $1$ and $\tau$ \cite{MR890960,MR808396}. 
The $\wp(z,\tau)$ is a meromorphic Jacobi form of weight $2$ and index $0$. 
We have
\begin{align}
\wp(z,\tau)&=-\frac{\partial}{\partial z}\zeta(z,\tau). 
\end{align}
One has \cite{MR1723749}
\begin{align}
P_2(z,\tau)-P_2(z+z',\tau)&=\frac{(q)_{\infty}^6 \theta(x^2x';q) \theta(x';q)}{\theta(x;q)^2\theta(xx';q)^2}. 
\end{align}

We also define 
\begin{align}
\label{P0}
P_0(z;\tau)=-\log(2\pi iz)+\sum_{k=0}^{\infty} G_k(\tau)\frac{(2\pi i z)^k}{k}, 
\end{align}
which satisfies 
\begin{align}
P_1(z,\tau)&=\frac{-1}{2\pi i} \frac{\partial}{\partial z}
P_0(z,\tau). 
\end{align}
The exponential is given by \cite{MR1723749}
\footnote{It is also known as the elliptic prime form \cite{MR2352717}  on the elliptic curve with modulus $\tau$ and as Hirzebruch's function 
\cite{MR981372,MR970278,MR1189136,MR1757003,MR1765709}. It is expressed as $\varphi(z)$ in \cite{MR1723749}. }
\begin{align}
\label{P0_1}
\exp\Bigl[-P_0(z,\tau)\Bigr]&=
\frac{i\vartheta_1(z;\tau)}{\eta(\tau)^3}
\nonumber\\
&=\exp\left[
-\frac{G_2(\tau)}{2}(2\pi iz)^2
\right]\sigma(z,\tau)
\nonumber\\
&=2\pi iz \cdot \exp \left[
-\sum_{k=1}^{\infty} \frac{G_{k}(\tau)}{k} (2\pi iz)^{k}
\right],
\end{align}
where 
\begin{align}
\label{Weier_s}
\sigma(z,\tau)=z\prod_{(m,n) \in \mathbb{Z}^2\setminus \{(0,0)\}}
\left(1-\frac{z}{m\tau+n} \right)e^{\frac{z}{m\tau+n}+\frac{z^2}{2(m\tau+n)^2}},
\end{align}
is the Weierstrass $\sigma$-function \cite{MR808396} 
which obeys
\begin{align}
\frac{\partial }{\partial z}\log \sigma(z,\tau)&=\zeta(z,\tau). 
\end{align}

For $k\ge 3$ the Weierstrass function $P_k(z,\tau)$ is a weak meromorphic Jacobi form of weight $k$ and index $0$ \cite{MR2796409}. 
One has the following identities \cite{MR1723749}
\begin{align}
P_4(z,\tau)&=(P_2(z,\tau)-G_2(\tau))^2-5G_4(\tau),\\
P_3(z,\tau)^2&=(P_2(z,\tau)-G_2(\tau))^3-15G_4(\tau)(P_2(z,\tau)-G_2(\tau))-35G_6(\tau). 
\end{align}
We have differential equations
\begin{align}
q\frac{\partial}{\partial q}P_1(z,\tau)&=P_3(z,\tau)-P_1(z,\tau)P_2(z,\tau),\\
q\frac{\partial}{\partial q}P_2(z,\tau)&=3P_4(z,\tau)-2P_1(z,\tau)P_3(z,\tau)-P_2(z,\tau)^2. 
\end{align}

\subsubsection{Twisted Weierstrass functions}
We define the twisted Weierstrass function by 
\footnote{The $P_k\left[\begin{smallmatrix}\theta\\ \phi\\ \end{smallmatrix}\right](z,\tau)$ defined here is the same as 
$P_k\left[\begin{smallmatrix}\theta\\ \phi\\ \end{smallmatrix}\right](2\pi iz,\tau)$ in \cite{Mason:2008zzb}. }
\begin{align}
\label{tP1}
P_1\left[
\begin{matrix}
\theta\\
\phi\\
\end{matrix}
\right](z,\tau)&=
-{\sum_{n\in \mathbb{Z}}}' \frac{x^{n+\lambda}}{1-\theta^{-1}q^{n+\lambda}},
\end{align}
and 
\begin{align}
\label{tPk}
P_k\left[
\begin{matrix}
\theta\\
\phi\\
\end{matrix}
\right](z,\tau)
&=
\frac{(-1)^k}{(k-1)!}
\frac{1}{(2\pi i)^{k-1}} 
\frac{\partial^{k-1}}{\partial z^{k-1}}
P_1\left[
\begin{matrix}
\theta\\
\phi\\
\end{matrix}
\right](z,\tau)
\nonumber\\
&=
\frac{(-1)^k}{(k-1)!} {\sum_{n\in \mathbb{Z}}}' 
\frac{(n+\lambda)^{k-1} x^{n+\lambda}}{1-\theta^{-1}q^{n+\lambda}},
\end{align}
where $\phi=e^{2\pi i\lambda}$. 

We have 
\begin{align}
P_1\left[
\begin{matrix}
\theta\\
\phi\\
\end{matrix}
\right](z,\tau)
&=\frac{1}{2\pi iz}-\sum_{k=1}^{\infty}
G_k\left[
\begin{matrix}
\theta\\
\phi\\
\end{matrix}
\right](\tau) (2\pi iz)^{k-1},\\
P_k\left[
\begin{matrix}
\theta\\
\phi\\
\end{matrix}
\right](z,\tau)
&=\frac{1}{(2\pi iz)^k}+(-1)^k \sum_{l=k}^{\infty}
\left(
\begin{matrix}
l-1\\
k-1\\
\end{matrix}
\right)
G_l\left[
\begin{matrix}
\theta\\
\phi\\
\end{matrix}
\right](\tau) (2\pi iz)^{l-k}, \\
P_k\left[
\begin{matrix}
1 \\
1 \\
\end{matrix}
\right](z,\tau)&=
P_k(z,\tau)+\frac12\delta_{k,1}. 
\end{align}

It follows that \cite{Mason:2008zzb}
\begin{align}
P_k\left[
\begin{matrix}
\theta\\
\phi\\
\end{matrix}
\right](z,\tau)&=\frac{1}{(2\pi i)^k}
\sum_{m\in \mathbb{Z}}\theta^m 
\left[
\sum_{n\in \mathbb{Z}}
\frac{\phi^n}{(z-m\tau-n)^k}
\right],
\end{align}
for $|q|<|x|<1$ and $\phi\neq 1$ 
and that 
\begin{align}
P_k\left[
\begin{matrix}
\theta\\
\phi\\
\end{matrix}
\right](z,\tau)&=\frac{1}{(2\pi i)^k}
\sum_{n\in \mathbb{Z}}\phi^n 
\left[
\sum_{m\in \mathbb{Z}}
\frac{\theta^m}{(z-m\tau-n)^k}
\right],
\end{align}
for $\theta\neq 1$. 

The twisted Weierstrass functions obey \cite{Dong:1997ea,Mason:2008zzb}
\begin{align}
\label{tPk_sym}
P_k\left[
\begin{matrix}
\theta^{-1}\\
\phi^{-1}\\
\end{matrix}
\right](-z,\tau)
&=
(-1)^k 
P_k\left[
\begin{matrix}
\theta\\
\phi\\
\end{matrix}
\right](z,\tau),\\
\label{tPk_sym2}
P_k\left[
\begin{matrix}
\theta\\
1\\
\end{matrix}
\right](z+\tau,\tau)
&=(-1)^k \theta 
P_k\left[
\begin{matrix}
\theta\\
1\\
\end{matrix}
\right](z,\tau). 
\end{align}

The twisted Weierstrass functions are expressible in terms of the theta functions 
as well as the ordinary Weierstrass functions. 
For example, we have
\begin{align}
\label{tP1_theta}
P_1\left[
\begin{matrix}
x\\
1\\
\end{matrix}
\right](\nu,\tau)&=
\frac{(q)_{\infty}^3 \theta(x^{-1}u;q)}{\theta(x^{-1};q)\theta(u;q)}, 
\\
\label{tP2_theta}
P_2\left[
\begin{matrix}
x\\
1\\
\end{matrix}
\right](\nu,\tau)
&=-uP_1\left[
\begin{matrix}
x\\
1\\
\end{matrix}
\right](\nu,\tau)
\left(
x^{-1}\frac{\theta'(x^{-1}u)}{\theta(x^{-1}u)}
-\frac{\theta'(u;q)}{\theta(u;q)}
\right)
\nonumber\\
&=
\frac{(q)_{\infty}^3 \theta(x^{-1}u;q)}{\theta(x^{-1};q)\theta(u;q)}
\left(
u\frac{\theta'(u;q)}{\theta(u;q)}
-
x^{-1} u \frac{\theta'(x^{-1}u;q)}{\theta(x^{-1}u;q)}
\right)
\nonumber\\
&=P_1\left[
\begin{matrix}
x\\
1\\
\end{matrix}
\right](\nu,\tau)
\left(
P_1(\nu,\tau)-P_1(\nu-z,\tau)
\right),\\
\label{tP3_theta}
P_3\left[
\begin{matrix}
x\\
1\\
\end{matrix}
\right](\nu,\tau)
&=\frac12 
P_1\left[
\begin{matrix}
x\\
1\\
\end{matrix}
\right](\nu,\tau)
\Bigl[
\left(
P_1(\nu,\tau)-P_1(\nu-z,\tau)
\right)^2
\nonumber\\
&+
\left(
P_2(\nu,\tau)-P_2(\nu-z,\tau)
\right)
\Bigr],\\
\label{tP4_theta}
P_4\left[
\begin{matrix}
x\\
1\\
\end{matrix}
\right](\nu,\tau)
&=\frac16 P_1\left[
\begin{matrix}
x\\
1\\
\end{matrix}
\right](\nu,\tau)
\Bigg[
\left(
P_1(\nu,\tau)-P_1(\nu-z,\tau)
\right)^3
\nonumber\\
&+3\left(
P_1(\nu,\tau)-P_1(\nu-z,\tau)
\right)
\left(
P_2(\nu,\tau)-P_2(\nu-z,\tau)
\right)
\nonumber\\
&+2\left(
P_3(\nu,\tau)-P_3(\nu-z,\tau)
\right)
\Bigg],\\
\label{tP5_theta}
P_5\left[
\begin{matrix}
x\\
1\\
\end{matrix}
\right](\nu,\tau)
&=\frac{1}{24}P_1\left[
\begin{matrix}
x\\
1\\
\end{matrix}
\right](\nu,\tau)
\Biggl[
\left(
P_1(\nu,\tau)-P_1(\nu-z,\tau)
\right)^4
\nonumber\\
&+6\left(
P_1(\nu,\tau)-P_1(\nu-z,\tau)
\right)^2
\left(
P_2(\nu,\tau)-P_2(\nu-z,\tau)
\right)
\nonumber\\
&+8\left(
P_1(\nu,\tau)-P_1(\nu-z,\tau)
\right)
\left(
P_3(\nu,\tau)-P_3(\nu-z,\tau)
\right)
\nonumber\\
&+
3\left(
P_2(\nu,\tau)-P_2(\nu-z,\tau)
\right)^2
+6
\left(
P_4(\nu,\tau)-P_4(\nu-z,\tau)
\right)
\Biggr], 
\end{align}
where $u=e^{2\pi i\nu}$. 

\section{Other analytic expressions}\label{app_other}
Here we show various closed-form expression of the $\mathcal{N}=2^*$ Schur index.
The derivation is based on the Fermi-gas formulation. 

\subsection{$N=2$}
We can write the $\mathcal{N}=2^{*}$ $U(2)$ Schur index as
\begin{align}
\label{u2_findex}
&\mathcal{I}^{U(2)}(\xi;q)=
\xi^2\frac{\theta(u;q)}{\theta(u\xi^{-2};q)}
\mathcal{Z}(2;u;\xi;q)
\nonumber\\
&=
\frac{\xi^2}{2} \Biggl(
\frac{(q)_{\infty}^6 \theta(\xi^{-1}u;q)^2}{\theta(\xi^{-1};q)^2\theta(u;q)\theta(\xi^{-2}u;q)}
+\xi^{-2}q^{-1}
\frac{(q)_{\infty}^3 \theta'(\xi^{-2}q^{-1}u;q)}{\theta(\xi^{-2}q^{-1};q)\theta(\xi^{-2}u;q)}
\nonumber\\
&
-\frac{(q)_{\infty}^3 \theta(\xi^{-2}q^{-1}u;q)\theta'(u;q)}{\theta(\xi^{-2}q^{-1};q)\theta(u;q)\theta(\xi^{-2}u;q)}
\Biggr). 
\end{align}
The R.H.S. of (\ref{u2_findex}) is expressed in terms of the $u$-dependent theta functions (\ref{theta_DEF}). 
Nevertheless, the index $\mathcal{I}^{U(2)}$ has no dependence on $u$. 
So we can simplify the expression (\ref{u2_findex}) by specializing $u$. 
Setting $u=\xi$, we find
\begin{align}
\label{u2_findex2}
\mathcal{I}^{U(2)}(\xi;q)&=
-\frac{\xi^2}{2}
\frac{(q)_{\infty}^3 \theta(\xi^{-1}q^{-1};q)}{\theta(\xi^{-2}q^{-1};q)\theta(\xi)}
\left(
\xi^{-2}q^{-1}
\frac{\theta'(\xi^{-1}q^{-1};q)}{\theta(\xi^{-1}q^{-1};q)}
-\frac{\theta'(\xi;q)}{\theta(\xi;q)}
\right). 
\end{align}
Using the identity
\begin{align}
\frac{\theta(\xi^{-1}q^{-1})}{\theta(\xi^{-2}q^{-1}) \theta(\xi)}
&=\frac{\xi}{\theta(\xi^2)},\\
\xi^{-2}q^{-1}\frac{\theta'(\xi^{-1}q^{-1})}{\theta(\xi^{-1}q^{-1})}
-\frac{\theta'(\xi)}{\theta(\xi)}
&=-\frac{1}{\xi}\left(-1+2\xi \frac{\theta'(\xi)}{\theta(\xi)}\right), 
\end{align}
we get 
\begin{align}
\label{u2_findex3}
\mathcal{I}^{U(2)}(\xi;q)
&=\frac{\xi^2}{2}
\frac{(q)_{\infty}^3}{\theta(\xi^2)}
\left(
-1+2\xi \frac{\theta'(\xi)}{\theta(\xi)}
\right). 
\end{align}
It can be also expressed in terms of the twisted Eisenstein series
\begin{align}
\label{u2_findex4}
\mathcal{I}^{U(2)}(\xi;q)
&=-\frac{\xi^2}{2}\frac{(q)_{\infty}^3}{\theta(\xi^2)}
\left(
2G_1(\zeta,\tau)+1
\right)
\nonumber\\
&=\frac{ie^{4\pi i\zeta}}{2}
\frac{\eta(\tau)^3}{\vartheta_1(2\zeta;\tau)}
\left(
1+2G_1(\zeta,\tau)
\right). 
\end{align}

\subsection{$N=3$}
Using the theta function (\ref{theta_DEF}) we can write the $\mathcal{N}=2^{*}$ $U(3)$ Schur index as
\begin{align}
\label{u3_findex}
&\mathcal{I}^{U(3)}(\xi;q)
=-\xi^{\frac92} \frac{\theta(u;q)}{\theta(u\xi^{-3};q)}\mathcal{Z}(3;u;\xi;q)
\nonumber\\
&=
-\frac{\xi^{9/2}}{6}
\Biggl[
\frac{(q)_{\infty}^9 \theta^3(\xi^{-1}u)}{\theta^3(\xi^{-1})\theta(\xi^{-3}u)\theta^2(u)}
+3
\frac{(q)_{\infty}^6 \theta(\xi^{-1}u)\theta(\xi^{-2}q^{-1}u)}{\theta(\xi^{-1})\theta(\xi^{-2}q^{-1})\theta(\xi^{-3}u)\theta(u)}
\nonumber\\
&\times
\left(
\xi^{-2}q^{-1}\frac{\theta'(\xi^{-2}q^{-1}u)}{\theta(\xi^{-2}q^{-1}u)}
-\frac{\theta'(u)}{\theta(u)}
\right) 
+\frac{(q)_{\infty}^3 \theta(\xi^{-3}q^{-2}u)}{\theta(\xi^{-3}q^{-2})\theta(\xi^{-3}u)} 
\Biggl\{
\left(
\xi^{-3}q^{-2}\frac{\theta'(\xi^{-3}q^{-2}u)}{\theta(\xi^{-3}q^{-2}u)}
-\frac{\theta'(u)}{\theta(u)}
\right)^2
\nonumber\\
&
+\xi^{-6}q^{-4} 
\frac{\theta''(\xi^{-3}q^{-2}u)\theta(\xi^{-3}q^{-2}u)-\theta'(\xi^{-3}q^{-2}u)^2}{\theta^2(\xi^{-3}q^{-2}u)}
-\frac{\theta''(u)\theta(u)-\theta'(u)^2}{\theta^2(u)}
\Biggr\}
\Biggr],
\end{align}
where $\theta(x)$ $=$ $\theta(x;q)$, $\theta'(x)$ $=$ $\frac{\partial}{\partial x}\theta(x;q)$ 
and $\theta''(x)$ $=$ $\frac{\partial^2}{\partial x^2}\theta(x;q)$. 
Since the index is independent of $u$, one can get expressions by fixing $u$.   
When we set $u=\xi^{3/2}$, we get
\begin{align}
\label{u3_findex2}
&\mathcal{I}^{U(3)}(\xi;q)
\nonumber\\
&=
-\frac{\xi^{9/2}}{6}
\Biggl[
\frac{(q)_{\infty}^9 \theta^3(\xi^{\frac12})}{\theta^3(\xi^{-1})\theta(\xi^{-\frac32})\theta^2(\xi^{\frac32})}
+3
\frac{(q)_{\infty}^6 \theta(\xi^{\frac12})\theta(\xi^{-\frac12}q^{-1})}{\theta(\xi^{-1})\theta(\xi^{-2}q^{-1})\theta(\xi^{-\frac32})\theta(\xi^{\frac32})}
\nonumber\\
&\times
\left(
\xi^{-2}q^{-1}\frac{\theta'(\xi^{-\frac12}q^{-1})}{\theta(\xi^{-\frac12}q^{-1})}
-\frac{\theta'(\xi^{\frac32})}{\theta(\xi^{\frac32})}
\right) 
+\frac{(q)_{\infty}^3 \theta(\xi^{-\frac32}q^{-2})}{\theta(\xi^{-3}q^{-2})\theta(\xi^{-\frac32})} 
\Biggl\{
\left(
\xi^{-3}q^{-2}\frac{\theta'(\xi^{-\frac32}q^{-2})}{\theta(\xi^{-\frac32}q^{-2})}
-\frac{\theta'(\xi^{\frac32})}{\theta(\xi^{\frac32})}
\right)^2
\nonumber\\
&
+\xi^{-6}q^{-4} 
\frac{\theta''(\xi^{-\frac32}q^{-2})\theta(\xi^{-\frac32}q^{-2})-\theta'(\xi^{-\frac32}q^{-2})^2}{\theta^2(\xi^{-\frac32}q^{-2})}
-\frac{\theta''(\xi^{\frac32})\theta(\xi^{\frac32})-\theta'(\xi^{\frac32})^2}{\theta^2(\xi^{\frac32})}
\Biggr\}
\Biggr]. 
\end{align}
There are other specializations of fugacity $u$ for which the expression gets simplified. 
When we take $u=\xi$, the expression (\ref{u3_findex}) becomes 
\begin{align}
\label{u3_findex3}
\mathcal{I}^{U(3)}(\xi;q)
&=
\frac{\xi^{13/2}}{6}\frac{(q)_{\infty}^3}{\theta(\xi^3)}
\Biggl[
\left(
\xi^{-3}q^{-2} \frac{\theta'(\xi^{-2}q^{-2})}{\theta(\xi^{-2}q^{-2})}
-\frac{\theta'(\xi)}{\theta(\xi)}
\right)^2
\nonumber\\
&
+\xi^{-6}q^{-4}
\frac{\theta''(\xi^{-2}q^{-2})\theta(\xi^{-2}q^{-2})-\theta'(\xi^{-2}q^{-2})^2 }{\theta^2(\xi^{-2}q^{-2})}
-\frac{\theta''(\xi) \theta(\xi)-\theta'(\xi)^2}{\theta^2(\xi)}
\Biggr],
\end{align}
where we have used the identity
\begin{align}
\frac{\theta(\xi^{-2}q^{-2})}{\theta(\xi^{-3}q^{-2})\theta(\xi^{-2})}&=-\frac{\xi^2}{\theta(\xi^3)}. 
\end{align}
We can also express the $U(3)$ Schur index (\ref{u3_findex3}) as
\begin{align}
\label{u3_findex4}
\mathcal{I}^{U(3)}(\xi;q)
&=
\frac{\xi^{9/2}}{6}
\frac{(q)_{\infty}^3}{\theta(\xi^3)}
\Biggl[
-2\Bigl(
G_2(2\zeta+2\tau,\tau)
-G_2(\zeta,\tau)
\Bigr)
\nonumber\\
&+\Bigl(
G_1(2\zeta+2\tau,\tau)+G_1(\zeta,\tau)
\Bigr)
\Bigl(
G_1(\zeta-\tau,\tau)+G_1(\zeta,\tau)
\Bigr)
\Biggr],
\end{align}
in terms of the twisted Eisenstein series. 

\subsection{$N=4$}
When we express the $\mathcal{N}=2^{*}$ $U(4)$ Schur index in terms of the theta function (\ref{theta_DEF}), 
it can be simplified by specializing the auxiliary fugacity $u$. 
When we set $u=\xi$, the non-trivial terms only appear from the Young diagrams which has no row of length $1$, that is $\tiny{\yng(2,2)}$ and $\tiny{\yng(4)}$. 
We find that 
\begin{align}
\label{u4_findex}
\mathcal{I}^{U(4)}(\xi;q)&=
-\frac{\xi^{10}}{8}
\Biggl[
\frac{(q)_{\infty}^6 \theta(\xi)}{\theta(\xi^{3})\theta(\xi^2)^2}
{A_1^{(1)}} (\xi;q)^2
\nonumber\\
&
-\frac{\xi^2}{3}
\frac{(q)_{\infty}^3}{\theta(\xi^4)}
\left(
A_3^{(1)}(\xi;q)^3
+3 A_{3}^{(1)}(\xi;q)A_3^{(2)}(\xi;q)
+A_{3}^{(3)}(\xi;q)
\right)
\Biggr],
\end{align}
where we have defined
\begin{align}
\label{A1}
A_l^{(1)}(\xi;q)&:=
\xi^{-1}(\xi^{-1}q^{-1})^l \frac{\theta'(\xi^{-l}q^{-l})}{\theta(\xi^{-l} q^{-l})}
-\frac{\theta'(\xi)}{\theta(\xi)}, \\
\label{F2}
A_l^{(2)}(\xi;q)&:=
\xi^{-2}(\xi^{-2}q^{-2})^{l}
\frac{\theta''(\xi^{-l}q^{-l}) \theta(\xi^{-l}q^{-l})-\theta'(\xi^{-l}q^{-l})^2}
{\theta(\xi^{-l}q^{-l})^2}
-\frac{\theta''(\xi)\theta(\xi)-\theta'(\xi)^2}{\theta^2(\xi)}, \\
\label{F3}
A_l^{(3)}(\xi;q)&:=
\xi^{-3}(\xi^{-3}q^{-3})^l 
\frac{1}
{\theta(\xi^{-l}q^{-q})^4}
\Biggl[
\theta'''(\xi^{-l}q^{-l}) \theta(\xi^{-l}q^{-l})^3 
\nonumber\\
&
- 3\theta''(\xi^{-l}q^{-l}) \theta'(\xi^{-l}q^{-l})\theta^2(\xi^{-l}q^{-l})
+2\theta'(\xi^{-l}q^{-l})^3 \theta(\xi^{-l}q^{-l})
\Biggr]
\nonumber\\
&-\frac{\theta'''(\xi)\theta(\xi)-3\theta''(\xi)\theta'(\xi)\theta^2(\xi)+2\theta'(\xi)^3\theta(\xi)}{\theta^4(\xi)}. 
\end{align}

\subsection{$N=5$}

While one can write the $\mathcal{N}=2^{*}$ $U(5)$ Schur index in terms of the twisted Weierstrass functions from the expression (\ref{fermi_pfn5_P}), 
it can be also expressed in terms of the theta functions (\ref{theta_DEF}) or the twisted Eisenstein series with fewer terms. 
When we specialize the fugacity as $u=\xi$, only two Young diagrams $\tiny{\yng(3,2)}$ and $\tiny{\yng(4)}$ contribute to the $\mathcal{N}=2^{*}$ $U(5)$ Schur index. 
We obtain
\begin{align}
\label{u5_findex}
\mathcal{I}^{U(5)}(\xi;q)&=
\frac{\xi^{25/2}}{120}
\Biggl[
10\frac{(q)_{\infty}^6}{\theta(\xi^3)\theta(\xi^4)}A_1^{(1)}(\xi;q)
\left(
A_2^{(1)}(\xi;q)^2+A_2^{(2)}(\xi;q)
\right)
\nonumber\\
&+\xi \frac{(q)_{\infty}^3}{\theta(\xi^5)}
\Biggl(
A_4^{(1)}(\xi;q)^4
+6A_4^{(1)}(\xi;q)^2A_2^{(2)}(\xi;q)
\Biggr)
\nonumber\\
&+4A_4^{(1)}(\xi;q)A_4^{(3)}(\xi;q)
+3A_4^{(2)}(\xi;q)^2
+A_4^{(4)}(\xi;q)
\Biggr],
\end{align}
where 
\begin{align}
A_l^{(4)}(\xi;q)&=\xi^{-4}(\xi^{-4}q^{-4})^l 
\frac{1}{\theta(\xi^{-l} q^{-l})^8}
\Biggl[
\theta''''(\xi^{-l}q^{-l})\theta(\xi^{-l}q^{-l})^7
-3\theta''(\xi^{-l}q^{-l})^2 \theta(\xi^{-l}q^{-l})^6
\nonumber\\
&-4\theta'''(\xi^{-l}q^{-l})\theta'(\xi^{-l}q^{-l})\theta(\xi^{-l}q^{-l})^6
+12\theta''(\xi^{-l}q^{-l})\theta'(\xi^{-l}q^{-l})^2\theta(\xi^{-l}q^{-l})^5
\nonumber\\
&-6\theta'(\xi^{-l}q^{-l})^4\theta(\xi^{-l}q^{-l})^4
\Biggr]
-\frac{1}{\theta(\xi;q)^8}
\Biggl[
\theta''''(\xi)\theta(\xi)^7
-3\theta''(\xi)^2\theta(\xi)^6
\nonumber\\
&-4\theta'''(\xi)\theta'(\xi)\theta(\xi)^6
+12\theta''(\xi)\theta'(\xi)^2\theta(\xi)^5
-6\theta'(\xi)^4\theta(\xi)^4
\Biggr]. 
\end{align}

\subsection{Higher ranks}
It is straihgtforward to extend to higher rank. Setting $u=\xi$,
we display the $\mathcal{N}=2^{*}$ $U(N)$ Schur indices for $N=6,7,8,9,$ and $10$
\begin{align}
\label{u6_findex}
\mathcal{I}^{U(6)}(\xi;q)&=
\frac{\xi^{18}}{144}
\frac{\theta(\xi)}{\theta(\xi^5)}
\left(
3Z_2^3-8Z_3^2-18Z_2 Z_4+24Z_6
\right), \\
\label{u7_findex}
\mathcal{I}^{U(7)}(\xi;q)&=
\frac{\xi^{49/2}}{840}\frac{\theta(\xi)}{\theta(\xi^6)}
\left(
35Z_2^2Z_3-70Z_3Z_4-84Z_2Z_5+120Z_7
\right),\\
\label{u8_findex}
\mathcal{I}^{U(8)}(\xi;q)&=
\frac{\xi^{32}}{5760}\frac{\theta(\xi)}{\theta(\xi^7)}
\Biggl(
-15Z_2^4+180Z_2^2Z_4+160Z_2(Z_3^2-3Z_6)
\nonumber\\
&-12
\left(
15Z_4^2+32Z_3Z_5-60Z_8
\right)
\Biggr),\\
\label{u9_findex}
\mathcal{I}^{U(9)}(\xi;q)
&=\frac{\xi^{81/2}}{45360}
\frac{\theta(\xi)}{\theta(\xi^8)}
\Biggl(
315Z_2^3Z_3-280Z_3^3-1890Z_2Z_3Z_4
\nonumber\\
&-1134Z_3^2Z_5+2268Z_4Z_5
+2520Z_3Z_6+3240Z_2Z_7
-5040Z_9
\Biggr),\\
\label{u10_findex}
\mathcal{I}^{U(10)}(\xi;q)
&=
\frac{\xi^{50}}{403200}
\frac{\theta(\xi)}{\theta(\xi^9)}
\Biggl(
105Z_2^5-2800Z_2^2Z_3^2-2100Z_2^3Z_4+5600Z_3^2Z_4
\nonumber\\
&
+6300Z_2Z_4^2+13440Z_2Z_3Z_5
-8064Z_5^2+8400Z_2^2Z_6-16800Z_4Z_6
\nonumber\\
&
-19200Z_3Z_7-25200Z_2Z_8
+40320Z_{10}
\Biggr),
\end{align}
where $Z_l$ are abbreviations for the specialized spectral zeta functions $Z_l(\xi;\xi;q)$ which are obtained from (\ref{Z_2})-(\ref{Z_5})  
and 
\begin{align}
\label{Z_6}
Z_6(u;\xi;q)&=
\frac{1}{60u^5}\Biggl(
12
P_2\left[
\begin{matrix}
q^5\xi^6\\
1\\
\end{matrix}
\right](\nu,\tau)
+50
P_4\left[
\begin{matrix}
q^5\xi^6\\
1\\
\end{matrix}
\right](\nu,\tau)
+105
P_5\left[
\begin{matrix}
q^5\xi^6\\
1\\
\end{matrix}
\right](\nu,\tau)
\nonumber\\
&+120
P_5\left[
\begin{matrix}
q^5\xi^6\\
1\\
\end{matrix}
\right](\nu,\tau)
+60
P_6\left[
\begin{matrix}
q^5\xi^6\\
1\\
\end{matrix}
\right](\nu,\tau)
\Biggr),
\end{align}
\begin{align}
\label{Z_7}
Z_7(u;\xi;q)&=\frac{1}{360u^6}
\Biggl(
60P_2\left[
\begin{matrix}
q^6\xi^7\\
1\\
\end{matrix}
\right](\nu,\tau)
+274P_3\left[
\begin{matrix}
q^6\xi^7\\
1\\
\end{matrix}
\right](\nu,\tau)
+675P_4\left[
\begin{matrix}
q^6\xi^7\\
1\\
\end{matrix}
\right](\nu,\tau)
\nonumber\\
&+1020P_5\left[
\begin{matrix}
q^6\xi^7\\
1\\
\end{matrix}
\right](\nu,\tau)
+900P_6\left[
\begin{matrix}
q^6\xi^7\\
1\\
\end{matrix}
\right](\nu,\tau)
+360P_7\left[
\begin{matrix}
q^6\xi^7\\
1\\
\end{matrix}
\right](\nu,\tau)
\Biggr),\\
\label{Z_8}
Z_8(u;\xi;q)&=
\frac{1}{210u^7}
\Biggl(
30P_2\left[
\begin{matrix}
q^7\xi^8\\
1\\
\end{matrix}
\right](\nu,\tau)
+147P_3\left[
\begin{matrix}
q^7\xi^8\\
1\\
\end{matrix}
\right](\nu,\tau)
+406P_3\left[
\begin{matrix}
q^7\xi^8\\
1\\
\end{matrix}
\right](\nu,\tau)
\nonumber\\
&+735P_5\left[
\begin{matrix}
q^7\xi^8\\
1\\
\end{matrix}
\right](\nu,\tau)
+875P_6\left[
\begin{matrix}
q^7\xi^8\\
1\\
\end{matrix}
\right](\nu,\tau)
+630P_7\left[
\begin{matrix}
q^7\xi^8\\
1\\
\end{matrix}
\right](\nu,\tau)
\nonumber\\
&+210P_8\left[
\begin{matrix}
q^7\xi^8\\
1\\
\end{matrix}
\right](\nu,\tau)
\Biggr),
\end{align}
\begin{align}
\label{Z_9}
Z_9(u;\xi;q)&=
\frac{1}{1680u^8}
\Biggl(
210P_2\left[
\begin{matrix}
q^8\xi^9\\
1\\
\end{matrix}
\right](\nu,\tau)
+1089P_3\left[
\begin{matrix}
q^8\xi^9\\
1\\
\end{matrix}
\right](\nu,\tau)
+3283P_4\left[
\begin{matrix}
q^8\xi^9\\
1\\
\end{matrix}
\right](\nu,\tau)
\nonumber\\
&
+6769P_5\left[
\begin{matrix}
q^8\xi^9\\
1\\
\end{matrix}
\right](\nu,\tau)
+9800P_6\left[
\begin{matrix}
q^8\xi^9\\
1\\
\end{matrix}
\right](\nu,\tau)
+9660P_7\left[
\begin{matrix}
q^8\xi^9\\
1\\
\end{matrix}
\right](\nu,\tau)
\nonumber\\
&+5880P_8\left[
\begin{matrix}
q^8\xi^9\\
1\\
\end{matrix}
\right](\nu,\tau)
+1680P_9\left[
\begin{matrix}
q^8\xi^9\\
1\\
\end{matrix}
\right](\nu,\tau)
\Biggr), \\
\label{Z_10}
Z_{10}(u;\xi;q)&=
\frac{1}{15120u^9}
\Biggl(
1680P_2\left[
\begin{matrix}
q^9\xi^{10}\\
1\\
\end{matrix}
\right](\nu,\tau)
+9132P_3\left[
\begin{matrix}
q^9\xi^{10}\\
1\\
\end{matrix}
\right](\nu,\tau)
\nonumber\\
&+29531P_4\left[
\begin{matrix}
q^9\xi^{10}\\
1\\
\end{matrix}
\right](\nu,\tau)
+67284P_5\left[
\begin{matrix}
q^9\xi^{10}\\
1\\
\end{matrix}
\right](\nu,\tau)
+112245P_6\left[
\begin{matrix}
q^9\xi^{10}\\
1\\
\end{matrix}
\right](\nu,\tau)
\nonumber\\
&+
136080P_7\left[
\begin{matrix}
q^9\xi^{10}\\
1\\
\end{matrix}
\right](\nu,\tau)
+114660P_8\left[
\begin{matrix}
q^9\xi^{10}\\
1\\
\end{matrix}
\right](\nu,\tau)
\nonumber\\
&+60480P_9\left[
\begin{matrix}
q^9\xi^{10}\\
1\\
\end{matrix}
\right](\nu,\tau)
+15120P_{10}\left[
\begin{matrix}
q^9\xi^{10}\\
1\\
\end{matrix}
\right](\nu,\tau)
\Biggr). 
\end{align}

\bibliographystyle{utphys}
\bibliography{ref}

\def\polhk#1{\setbox0=\hbox{#1}{\ooalign{\hidewidth
  \lower1.5ex\hbox{`}\hidewidth\crcr\unhbox0}}} \def\cprime{$'$}
\providecommand{\href}[2]{#2}\begingroup\raggedright\begin{thebibliography}{10}

\bibitem{Romelsberger:2005eg}
C.~Romelsberger, ``{Counting chiral primaries in N = 1, d=4 superconformal
  field theories},''
  \href{http://dx.doi.org/10.1016/j.nuclphysb.2006.03.037}{{\em Nucl. Phys.}
  {\bfseries B747} (2006) 329--353},
\href{http://arxiv.org/abs/hep-th/0510060}{{\ttfamily arXiv:hep-th/0510060
  [hep-th]}}.

\bibitem{Kinney:2005ej}
J.~Kinney, J.~M. Maldacena, S.~Minwalla, and S.~Raju, ``{An Index for 4
  dimensional super conformal theories},''
  \href{http://dx.doi.org/10.1007/s00220-007-0258-7}{{\em Commun. Math. Phys.}
  {\bfseries 275} (2007) 209--254},
\href{http://arxiv.org/abs/hep-th/0510251}{{\ttfamily arXiv:hep-th/0510251
  [hep-th]}}.

\bibitem{Assel:2015nca}
B.~Assel, D.~Cassani, L.~Di~Pietro, Z.~Komargodski, J.~Lorenzen, and
  D.~Martelli, ``{The Casimir Energy in Curved Space and its Supersymmetric
  Counterpart},'' \href{http://dx.doi.org/10.1007/JHEP07(2015)043}{{\em JHEP}
  {\bfseries 07} (2015) 043}, \href{http://arxiv.org/abs/1503.05537}{{\ttfamily
  arXiv:1503.05537 [hep-th]}}.

\bibitem{Gadde:2011ik}
A.~Gadde, L.~Rastelli, S.~S. Razamat, and W.~Yan, ``{The 4d Superconformal
  Index from q-deformed 2d Yang-Mills},''
  \href{http://dx.doi.org/10.1103/PhysRevLett.106.241602}{{\em Phys. Rev.
  Lett.} {\bfseries 106} (2011) 241602},
  \href{http://arxiv.org/abs/1104.3850}{{\ttfamily arXiv:1104.3850 [hep-th]}}.

\bibitem{Gadde:2011uv}
A.~Gadde, L.~Rastelli, S.~S. Razamat, and W.~Yan, ``{Gauge Theories and
  Macdonald Polynomials},''
  \href{http://dx.doi.org/10.1007/s00220-012-1607-8}{{\em Commun. Math. Phys.}
  {\bfseries 319} (2013) 147--193},
  \href{http://arxiv.org/abs/1110.3740}{{\ttfamily arXiv:1110.3740 [hep-th]}}.

\bibitem{Gadde:2009kb}
A.~Gadde, E.~Pomoni, L.~Rastelli, and S.~S. Razamat, ``{S-duality and 2d
  Topological QFT},'' \href{http://dx.doi.org/10.1007/JHEP03(2010)032}{{\em
  JHEP} {\bfseries 03} (2010) 032},
  \href{http://arxiv.org/abs/0910.2225}{{\ttfamily arXiv:0910.2225 [hep-th]}}.

\bibitem{Beem:2013sza}
C.~Beem, M.~Lemos, P.~Liendo, W.~Peelaers, L.~Rastelli, and B.~C. van Rees,
  ``{Infinite Chiral Symmetry in Four Dimensions},''
  \href{http://dx.doi.org/10.1007/s00220-014-2272-x}{{\em Commun. Math. Phys.}
  {\bfseries 336} no.~3, (2015) 1359--1433},
  \href{http://arxiv.org/abs/1312.5344}{{\ttfamily arXiv:1312.5344 [hep-th]}}.

\bibitem{Razamat:2012uv}
S.~S. Razamat, ``{On a modular property of N=2 superconformal theories in four
  dimensions},'' \href{http://dx.doi.org/10.1007/JHEP10(2012)191}{{\em JHEP}
  {\bfseries 10} (2012) 191}, \href{http://arxiv.org/abs/1208.5056}{{\ttfamily
  arXiv:1208.5056 [hep-th]}}.

\bibitem{Beem:2017ooy}
C.~Beem and L.~Rastelli, ``{Vertex operator algebras, Higgs branches, and
  modular differential equations},''
  \href{http://dx.doi.org/10.1007/JHEP08(2018)114}{{\em JHEP} {\bfseries 08}
  (2018) 114}, \href{http://arxiv.org/abs/1707.07679}{{\ttfamily
  arXiv:1707.07679 [hep-th]}}.

\bibitem{MR3890204}
T.~Arakawa and K.~Kawasetsu, ``Quasi-lisse vertex algebras and modular linear
  differential equations,'' in {\em Lie groups, geometry, and representation
  theory}, vol.~326 of {\em Progr. Math.}, pp.~41--57.
\newblock Birkh\"{a}user/Springer, Cham, 2018.

\bibitem{Pan:2021mrw}
Y.~Pan and W.~Peelaers, ``{The exact Schur index in closed form},''
  \href{http://arxiv.org/abs/2112.09705}{{\ttfamily arXiv:2112.09705
  [hep-th]}}.

\bibitem{Gaiotto:2020vqj}
D.~Gaiotto and J.~Abajian, ``{Twisted M2 brane holography and sphere
  correlation functions},'' \href{http://arxiv.org/abs/2004.13810}{{\ttfamily
  arXiv:2004.13810 [hep-th]}}.

\bibitem{Bourdier:2015wda}
J.~Bourdier, N.~Drukker, and J.~Felix, ``{The exact Schur index of
  $\mathcal{N}=4$ SYM},'' \href{http://dx.doi.org/10.1007/JHEP11(2015)210}{{\em
  JHEP} {\bfseries 11} (2015) 210},
  \href{http://arxiv.org/abs/1507.08659}{{\ttfamily arXiv:1507.08659
  [hep-th]}}.

\bibitem{Beem:2021zvt}
C.~Beem, S.~S. Razamat, and P.~Singh, ``{Schur indices of class S and
  quasimodular forms},''
  \href{http://dx.doi.org/10.1103/PhysRevD.105.085009}{{\em Phys. Rev. D}
  {\bfseries 105} no.~8, (2022) 085009},
  \href{http://arxiv.org/abs/2112.10715}{{\ttfamily arXiv:2112.10715
  [hep-th]}}.

\bibitem{zbMATH02706826}
Kronecker, ``On the theory of the elliptic functions.'' {\em Berl. Monatsber.}
  {\bfseries 1881} (1881) 1165--1172.

\bibitem{MR1723749}
A.~Weil, {\em Elliptic functions according to {E}isenstein and {K}ronecker}.
\newblock Classics in Mathematics. Springer-Verlag, Berlin, 1999.
\newblock Reprint of the 1976 original.

\bibitem{MR1106744}
D.~Zagier, ``Periods of modular forms and {J}acobi theta functions,''
  \href{http://dx.doi.org/10.1007/BF01245085}{{\em Invent. Math.} {\bfseries
  104} no.~3, (1991) 449--465}. \url{https://doi.org/10.1007/BF01245085}.

\bibitem{Dong:1997ea}
C.-y. Dong, H.-s. Li, and G.~Mason, ``{Modular invariance of trace functions in
  orbifold theory},'' \href{http://dx.doi.org/10.1007/s002200000242}{{\em
  Commun. Math. Phys.} {\bfseries 214} (2000) 1--56},
  \href{http://arxiv.org/abs/q-alg/9703016}{{\ttfamily arXiv:q-alg/9703016}}.

\bibitem{Mason:2008zzb}
G.~Mason, M.~P. Tuite, and A.~Zuevsky, ``{Torus n-point functions for R-graded
  vertex operator superalgebras and continuous fermion orbifolds},''
  \href{http://dx.doi.org/10.1007/s00220-008-0510-9}{{\em Commun. Math. Phys.}
  {\bfseries 283} (2008) 305--342},
  \href{http://arxiv.org/abs/0708.0640}{{\ttfamily arXiv:0708.0640 [math.QA]}}.

\bibitem{MR2796409}
A.~Libgober, ``Elliptic genera, real algebraic varieties and quasi-{J}acobi
  forms,'' in {\em Topology of stratified spaces}, vol.~58 of {\em Math. Sci.
  Res. Inst. Publ.}, pp.~95--120.
\newblock Cambridge Univ. Press, Cambridge, 2011.

\bibitem{Buican:2020moo}
M.~Buican and T.~Nishinaka, ``{$ \mathcal{N} $ = 4 SYM, Argyres-Douglas
  theories, and an exact graded vector space isomorphism},''
  \href{http://dx.doi.org/10.1007/JHEP04(2022)028}{{\em JHEP} {\bfseries 04}
  (2022) 028}, \href{http://arxiv.org/abs/2012.13209}{{\ttfamily
  arXiv:2012.13209 [hep-th]}}.

\bibitem{Kang:2021lic}
M.~J. Kang, C.~Lawrie, and J.~Song, ``{Infinitely many 4D N=2 SCFTs with a=c
  and beyond},'' \href{http://dx.doi.org/10.1103/PhysRevD.104.105005}{{\em
  Phys. Rev. D} {\bfseries 104} no.~10, (2021) 105005},
  \href{http://arxiv.org/abs/2106.12579}{{\ttfamily arXiv:2106.12579
  [hep-th]}}.

\bibitem{DelZotto:2015rca}
M.~Del~Zotto, C.~Vafa, and D.~Xie, ``{Geometric engineering, mirror symmetry
  and $ 6{\mathrm{d}}_{\left(1,0\right)}\to
  4{\mathrm{d}}_{\left(\mathcal{N}=2\right)} $},''
  \href{http://dx.doi.org/10.1007/JHEP11(2015)123}{{\em JHEP} {\bfseries 11}
  (2015) 123}, \href{http://arxiv.org/abs/1504.08348}{{\ttfamily
  arXiv:1504.08348 [hep-th]}}.

\bibitem{Xie:2016evu}
D.~Xie, W.~Yan, and S.-T. Yau, ``{Chiral algebra of the Argyres-Douglas theory
  from M5 branes},'' \href{http://dx.doi.org/10.1103/PhysRevD.103.065003}{{\em
  Phys. Rev. D} {\bfseries 103} no.~6, (2021) 065003},
  \href{http://arxiv.org/abs/1604.02155}{{\ttfamily arXiv:1604.02155
  [hep-th]}}.

\bibitem{Buican:2016arp}
M.~Buican and T.~Nishinaka, ``{Conformal Manifolds in Four Dimensions and
  Chiral Algebras},''
  \href{http://dx.doi.org/10.1088/1751-8113/49/46/465401}{{\em J. Phys. A}
  {\bfseries 49} no.~46, (2016) 465401},
  \href{http://arxiv.org/abs/1603.00887}{{\ttfamily arXiv:1603.00887
  [hep-th]}}.

\bibitem{Closset:2020scj}
C.~Closset, S.~Schafer-Nameki, and Y.-N. Wang, ``{Coulomb and Higgs Branches
  from Canonical Singularities: Part 0},''
  \href{http://dx.doi.org/10.1007/JHEP02(2021)003}{{\em JHEP} {\bfseries 02}
  (2021) 003}, \href{http://arxiv.org/abs/2007.15600}{{\ttfamily
  arXiv:2007.15600 [hep-th]}}.

\bibitem{Closset:2020afy}
C.~Closset, S.~Giacomelli, S.~Schafer-Nameki, and Y.-N. Wang, ``{5d and 4d
  SCFTs: Canonical Singularities, Trinions and S-Dualities},''
  \href{http://dx.doi.org/10.1007/JHEP05(2021)274}{{\em JHEP} {\bfseries 05}
  (2021) 274}, \href{http://arxiv.org/abs/2012.12827}{{\ttfamily
  arXiv:2012.12827 [hep-th]}}.

\bibitem{MR2766985}
B.~L. Feigin and I.~Y. Tipunin,
  \href{http://dx.doi.org/10.1142/9789814324373\_0003}{``Characters of
  coinvariants in {$(1,p)$} logarithmic models,''} in {\em New trends in
  quantum integrable systems}, pp.~35--60.
\newblock World Sci. Publ., Hackensack, NJ, 2011.
\newblock \url{https://doi.org/10.1142/9789814324373_0003}.

\bibitem{Feigin:2007sp}
B.~Feigin, E.~Feigin, and I.~Tipunin, ``{Fermionic formulas for (1,p)
  logarithmic model characters in Phi{2,1} quasiparticle realisation},''
  \href{http://arxiv.org/abs/0704.2464}{{\ttfamily arXiv:0704.2464 [hep-th]}}.

\bibitem{MR1576612}
P.~A. MacMahon, ``Divisors of {N}umbers and their {C}ontinuations in the
  {T}heory of {P}artitions,''
  \href{http://dx.doi.org/10.1112/plms/s2-19.1.75}{{\em Proc. London Math. Soc.
  (2)} {\bfseries 19} no.~1, (1920) 75--113}.
  \url{https://doi.org/10.1112/plms/s2-19.1.75}.

\bibitem{MR1363056}
M.~Kaneko and D.~Zagier,
  \href{http://dx.doi.org/10.1007/978-1-4612-4264-2\_6}{``A generalized
  {J}acobi theta function and quasimodular forms,''} in {\em The moduli space
  of curves ({T}exel {I}sland, 1994)}, vol.~129 of {\em Progr. Math.},
  pp.~165--172.
\newblock Birkh\"{a}user Boston, Boston, MA, 1995.
\newblock \url{https://doi.org/10.1007/978-1-4612-4264-2_6}.

\bibitem{MR2034322}
S.~Corteel and J.~Lovejoy, ``Overpartitions,''
  \href{http://dx.doi.org/10.1090/S0002-9947-03-03328-2}{{\em Trans. Amer.
  Math. Soc.} {\bfseries 356} no.~4, (2004) 1623--1635}.
  \url{https://doi.org/10.1090/S0002-9947-03-03328-2}.

\bibitem{Honda:2022hvy}
M.~Honda and T.~Yoda, ``{String theory, $\mathcal{N}=4$ SYM and Riemann
  hypothesis},'' \href{http://arxiv.org/abs/2203.17091}{{\ttfamily
  arXiv:2203.17091 [hep-th]}}.

\bibitem{MR62781}
G.~Meinardus, ``Asymptotische {A}ussagen \"{u}ber {P}artitionen,''
  \href{http://dx.doi.org/10.1007/BF01180268}{{\em Math. Z.} {\bfseries 59}
  (1954) 388--398}. \url{https://doi.org/10.1007/BF01180268}.

\bibitem{Gang:2012yr}
D.~Gang, E.~Koh, and K.~Lee, ``{Line Operator Index on $S^{1}\times S^{3}$},''
  \href{http://dx.doi.org/10.1007/JHEP05(2012)007}{{\em JHEP} {\bfseries 05}
  (2012) 007},
\href{http://arxiv.org/abs/1201.5539}{{\ttfamily arXiv:1201.5539 [hep-th]}}.

\bibitem{Cordova:2016uwk}
C.~Cordova, D.~Gaiotto, and S.-H. Shao, ``{Infrared Computations of Defect
  Schur Indices},'' \href{http://dx.doi.org/10.1007/JHEP11(2016)106}{{\em JHEP}
  {\bfseries 11} (2016) 106},
\href{http://arxiv.org/abs/1606.08429}{{\ttfamily arXiv:1606.08429 [hep-th]}}.

\bibitem{Neitzke:2017cxz}
A.~Neitzke and F.~Yan, ``{Line defect Schur indices, Verlinde algebras and
  $U(1)_r$ fixed points},''
  \href{http://dx.doi.org/10.1007/JHEP11(2017)035}{{\em JHEP} {\bfseries 11}
  (2017) 035}, \href{http://arxiv.org/abs/1708.05323}{{\ttfamily
  arXiv:1708.05323 [hep-th]}}.

\bibitem{Hatsuda:2021oxa}
Y.~Hatsuda and T.~Okazaki, ``{Fermi-gas correlators of ADHM theory and triality
  symmetry},'' \href{http://dx.doi.org/10.21468/SciPostPhys.12.1.005}{{\em
  SciPost Phys.} {\bfseries 12} (2022) 005},
  \href{http://arxiv.org/abs/2107.01924}{{\ttfamily arXiv:2107.01924
  [hep-th]}}.

\bibitem{HOline:2022}
Y.~Hatsuda and T.~Okazaki, ``{$\mathcal{N}=2^{*}$ Schur correlators},'' {\em To
  appear} .

\bibitem{Gaberdiel:2009vs}
M.~R. Gaberdiel and C.~A. Keller, ``{Differential operators for elliptic
  genera},'' \href{http://dx.doi.org/10.4310/CNTP.2009.v3.n4.a1}{{\em Commun.
  Num. Theor. Phys.} {\bfseries 3} (2009) 593--618},
  \href{http://arxiv.org/abs/0904.1831}{{\ttfamily arXiv:0904.1831 [hep-th]}}.

\bibitem{MR4281261}
J.-W. van Ittersum, G.~Oberdieck, and A.~Pixton, ``Gromov-{W}itten theory of
  {K}3 surfaces and a {K}aneko-{Z}agier equation for {J}acobi forms,''
  \href{http://dx.doi.org/10.1007/s00029-021-00673-y}{{\em Selecta Math.
  (N.S.)} {\bfseries 27} no.~4, (2021) Paper No. 64, 30}.
  \url{https://doi.org/10.1007/s00029-021-00673-y}.

\bibitem{Okazaki:2022sxo}
T.~Okazaki, ``{M2-branes and plane partitions},''
  \href{http://dx.doi.org/10.1007/JHEP07(2022)028}{{\em JHEP} {\bfseries 07}
  (2022) 028}, \href{http://arxiv.org/abs/2204.01973}{{\ttfamily
  arXiv:2204.01973 [hep-th]}}.

\bibitem{Hayashi:2022ldo}
H.~Hayashi, T.~Nosaka, and T.~Okazaki, ``{Dualities and flavored indices of
  M2-brane SCFTs},'' \href{http://arxiv.org/abs/2206.05362}{{\ttfamily
  arXiv:2206.05362 [hep-th]}}.

\bibitem{Frobenius:1882uber}
G.~Frobenius, ``{\"{U}ber die elliptischen Funktionen zweiter},'' {\em Art, J.
  Reine Angew. Math} {\bfseries 93} (1882) 53--68.

\bibitem{MR0335789}
J.~D. Fay, {\em Theta functions on {R}iemann surfaces}.
\newblock Lecture Notes in Mathematics, Vol. 352. Springer-Verlag, Berlin-New
  York, 1973.

\bibitem{https://doi.org/10.48550/arxiv.2109.10394}
K.~Bringmann, W.~Craig, J.~Males, and K.~Ono, ``Distributions on partitions
  arising from hilbert schemes and hook lengths,'' 2021.
\newblock \url{https://arxiv.org/abs/2109.10394}.

\bibitem{Eleftheriou:2022kkv}
G.~Eleftheriou, ``{Root of unity asymptotics for Schur indices of 4d Lagrangian
  theories},'' \href{http://arxiv.org/abs/2207.14271}{{\ttfamily
  arXiv:2207.14271 [hep-th]}}.

\bibitem{Gaiotto:2019jvo}
D.~Gaiotto and T.~Okazaki, ``{Dualities of Corner Configurations and
  Supersymmetric Indices},''
  \href{http://dx.doi.org/10.1007/JHEP11(2019)056}{{\em JHEP} {\bfseries 11}
  (2019) 056},
\href{http://arxiv.org/abs/1902.05175}{{\ttfamily arXiv:1902.05175 [hep-th]}}.

\bibitem{Dedushenko:2019yiw}
M.~Dedushenko and M.~Fluder, ``{Chiral Algebra, Localization, Modularity,
  Surface defects, And All That},''
  \href{http://dx.doi.org/10.1063/5.0002661}{{\em J. Math. Phys.} {\bfseries
  61} no.~9, (2020) 092302}, \href{http://arxiv.org/abs/1904.02704}{{\ttfamily
  arXiv:1904.02704 [hep-th]}}.

\bibitem{Pan:2019bor}
Y.~Pan and W.~Peelaers, ``{Schur correlation functions on $S^3\times S^1$},''
  \href{http://dx.doi.org/10.1007/JHEP07(2019)013}{{\em JHEP} {\bfseries 07}
  (2019) 013}, \href{http://arxiv.org/abs/1903.03623}{{\ttfamily
  arXiv:1903.03623 [hep-th]}}.

\bibitem{Jeong:2019pzg}
S.~Jeong, ``{SCFT/VOA correspondence via $\Omega$-deformation},''
  \href{http://dx.doi.org/10.1007/JHEP10(2019)171}{{\em JHEP} {\bfseries 10}
  (2019) 171}, \href{http://arxiv.org/abs/1904.00927}{{\ttfamily
  arXiv:1904.00927 [hep-th]}}.

\bibitem{Corley:2001zk}
S.~Corley, A.~Jevicki, and S.~Ramgoolam, ``{Exact correlators of giant
  gravitons from dual N=4 SYM theory},''
  \href{http://dx.doi.org/10.4310/ATMP.2001.v5.n4.a6}{{\em Adv. Theor. Math.
  Phys.} {\bfseries 5} (2002) 809--839},
  \href{http://arxiv.org/abs/hep-th/0111222}{{\ttfamily arXiv:hep-th/0111222}}.

\bibitem{Gaiotto:2021xce}
D.~Gaiotto and J.~H. Lee, ``{The Giant Graviton Expansion},''
  \href{http://arxiv.org/abs/2109.02545}{{\ttfamily arXiv:2109.02545
  [hep-th]}}.

\bibitem{Arai:2020qaj}
R.~Arai, S.~Fujiwara, Y.~Imamura, and T.~Mori, ``{Schur index of the ${\cal
  N}=4$ $U(N)$ supersymmetric Yang-Mills theory via the AdS/CFT
  correspondence},'' \href{http://dx.doi.org/10.1103/PhysRevD.101.086017}{{\em
  Phys. Rev. D} {\bfseries 101} no.~8, (2020) 086017},
  \href{http://arxiv.org/abs/2001.11667}{{\ttfamily arXiv:2001.11667
  [hep-th]}}.

\bibitem{Imamura:2021ytr}
Y.~Imamura, ``{Finite-N superconformal index via the AdS/CFT correspondence},''
  \href{http://dx.doi.org/10.1093/ptep/ptab141}{{\em PTEP} {\bfseries 2021}
  no.~12, (2021) 123B05}, \href{http://arxiv.org/abs/2108.12090}{{\ttfamily
  arXiv:2108.12090 [hep-th]}}.

\bibitem{Murthy:2022ien}
S.~Murthy, ``{Unitary matrix models, free fermion ensembles, and the giant
  graviton expansion},'' \href{http://arxiv.org/abs/2202.06897}{{\ttfamily
  arXiv:2202.06897 [hep-th]}}.

\bibitem{Lee:2022vig}
J.~H. Lee, ``{Exact Stringy Microstates from Gauge Theories},''
  \href{http://arxiv.org/abs/2204.09286}{{\ttfamily arXiv:2204.09286
  [hep-th]}}.

\bibitem{Imamura:2022aua}
Y.~Imamura, ``{Analytic continuation for giant gravitons},''
  \href{http://arxiv.org/abs/2205.14615}{{\ttfamily arXiv:2205.14615
  [hep-th]}}.

\bibitem{McGreevy:2000cw}
J.~McGreevy, L.~Susskind, and N.~Toumbas, ``{Invasion of the giant gravitons
  from Anti-de Sitter space},''
  \href{http://dx.doi.org/10.1088/1126-6708/2000/06/008}{{\em JHEP} {\bfseries
  06} (2000) 008}, \href{http://arxiv.org/abs/hep-th/0003075}{{\ttfamily
  arXiv:hep-th/0003075}}.

\bibitem{Grisaru:2000zn}
M.~T. Grisaru, R.~C. Myers, and O.~Tafjord, ``{SUSY and goliath},''
  \href{http://dx.doi.org/10.1088/1126-6708/2000/08/040}{{\em JHEP} {\bfseries
  08} (2000) 040}, \href{http://arxiv.org/abs/hep-th/0008015}{{\ttfamily
  arXiv:hep-th/0008015}}.

\bibitem{Hashimoto:2000zp}
A.~Hashimoto, S.~Hirano, and N.~Itzhaki, ``{Large branes in AdS and their field
  theory dual},'' \href{http://dx.doi.org/10.1088/1126-6708/2000/08/051}{{\em
  JHEP} {\bfseries 08} (2000) 051},
  \href{http://arxiv.org/abs/hep-th/0008016}{{\ttfamily arXiv:hep-th/0008016}}.

\bibitem{MR2445243}
G.~H. Hardy and E.~M. Wright, {\em An introduction to the theory of numbers}.
\newblock Oxford University Press, Oxford, sixth~ed., 2008.
\newblock Revised by D. R. Heath-Brown and J. H. Silverman, With a foreword by
  Andrew Wiles.

\bibitem{MR2280879}
G.~H. Hardy and S.~Ramanujan, ``Asymptotic formul\ae in combinatory analysis
  [{P}roc. {L}ondon {M}ath. {S}oc. (2) {\bf 17} (1918), 75--115],'' in {\em
  Collected papers of {S}rinivasa {R}amanujan}, pp.~276--309.
\newblock AMS Chelsea Publ., Providence, RI, 2000.

\bibitem{MR0004860}
G.~H. Hardy, {\em Ramanujan. {T}welve lectures on subjects suggested by his
  life and work}.
\newblock Cambridge University Press, Cambridge, England; Macmillan Company,
  New York, 1940.

\bibitem{Kawai:2000px}
T.~Kawai and K.~Yoshioka, ``{String partition functions and infinite
  products},'' \href{http://dx.doi.org/10.4310/ATMP.2000.v4.n2.a7}{{\em Adv.
  Theor. Math. Phys.} {\bfseries 4} (2000) 397--485},
  \href{http://arxiv.org/abs/hep-th/0002169}{{\ttfamily arXiv:hep-th/0002169}}.

\bibitem{Krauel:2013lra}
M.~Krauel and G.~Mason, ``{Jacobi trace functions in the theory of vertex
  operator algebras},''
  \href{http://dx.doi.org/10.4310/CNTP.2015.v9.n2.a2}{{\em Commun. Num. Theor.
  Phys.} {\bfseries 09} (2015) 273--305},
  \href{http://arxiv.org/abs/1309.5720}{{\ttfamily arXiv:1309.5720 [math.QA]}}.

\bibitem{MR3956895}
G.~Oberdieck and A.~Pixton, ``Gromov-{W}itten theory of elliptic fibrations:
  {J}acobi forms and holomorphic anomaly equations,''
  \href{http://dx.doi.org/10.2140/gt.2019.23.1415}{{\em Geom. Topol.}
  {\bfseries 23} no.~3, (2019) 1415--1489}.
  \url{https://doi.org/10.2140/gt.2019.23.1415}.

\bibitem{Bringmann:2020ndq}
K.~Bringmann, M.~Krauel, and M.~Tuite, ``{Zhu reduction for Jacobi
  \ensuremath{\mathit{n}}-point functions and applications},''
  \href{http://dx.doi.org/10.1090/tran/8013}{{\em Trans. Am. Math. Soc.}
  {\bfseries 373} no.~5, (2020) 3261--3293}.

\bibitem{MR2352717}
D.~Mumford, \href{http://dx.doi.org/10.1007/978-0-8176-4578-6}{{\em Tata
  lectures on theta. {I}}}.
\newblock Modern Birkh\"auser Classics. Birkh\"auser Boston, Inc., Boston, MA,
  2007.
\newblock \url{http://dx.doi.org/10.1007/978-0-8176-4578-6}.
\newblock With the collaboration of C. Musili, M. Nori, E. Previato and M.
  Stillman, Reprint of the 1983 edition.

\bibitem{MR3028756}
G.~E. Andrews and S.~C.~F. Rose, ``Mac{M}ahon's sum-of-divisors functions,
  {C}hebyshev polynomials, and quasi-modular forms,''
  \href{http://dx.doi.org/10.1515/crelle.2011.179}{{\em J. Reine Angew. Math.}
  {\bfseries 676} (2013) 97--103}.
  \url{https://doi.org/10.1515/crelle.2011.179}.

\bibitem{MR4286926}
E.~T. Whittaker and G.~N. Watson, {\em A course of modern analysis---an
  introduction to the general theory of infinite processes and of analytic
  functions with an account of the principal transcendental functions}.
\newblock Cambridge University Press, Cambridge, fifth~ed., 2021.
\newblock Edited by Victor H. Moll, With a foreword by S. J. Patterson, For an
  unaltered reprint of the fourth edition see [ 0178117 ].

\bibitem{MR1634067}
G.~E. Andrews, {\em The theory of partitions}.
\newblock Cambridge Mathematical Library. Cambridge University Press,
  Cambridge, 1998.
\newblock Reprint of the 1976 original.

\bibitem{MR1117903}
B.~C. Berndt, \href{http://dx.doi.org/10.1007/978-1-4612-0965-2}{{\em
  Ramanujan's notebooks. {P}art {III}}}.
\newblock Springer-Verlag, New York, 1991.
\newblock \url{https://doi.org/10.1007/978-1-4612-0965-2}.

\bibitem{MR2180457}
S.~H. Chan, ``Generalized {L}ambert series identities,''
  \href{http://dx.doi.org/10.1112/S0024611505015364}{{\em Proc. London Math.
  Soc. (3)} {\bfseries 91} no.~3, (2005) 598--622}.
  \url{https://doi.org/10.1112/S0024611505015364}.

\bibitem{Huang:2022bry}
M.-x. Huang, ``{Modular Anomaly Equation for Schur Index of $\mathcal{N}=4$
  Super-Yang-Mills},'' \href{http://arxiv.org/abs/2205.00818}{{\ttfamily
  arXiv:2205.00818 [hep-th]}}.

\bibitem{MR4254766}
W.~J. Keith, ``Restricted {$k$}-color partitions, {II},''
  \href{http://dx.doi.org/10.1142/S1793042120400151}{{\em Int. J. Number
  Theory} {\bfseries 17} no.~3, (2021) 591--601}.
  \url{https://doi.org/10.1142/S1793042120400151}.

\bibitem{MR4364153}
M.~Merca, ``Overpartitions as sums over partitions,'' {\em Proc. Rom. Acad.
  Ser. A Math. Phys. Tech. Sci. Inf. Sci.} {\bfseries 22} no.~4, (2021)
  327--333.

\bibitem{MR2195564}
J.-F. Fortin, P.~Jacob, and P.~Mathieu, ``Jagged partitions,''
  \href{http://dx.doi.org/10.1007/s11139-005-4848-8}{{\em Ramanujan J.}
  {\bfseries 10} no.~2, (2005) 215--235}.
  \url{https://doi.org/10.1007/s11139-005-4848-8}.

\bibitem{Cardy:1991kr}
J.~L. Cardy, ``{Operator content and modular properties of higher dimensional
  conformal field theories},''
  \href{http://dx.doi.org/10.1016/0550-3213(91)90024-R}{{\em Nucl. Phys. B}
  {\bfseries 366} (1991) 403--419}.

\bibitem{MR1027834}
T.~M. Apostol, \href{http://dx.doi.org/10.1007/978-1-4612-0999-7}{{\em Modular
  functions and {D}irichlet series in number theory}}, vol.~41 of {\em Graduate
  Texts in Mathematics}.
\newblock Springer-Verlag, New York, second~ed., 1990.
\newblock \url{https://doi.org/10.1007/978-1-4612-0999-7}.

\bibitem{MR2280861}
S.~Ramanujan, \href{http://dx.doi.org/10.1016/s0164-1212(00)00033-9}{``On
  certain arithmetical functions [{T}rans. {C}ambridge {P}hilos. {S}oc. {\bf
  22} (1916), no. 9, 159--184],''} in {\em Collected papers of {S}rinivasa
  {R}amanujan}, pp.~136--162.
\newblock AMS Chelsea Publ., Providence, RI, 2000.
\newblock \url{https://doi.org/10.1016/s0164-1212(00)00033-9}.

\bibitem{MR1837822}
\href{http://dx.doi.org/10.1007/b76882}{{\em Introduction to algebraic
  independence theory}}, vol.~1752 of {\em Lecture Notes in Mathematics}.
\newblock Springer-Verlag, Berlin, 2001.
\newblock \url{https://doi.org/10.1007/b76882}.
\newblock With contributions from F. Amoroso, D. Bertrand, W. D. Brownawell, G.
  Diaz, M. Laurent, Yuri V. Nesterenko, K. Nishioka, Patrice Philippon, G.
  R\'{e}mond, D. Roy and M. Waldschmidt, Edited by Nesterenko and Philippon.

\bibitem{https://doi.org/10.48550/arxiv.1209.5628}
G.~Oberdieck, ``A serre derivative for even weight jacobi forms,'' 2012.
\newblock \url{https://arxiv.org/abs/1209.5628}.

\bibitem{MR1317233}
Y.~Zhu, ``Modular invariance of characters of vertex operator algebras,''
  \href{http://dx.doi.org/10.1090/S0894-0347-96-00182-8}{{\em J. Amer. Math.
  Soc.} {\bfseries 9} no.~1, (1996) 237--302}.
  \url{https://doi.org/10.1090/S0894-0347-96-00182-8}.

\bibitem{MR808396}
K.~Chandrasekharan, \href{http://dx.doi.org/10.1007/978-3-642-52244-4}{{\em
  Elliptic functions}}, vol.~281 of {\em Grundlehren der mathematischen
  Wissenschaften [Fundamental Principles of Mathematical Sciences]}.
\newblock Springer-Verlag, Berlin, 1985.
\newblock \url{https://doi.org/10.1007/978-3-642-52244-4}.

\bibitem{MR890960}
S.~Lang, \href{http://dx.doi.org/10.1007/978-1-4612-4752-4}{{\em Elliptic
  functions}}, vol.~112 of {\em Graduate Texts in Mathematics}.
\newblock Springer-Verlag, New York, second~ed., 1987.
\newblock \url{https://doi.org/10.1007/978-1-4612-4752-4}.
\newblock With an appendix by J. Tate.

\bibitem{MR981372}
F.~Hirzebruch, ``Elliptic genera of level {$N$} for complex manifolds,'' in
  {\em Differential geometrical methods in theoretical physics ({C}omo, 1987)},
  vol.~250 of {\em NATO Adv. Sci. Inst. Ser. C: Math. Phys. Sci.}, pp.~37--63.
\newblock Kluwer Acad. Publ., Dordrecht, 1988.

\bibitem{MR970278}
P.~S. Landweber, ed., \href{http://dx.doi.org/10.1007/BFb0078035}{{\em Elliptic
  curves and modular forms in algebraic topology}}, vol.~1326 of {\em Lecture
  Notes in Mathematics}.
\newblock Springer-Verlag, Berlin, 1988.
\newblock \url{https://doi.org/10.1007/BFb0078035}.

\bibitem{MR1189136}
F.~Hirzebruch, T.~Berger, and R.~Jung,
  \href{http://dx.doi.org/10.1007/978-3-663-14045-0}{{\em Manifolds and modular
  forms}}.
\newblock Aspects of Mathematics, E20. Friedr. Vieweg \& Sohn, Braunschweig,
  1992.
\newblock \url{https://doi.org/10.1007/978-3-663-14045-0}.
\newblock With appendices by Nils-Peter Skoruppa and by Paul Baum.

\bibitem{MR1757003}
L.~A. Borisov and A.~Libgober, ``Elliptic genera of toric varieties and
  applications to mirror symmetry,''
  \href{http://dx.doi.org/10.1007/s002220000058}{{\em Invent. Math.} {\bfseries
  140} no.~2, (2000) 453--485}. \url{https://doi.org/10.1007/s002220000058}.

\bibitem{MR1765709}
B.~Totaro, ``Chern numbers for singular varieties and elliptic homology,''
  \href{http://dx.doi.org/10.2307/121047}{{\em Ann. of Math. (2)} {\bfseries
  151} no.~2, (2000) 757--791}. \url{https://doi.org/10.2307/121047}.

\end{thebibliography}\endgroup

\end{document}